\documentclass{aa}  

\usepackage{graphicx}
\usepackage{txfonts}
\usepackage{graphicx}
\usepackage{adjustbox}
\usepackage{amsmath}
\usepackage{color}
\usepackage[normalem]{ulem}
\usepackage{booktabs}
\usepackage{multirow} 
\usepackage[flushleft]{threeparttable}
\usepackage[colorlinks=true,linkcolor=blue, allcolors=blue]{hyperref}
\usepackage{enumitem}
\usepackage{caption}
\usepackage{subcaption}

\begin{document}

    \title{Compact stellar systems hosting an intermediate mass black hole:\\ magnetohydrodynamic study of inflow-outflow dynamics}

    \titlerunning{Compact stellar systems hosting an intermediate mass black hole}
    
    \author{Mat\'{u}\v{s} Labaj
          \inst{1}         
          \and
          Sean M. Ressler \inst{2}
          \and
          Michal Zaja\v{c}ek \inst{1}
          \and
          Tomáš Plšek \inst{1}
          \and 
          Bart Ripperda \inst{2,3,4,5}
          \and
          Florian Pei{\ss}ker \inst{6}          
          }

    \institute{Department of Theoretical Physics and Astrophysics, Masaryk University, Kotlářska 2, 611 37 Brno, Czech Republic
    \and
        Canadian Institute for Theoretical Astrophysics, University of Toronto, 60 St. George St, Toronto, ON M5S 3H8, Canada
    \and
        Department of Physics, University of Toronto, 60 St. George St, Toronto, ON M5S 1A7, Canada
    \and
        David A. Dunlap Department of Astronomy \& Astrophysics, University of Toronto, Toronto, ON M5S 3H4, Canada
    \and
        Perimeter Institute for Theoretical Physics, 31 Caroline St. North, Waterloo, ON N2L 2Y5, Canada
    \and
        I.Physikalisches Institut der Universität zu Köln, Zülpicher Str. 77, D-50937 Köln, Germany
    \\
    \email{labaj@physics.muni.cz}
            }

    \date{Received 1 April 2025; accepted 24 August 2025}

  \abstract
  {Intermediate-mass black holes (IMBHs) have remained elusive in observations, with only a handful of tentative detections. One promising location for the existence of IMBHs is in dense stellar clusters such as IRS 13E in the Galactic Center. Such systems are thought to be fed by stellar winds from massive Wolf--Rayet (WR) stars. Understanding the accretion dynamics in wind-fed IMBHs in such clusters is important for predicting observational signatures. These systems are, however, relatively unexplored theoretically. Understanding the interplay between IMBHs and the surrounding stars—especially through the high-velocity, dense stellar winds of WR stars—is essential for clarifying IMBH dynamics within such environments and explaining why these objects continue to evade unambiguous detection.}
   {Inspired by the IRS 13E stellar association near the Galactic center, we examine how high wind velocities, magnetic fields, and metallicity-dependent radiative cooling influence the fraction of stellar wind material captured by the black hole, the formation and survival of dense clumps, and the resulting high-energy emission. We also compare isotropic and disk-like stellar distributions to see how the flow structure and IMBH detectability may vary.}
   {We conduct three-dimensional magnetohydrodynamic and hydrodynamic simulations in which each WR star represents a source term in mass, momentum, energy, and magnetic field. A metallicity-dependent cooling function accounts for radiative energy losses. By varying the cluster's geometry, magnetization, and chemical abundance, we characterize the resulting flow structures, accretion rates, and X-ray luminosity corresponding to the cluster and the IMBH.}
   {In all configurations, the accretion rate onto the IMBH is up to five orders of magnitude lower than the total mass-loss rate from the cluster's WR stars. High-velocity wind–wind collisions generate turbulent, shock-heated outflows that expel most injected gas. While enhanced cooling in high-metallicity runs fosters dense clump formation, these clumps typically do not reach the black hole. The integrated X-ray emission is dominated by colliding stellar winds, rendering the IMBH's radiative signature elusive. Intermittent, quasi-periodic variations in inflow rates are driven by close stellar passages, leading to short-lived accretion enhancements or flares that nonetheless remain difficult to detect against the dominant wind emission.}
   {Despite continuous mass injection from dense stellar winds, these compact systems exhibit outflow-dominated flows and weak, though strongly variable black-hole accretion, naturally explaining the low detectability of IMBHs in such settings. The observed quasi-periodic or flaring accretion episodes are overwhelmed by the luminous shock-heated winds, making unambiguous observational identification of the IMBH challenging with current X-ray instruments. Nevertheless, our results provide a framework for interpreting current data and developing future observational strategies to unveil IMBHs in dense stellar systems.}


   \maketitle

\section{Introduction}

Intermediate-mass black holes (IMBHs) with masses in the range $M_{\bullet}\sim 10^2-10^5$ solar masses ($M_{\odot}$) have largely eluded detection, leaving a noticeable gap in the mass distribution of black holes \citep[see][for a review]{2020ARA&A..58..257G}. At the same time, comprehending their formation, evolution, and occurrence across cosmic history is crucial for understanding black-hole growth and galaxy evolution. So far, there have been a few confirmed detections towards the upper IMBH mass range in lower-mass galaxies \citep[e.g., NGC 4395 with $M_{\bullet}\sim 10^4-10^5\,M_{\odot}$;][]{2003ApJ...588L..13F,2005ApJ...632..799P,2024ApJ...976..116P}. Toward the lower mass range, there was the first gravitational wave detection of the formation of a $\sim 140\,M_{\odot}$ black hole \citep[GW190521;][]{2020ApJ...900L..13A}, which falls in the pair-instability mass gap \citep{2021ApJ...912L..31W}. The most massive of the gravitational-wave merger products has been a $\sim 225\,M_{\odot}$ IMBH \citep[GW231123;][]{2025arXiv250708219T}, which formed the merger of a $\sim 100\,M_{\odot}$ and $\sim 140\,M_{\odot}$ black hole, both high-spinning. This finding supports the hierarchical merger channel for producing black holes of several $\sim 100\,M_{\odot}$.   

There are several ways that an IMBH could form, depending on the cosmic time at formation and the surrounding environment, both of which influence its subsequent growth. Three formation channels can be distinguished: (i) primordial/cosmological origin due to the collapse of population III stars \citep{2001ApJ...551L..27M} or a direct gas cloud collapse \citep{1994ApJ...432...52L,1995ApJ...443...11E,2006MNRAS.370..289B,2013MNRAS.436.2989L}; (ii) consecutive mergers of stellar-mass black holes in dense stellar systems, such as globular clusters \citep{2002MNRAS.330..232C,2004ApJ...616..221G,2024Sci...384.1488F}; and (iii) runaway collisions and mergers of massive stars in dense stellar clusters that subsequently collapse into an IMBH \citep{2002ApJ...576..899P}. Despite the uncertainties, several theoretical models support (iii), in particular for globular clusters \citep{2002MNRAS.330..232C,2002ApJ...576..899P,2004ApJ...616..221G,2006MNRAS.368..141F}. Once a sufficiently massive IMBH seed forms \citep[$>50\,M_{\odot}$;][]{2002MNRAS.330..232C}, it is expected to be retained within the cluster and undergo only a low-amplitude Brownian-like motion \citep{2002PhRvL..88l1103C,2002ApJ...572..371C}.

The scenarios above imply a plausible association of IMBHs with dense stellar systems at the time of their formation. For a galaxy as a whole, the central supermassive black hole mass is known to correlate with the stellar mass contained within the spheroidal component of that galaxy \citep[the $M_{\bullet}-M_{\rm spheroid}$ or \textit{Magorrian} relation;][]{1998AJ....115.2285M,2003ApJ...589L..21M} as well as with the stellar velocity dispersion within the spheroid \citep[$M_{\bullet}-\sigma_{\star}$ relation;][]{2000ApJ...539L...9F,2000ApJ...539L..13G}. 
If we assume that a similar relation holds for relaxed stellar systems with $M_{\rm spheroid} \sim 10^6 - 10^7 \, M_{\odot}$ (particularly globular clusters) with a typical mass ratio $M_{\bullet}/M_{\rm spheroid} \sim 2 \times 10^{-3}$ \citep{2003ApJ...589L..21M}, then these stellar systems would be expected to host an IMBH of mass $M_{\bullet} \sim 10^3 - 10^4 \, M_{\odot}$. 

Even granting those assumptions, however, both the occupancy fraction and mass ratio for clusters are uncertain \citep[see e.g.][]{2019MNRAS.484.2974S} due to uncertainties in the initial cluster concentration \citep{2015MNRAS.454.3150G}, which influences whether the hypothetical IMBH formation proceeds rapidly or more gradually. Nevertheless, several pieces of observational evidence support that some massive stellar clusters host an IMBH. For instance, the observed off-nuclear X-ray sources and transients in galactic halos with peak X-ray luminosities of $\sim 10^{43}\,{\rm erg\,s^{-1}}$ are consistent with the emission of IMBHs (with masses on the order of $\sim 10^4\,M_{\odot}-10^5\,M_{\odot}$) powered by the tidal disruption of stars, which supports their association with lower-mass stellar systems like stellar clusters or dwarf galaxies \citep{2018NatAs...2..656L,2025arXiv250109580J,2025NatAs...Zhang}. There is strong evidence that at least two globular clusters in the Milky Way host IMBHs based on dynamical grounds: $\omega$ Centauri \citep{2024Natur.631..285H} and 47 Tucanae (NGC 104, \citealt{2017Natur.542..203K}), with IMBH masses of $\gtrsim 8200\,M_{\odot}$ and $2300_{-850}^{+1500}\,M_{\odot}$, respectively, though several studies point out potential caveats and systematic effects that can influence the central mass constraints \citep{2017MNRAS.468.4429Z,2019MNRAS.482.4713Z,2025A&A...693A.104B}.  

Like the Galactic Center in our own galaxy, Galactic nuclei are especially promising regions for detecting IMBHs. 
Due to the short dynamical relaxation timescale, as many as $\gtrsim 50(M_{\bullet}/150\,M_{\odot})^{1/2}$ IMBHs could be concentrated towards the supermassive black hole (SMBH, \citealt{2001ApJ...551L..27M}).
These black holes would have been formed primordially or in globular clusters before undergoing a rapid inspiral caused by dynamical friction.
The infall rate is expected to decrease with cosmic time as $\propto t^{-1/2}$ and the current rate is estimated to be $\sim 2$ IMBHs per gigayear per galaxy \citep{2001ApJ...551L..27M}.

In addition, IMBHs up to $\lesssim 10^4\,M_{\odot}$ can also form in situ within the dense nuclear star clusters (NSCs) surrounding SMBHs \citep{2020A&ARv..28....4N} via black hole--star collisions \citep{2022ApJ...929L..22R,2025arXiv250315598H} and black hole-black hole mergers \citep{2022ApJ...927..231F}. Such IMBHs may be detected indirectly when they perturb the inner accretion flow, causing quasiperiodic accretion and outflow phenomena \citep{2021ApJ...917...43S,2024SciA...10J8898P,2024arXiv241012090Z}, through spin-orbit coupling-induced jet precession \citep{2023ApJ...951..106B,2023A&A...672L...5V,Ressler2024}, or occasionally as X-ray sources when they disrupt a star \citep{2018NatAs...2..656L,2025arXiv250109580J} or encounter a denser molecular cloud \citep{2022MNRAS.515.2110S,2024MNRAS.527.1062H}. Furthermore, recent observations suggest that some high-velocity compact clouds (HVCCs) in the Galactic center, such as CO–0.40–0.22, may be influenced by gravitational interactions with inactive IMBHs, highlighting their potential role in the dynamics of molecular clouds \citep{2017ApJ...843L..11T}. The existence of a past SMBH-IMBH binary in the Galactic center $\gtrsim 10$ million years ago is also supported by the distribution of distances and velocities of hypervelocity stars in the Galactic halo \citep{2025ApJ...982L..37C}. The timing of the Sgr A*-IMBH binary also agrees with the merger scenario, which can explain the origin of fast-moving S stars. The SMBH-IMBH merger naturally creates an apse-aligned eccentric disk due to the gravitational-wave recoil kick of the SMBH after the merger. The eccentric disk drives the inner stars into highly eccentric, inclined orbits, consistent with the orbital properties of S stars. According to \citet{2025ApJ...987L..27A} Sgr~A* could have merged with the IMBH of $2^{+3}_{-1.2}\times 10^5\,M_{\odot}$ within the last 10 million years.   

In the case of the Galactic center, current IMBHs could also be detected in less massive, dense stellar associations. Two candidate stellar associations have been identified -- IRS 13E \citep{2004A&A...423..155M,2005ApJ...625L.111S,2017ApJ...850L...5T,2019PASJ...71..105T,2023ApJ...956...70P,2024ApJ...970...74P} and IRS 1W \citep{2024ApJ...975..261H} at projected distances from Sgr~A* of $\sim 0.13\,{\rm pc}$ and $\sim 0.24$ pc, respectively -- both of which could host an IMBH of $\sim 10^{4}\,M_{\odot}$ that would make them more stable against tidal disruption by the central supermassive black hole, although some studies tend to disfavor the IMBH presence \citep{2005ApJ...625L.111S,2010ApJ...721..395F,2024A&A...692A.104P} or propose alternative models for the dark mass, such as a dark star cluster of stellar-mass black holes \citep{2011ApJ...741L..12B}. These IMBHs, assuming their presence for this study, are expected to be in the low-luminosity quiescent state characterized by a radiatively inefficient accretion flow (RIAF) with the spectral energy distribution peaking in the mid-infrared domain \citep{2024ApJ...970...74P,2024ApJ...975..261H}. A hot RIAF surrounding massive black holes is characteristic of wind-fed accretion provided by host stellar clusters, such as for the case of Sgr~A* embedded within the dense nuclear star cluster \citep{Ressler_Quataert_Stone_2018, Ressler_Quataert_Stone_2019}. On the other hand, an IMBH could host a dense, optically thick accretion disk if it has recently disrupted a star; such a disk would be revealed by an intense thermal soft X-ray emission with the luminosity reaching a fraction of the Eddington luminosity, 
\begin{equation}
   L_{\rm X}\approx \frac{\lambda_{\rm Edd}}{\kappa_{\rm bol}}L_{\rm Edd}\sim 1.26 \times 10^{41}\,\left(\frac{M}{10^4\,M_{\odot}} \right) \left(\frac{\lambda_{\rm Edd}}{1} \right) \left(\frac{\kappa_{\rm bol}}{10} \right){\rm erg\,s^{-1}}\,,  
\end{equation}
where $L_{\rm Edd}$ is the Eddington luminosity, $\lambda_{\rm Edd}$ is the Eddington ratio (here set to unity), and $\kappa_{\rm bol}$ is the bolometric correction \citep[here scaled to 10;][]{2006ApJS..166..470R,2019MNRAS.488.5185N}.
Detecting IMBHs in this ``high/soft'' state is another way to identify them, though it is expected to be rather transient and thus limited to only a small fraction of IMBHs \citep{2018NatAs...2..656L,2025arXiv250109580J,2025NatAs...Zhang}.

In this paper, we study the scenario of a compact stellar association bound to an IMBH, e.g.,  a stellar cluster inspiralling towards the Galactic center \citep{2003ApJ...593L..77H}. Such a cluster undergoes a continuous tidal disruption as it approaches the SMBH due to dynamical friction and may be representative of  IRS 13E, a concentration of a handful of massive early-type stars close to Sgr~A* \citep{2023ApJ...956...70P,2024ApJ...970...74P}. The constraints on the IMBH mass and IRS 13E distance from the SMBH limit the size of the stable region against tidal disruption. This size can be estimated by the tidal (Hill) radius, $r_{\rm H}\simeq d_{\rm IRS13E} (M_{\bullet}/3M_{\rm SgrA*})^{1/3}\sim 17.6 \times 10^{-3}\,{\rm pc} (d_{\rm IRS13E}/0.13\,{\rm pc}) (M_{\bullet}/3\times 10^4\,M_{\odot})^{1/3} (M_{\rm SgrA*}/4\times 10^6\,M_{\odot})^{-1/3}$, which is comparable to the projected radii of six early-type stars from the center of the IRS 13E cluster \citep[associated with the E3 source;][]{2023ApJ...956...70P}. Despite a small number of stars, the region has a very high mean stellar density of $\overline{n}_{\star}\sim 3 \times 10^5\,{\rm pc^{-3}}$, implying a mean mutual distance among stars of only $\sim 3000$ au, which is expected to result in ongoing wind-wind collisions, leading to shocks and production of hot, X-ray emitting gas. Therefore, such a stellar system motivates a detailed exploration of the observability of the hypothetical, embedded IMBH and general properties of the cluster inflow-outflow dynamics. 

A cluster bound by an IMBH in the central parsec of the galaxy would have a lifetime $\lesssim$ $100{,}000$ years due to a short dynamical friction timescale \citep{2024ApJ...975..261H}. An inspiralling cluster is generally unstable and is expected to disrupt with or without the IMBH while interacting with the SMBH and other stars in the NSC \citep{2024A&A...692A.104P}. However, the scenario of a compact group of stars surrounding the IMBH is still highly relevant since the IMBH, on its way towards the Galactic center, can also tidally capture stars from its surroundings \citep{2001ApJ...551L..27M}, even while it is losing some of them. Furthermore, the scenario not only captures the physics of IRS 13E-like associations in the Galactic center but is also related to the very cores of globular clusters where a handful of stars is expected to orbit hypothetical IMBHs on tight orbits \citep{2024Natur.631..285H}.   

The mutual interplay of massive stellar winds has been extensively studied in binary systems using hydrodynamical simulations, where high-density outflows create strong shocks and radiative cooling layers \citep{1992ApJ...386..265S,2008MNRAS.388.1047P,Pittard2009,2011ApJS..194....8P,2011A&A...530A.119P,Lamberts2011,2012A&A...546A..60L,vanMarle2011,Kee2014,2016MNRAS.460.3975H,2020MNRAS.493..447C}. In particular, the wind-wind interface is prone to a thin-shell cooling instability that drives dense clump formation. These clumps can be intermittently stripped or disrupted before reaching the companion star. A special class of binary systems is X-ray binaries, where a wind-blowing star provides the material for accretion by a compact companion. This compact companion then focuses the stellar wind into a gaseous tail due to its gravitational field \citep{2012A&A...542A..42H,2015A&A...575A...5C,2015MNRAS.450.2410C}. 

On far larger scales, wind-fed accretion scenarios have long been proposed to explain the moderate radiative flux observed toward some supermassive black holes, especially Sgr A* in the Galactic Center \citep{Loeb2004,2004ApJ...613..322Q}. Observational data (e.g., \citealt{1995ApJ...447L..95K,2006ApJ...643.1011P, 2009ApJ...697.1741B,Martins2007,Lu2009,2013ApJ...764..154D,YusefZadeh2015}) have identified a sizable population of massive WR and OB-spectral type stars (massive, hot, and luminous stars) within sub-parsec distances of Sgr~A*, each injecting mass and momentum into the hot ambient medium via their high-velocity outflows. Analytical \citep{2004ApJ...613..322Q,Shcherbakov2010} and numerical studies \citep{Rockefeller2004,Cuadra2005,Cuadra2006,Cuadra2008,cuadra2015,Ressler_Quataert_Stone_2018,Ressler_Quataert_Stone_2019a,Ressler_Quataert_Stone_2019,2020ApJ...888L...2C,2020ApJ...896L...6R,2022ApJ...932L..21M,2023ApJ...953...22S,2023MNRAS.521.4277R,2025A&A...693A.180C} have shown that these stellar winds, when tracked in three-dimensional (3D) hydrodynamic (HD) or magnetohydrodynamic (MHD) detail, can provide a viable, quasi-steady source of gas for low-luminosity accretion—successfully reproducing the broad X-ray and radio flux levels of Sgr~A*. 
While the stellar winds in the Galactic Center do not typically pass close enough to each other to form wind-wind shocks subject to the thin-shell cooling instability seen in WR binary simulations (e.g., \citealt{Calderon2016}), several simulations have still shown significant clump formation (e.g., \citealt{Russell2017,2020ApJ...888L...2C,2025A&A...693A.180C}) through the collision of winds with the collective material provided by all the winds. Depending on the composition of the gas, these clumps can collect into a cool accretion disk, increase the accretion rate onto the black hole, and have important consequences for the global emission properties observed in the Galactic center environment (e.g., the cool disk observed by ALMA, \citealt{Murchikova2019}).

Numerical modeling seldom targets compact clusters with an embedded IMBH between these two regimes. Compared to binary systems, these clusters feature overlapping wind–wind collision fronts and more complex orbital configurations. Nevertheless, their sub-pc extent and smaller black-hole masses clearly set them apart from the parsec-scale environment of Sgr~A*. Investigating this regime fills a key gap, determining whether wind-fed accretion processes remain similarly inefficient and clumpy and whether the accretion onto a possible IMBH candidate might be detectable above the emission from the colliding stellar winds.

This manuscript is organized as follows. In Sect.~\ref{sec_methods} we describe the numerical scheme applied to study the HD and MHD of interacting stellar winds around the IMBH, including the radiative cooling function, chemical abundances, and the detailed simulation setup. Subsequently, in Sect.~\ref{sec_results}, we present the main results of 3D MHD and HD simulations, including plane-parallel cuts of different simulation runs and their associated radial profiles of fluid quantities. Sect.~\ref{sec:obs} follows, where we analyze accretion flow variability and periodicities, and compare the predicted X-ray surface-brightness images with observations of IRS 13E, a candidate stellar system potentially hosting an IMBH close to the Galactic center. We put the presented simulation results in the context of previously published models of wind-fed accretion and discuss IMBH accretion regimes in Sect.~\ref{sec_discussion}. We also discuss the observability of IMBHs fed by stellar winds within their host clusters. Finally, we conclude with Sect.~\ref{sec_conclusions}.

\section{Methods}
\label{sec_methods}
We conduct our simulations with \textsc{athena++}, a 3D grid-based scheme that solves the equations of conservative MHD \citep{Stone_2020}. Our simulations are based on the HD stellar wind model described in \citet{Ressler_Quataert_Stone_2018} and extended in \citet{Ressler_Quataert_Stone_2019a,Ressler_Quataert_Stone_2019} to include magnetized winds. The stellar winds are treated as source terms in mass, momentum, energy, and magnetic field moving on fixed Keplerian orbits. The hydrodynamic properties of the stellar winds characterized their mass-loss rates, $\dot{M_\mathrm{w}}$, and terminal stellar-wind velocities, $v_\mathrm{w}$. Furthermore, we include magnetization of the winds in some of our simulations. The magnetic fields are solely toroidal with respect to the spin axes of the stars (unknown but chosen randomly), and their magnitude is expressed through the $\beta_\mathrm{w}$ parameter, defined as the ratio between the wind ram pressure and its magnetic pressure at the equator. In the general scenario involving magnetized winds, the simulation solves the following set of equations:
\begin{align*}
    \frac{\partial \rho}{\partial t} + \nabla \cdot (\rho \mathbf{v}) &= f \dot{\rho}_\mathrm{w} \\
    \frac{\partial (\rho \mathbf{v})}{\partial t} + \nabla \cdot \bigg(P_\mathrm{tot}\mathbf{I} + \rho \mathbf{v}\mathbf{v} - \mathbf{B}\mathbf{B} \bigg) &= -\frac{\rho GM_{\mathrm{BH}}}{r^2} \hat{\mathbf{r}} \\
    \frac{\partial E}{\partial t} + \nabla \cdot [(E+P_\mathrm{tot})\mathbf{v}-\mathbf{v}\cdot\mathbf{B}\mathbf{B}] &= -\frac{\rho GM_{\mathrm{BH}}}{r}\mathbf{v}\cdot \hat{\mathbf{r}} + \langle \dot{E}_\mathrm{B} \rangle \\
    &\quad +\frac{1}{2}f\dot{\rho}_\mathrm{w}\langle | \mathbf{v}_\mathrm{w,net} |^2 \rangle - q_{-} \\
    \frac{\partial \mathbf{B}}{\partial t} - \nabla \times (\mathbf{v} \times \mathbf{B})  &= \nabla \times (\mathbf{\Tilde{E}}_\mathrm{w}),
\end{align*}
where $\rho$ is the mass density, $\mathbf{v}$ is the velocity vector, $P_\mathrm{tot}=P+B^2/2$ is the total pressure with both thermal and magnetic contributions, $\mathbf{I}$ and $\hat{\mathbf{r}}$ represent the unit matrix and the unit vector respectively, $\mathbf{B}$ is the magnetic field vector, $E = 1/2\rho v^2 + P/(\gamma - 1) + B^2/2$ represents the total energy with $\gamma = 5/3$ being the non--relativistic adiabatic index of the gas, $q_-$ is the cooling rate per unit volume due to radiative losses, $f$ is the fraction of the cell by volume contained in the wind, $\dot{\rho}_\mathrm{w}=\dot{M}_\mathrm{w}/V_\mathrm{w}$ with $V_\mathrm{w} = 4/3(\pi r_{\rm w}^3)$, $\mathbf{v}_\mathrm{w, net}$ is the wind speed in the fixed frame of the grid, $\langle \rangle$ denotes a volume average of a quantity over the cell, $\dot{E}_\mathrm{B}$ is the magnetic energy source term generated by the winds, $\mathbf{\Tilde{E}}_\mathrm{w}$ is the average of the wind source electric field $\mathbf{E}_\mathrm{w}$ over the appropriate cell edge. In the simulation domain, each `wind' has a radius $r_\mathrm{w} \approx 2\sqrt{3}\Delta x$, where $\Delta x$ is the edge length of the cell which contains the center of the star. 
Note that here and throughout we use Lorentz-Heaviside units so that a factor of $\sqrt{4 \mathrm{\pi}}$ has been absorbed into the definition of $\mathbf{B}$ and the magnetic pressure is $P_{B} = B^2/2$.  
\subsection{Cooling function and chemical abundance models} \label{subsec:cooling}
Due to the proximity of the stellar winds, they interact and shock-heat, resulting in significant radiative losses due to optically thin bremsstrahlung and line cooling. To calculate the cooling function, $\Lambda$, for arbitrary hydrogen, helium, and metal mass fractions ($X$, $Y$, $Z$ respectively), we first determine the cooling curve for the solar abundances $X_\odot = 0.7491,\, Y_\odot = 0.2377,\, Z_\odot =  0.0133$, presented in \citet{Lodders_2003}. The calculation is done using the spectral analysis code \textsc{spex} \citep{kaastra1996uv}, following the methodology outlined in \citet{2009A&A...508..751S} to calculate the separate contributions of individual elements to the cooling function, 
\begin{equation}
    \Lambda = \frac{X}{X_\odot}\Lambda_{\mathrm{H},\odot} + \frac{Y}{Y_\odot}\Lambda_{\mathrm{He},\odot} + \frac{Z}{Z_\odot}\Lambda_{\mathrm{Z},\odot}\,.
\end{equation}
The mean molecular weight per electron, $\mu_{\rm e}$, and the mean molecular weight per particle, $\mu$, depend on $X$ and $Z$ as described in \citet{2009ApJS..181..391T}
\begin{align}
    \begin{split}
        \mu_{\rm e} &= \frac{2}{1+X} \\
        \mu &= \frac{1}{2X+3(1-X-Z)/4+Z/2},
    \end{split}
\end{align}
Here, we assume that oxygen contributes most of the mass in metals and treat both $\mu_{\rm e}$ and $\mu$ as constants. The cooling function $\Lambda$ is implemented by calculating the cooling rate per unit volume using a piece-wise power law approximation over a range of temperatures from $10^4$ to $10^9$ K. For additional details on the computation and implementation of the cooling function in the simulations, refer to Sect. 2.2 in \citet{Ressler_Quataert_Stone_2018}. 

The real chemical composition of WR stars in compact stellar clusters remains uncertain. We, therefore, use three different chemical abundance models of WR stars used previously in Galactic center simulations \citep{Cuadra2008, Calderon2016, Russell2017}. First, we consider hydrogen-depleted stars ($X=0$) with higher metallicity ($Z = 3Z_\odot$), referred to as Model I in Table \ref{tab:abundance}. This model motivated the fact that WR stars typically lose their outer hydrogen later in previous stages of stellar evolution. Additionally, we explore two alternative chemical abundance models: one with a higher hydrogen content (Model II), which aligns with a realistic expectation that the stars are not completely starved of hydrogen, and another with significantly elevated metallicity (Model III), to investigate effects such as enhanced cooling efficiency and the potential formation of cool clumps due to stellar wind interactions. The accretion flow in Galactic center simulations can be quite sensitive to these mass fractions, as was recently shown by \citet{2025A&A...693A.180C}.

\subsection{Simulation Setup}
We assume a compact stellar association consisting of six Wolf--Rayet (WR) stars and a central IMBH, numerically realized as a system of moving source terms in mass, momentum, energy, and magnetic field in the point source Newtonian gravitational potential of the central black hole $GM_\bullet/r$ ($G $ being the gravitational constant). Magnetic fields of the stars are parametrized using the ratio between the stellar wind ram pressure and magnetic pressure $\beta_\mathrm{w} \equiv 2\rho v_\mathrm{w}^2/B^2$, evaluated in the equatorial plane of each star. We choose $\beta_{\rm w} = 100$, so the magnetic field is modest but non-negligible. 

Our setup is loosely based on the stellar association IRS 13E within the Milky Way nuclear stellar cluster near Sgr A*. The six WR stars in our simulation are placed at fixed distances from the central black hole, matching the observed projected distances of these stars from the hypothetical IMBH with an approximate mass of $\sim 3 \times 10^4 M_\odot$, associated with the IRS 13E3 source \citep[see][for more details]{2024ApJ...970...74P}. That said, we use IRS 13E primarily as a reference for the stellar properties, taking a more flexible approach to other key properties, such as the spatial distribution of the stars, their chemical abundance models, and the inclusion of wind magnetization. 

\begin{table}
    \centering
    \begin{threeparttable}
    \caption{Stellar chemical abundance models used in our simulations.}
    \label{tab:abundance}
    \begin{tabular}{@{} l *4c @{}}
        \toprule \midrule
        Model & ID & $X$  & $Y$  & $Z$  \\ 
        \midrule
        I & $Z = 3Z_\odot$ & 0 & 0.9601 & 0.039 \\
        II & WN8-9 & 0.115 & 0.824 & 0.061 \\
        III & WC8-9 & 0 & 0.6 & 0.4 \\
        \bottomrule
    \end{tabular} 
    \begin{tablenotes}
      \small
      \item \textit{Notes.} WN8-9: \citet{Herald2001}; WC8-9: \citet{2007ARA&A..45..177C}.
    \end{tablenotes}
    \end{threeparttable}
\end{table}

\begin{table}
    \centering
    \begin{threeparttable}
    \caption{Overview of the simulations presented in the work.}
    \label{tab:simulations}
    \begin{tabular}{@{} l c c c c c @{}}
        \toprule \midrule
        Sim. & Chem. model & Stellar distr. & $\beta_\mathrm{w}$ & Res. & SMR \\
        \midrule
        1 & I   & sph.\ A & $\infty$ & $128^3$ & 5 \\
        2 & II  & sph.\ B & $\infty$ & $256^3$ & 5 \\
        3 & II  & sph.\ B & 100      & $256^3$ & 5 \\
        4 & III & sph.\ A & $\infty$ & $256^3$ & 5 \\
        5 & I   & disk    & $\infty$ & $128^3$ & 5 \\
        6 & III & disk    & $\infty$ & $256^3$ & 5 \\
        \bottomrule
    \end{tabular}
    \begin{tablenotes}
      \small
      \item \textit{Notes.} A and B in the spherical stellar distributions denote different random realizations.
    \end{tablenotes}
    \end{threeparttable}
\end{table}

\begin{figure*}[t!]
    \begin{subfigure}[b]{1\columnwidth}
    \centering
    \includegraphics[width=\columnwidth]{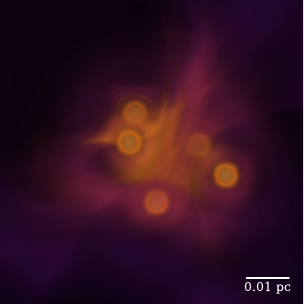}
    \end{subfigure}
    \hfill
    \begin{subfigure}[b]{1\columnwidth}
        \centering
    \includegraphics[width=\columnwidth]{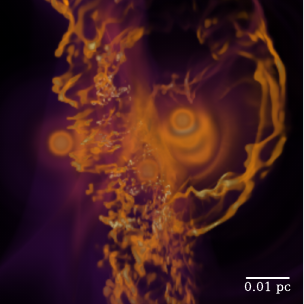}
    \end{subfigure}
    \caption{3D volume rendering of the density of the simulated stellar cluster. \textit{Left:} Colliding stellar winds 620 years into the Model II spherical simulation. \textit{Right:} Outbursts of cool clumps 815 years into the Model III spherical simulation, driven by the gas’s higher metal content, which enhances cooling efficiency and amplifies the development of the thin-shell instability.}
    \label{fig:snap}
\end{figure*}

\begin{table}
    \centering
    \begin{threeparttable}
    \caption{Orbital parameters of the simulated stars.}
    \label{tab:distances}
    \begin{tabular}{lccc}
        \toprule \midrule
        Source ID & $r$ [$10^{-3}$pc] & $v$ [km\,s$^{-1}$] & $P$ [yr] \\
        \midrule
        E1   & 10.52 & 110.75 & 584 \\
        E2   & 8.35  & 124.31 & 413 \\
        E4   & 4.06  & 178.27 & 140 \\
        E5.1 & 9.20  & 118.43 & 477 \\
        E5.2 & 9.96  & 113.82 & 538 \\
        E7   & 18.37 & 83.81  & 1347 \\
        \bottomrule
    \end{tabular}
    \begin{tablenotes}
        \small
        \item \textit{Notes.} Setup inspired by the IRS~13E association; $r$ from projected separations to the putative IMBH (IRS~13E3). Source IDs from \citet{2023ApJ...956...70P}.
    \end{tablenotes}
    \end{threeparttable}
\end{table}

Table \ref{tab:simulations} provides a summary of the simulation runs analyzed in this study. In our simulation, we assume that all the stars are identical WR stars, each characterized by a wind mass-loss rate of $\dot{M}_\text{w} = 10^{-5}\,M_\odot\,\mathrm{yr}^{-1}$ and a stellar wind speed of $v_\mathrm{w}~=~1000\,\mathrm{km\,s}^{-1}$, which are typical for the WR stars in the Galactic center region \citep{1995ApJ...447L..95K,2005A&A...443..163M,2020MNRAS.492.2481W}. These winds are moving on fixed Keplerian orbits around the center of the simulation domain. The stars are on circular orbits, with the orbital radius taken from projected distances (as listed in Table \ref{tab:distances}) and initial positions along the orbits generated randomly. We randomize the positions for a spherically isotropic stellar distribution entirely, except for the fixed orbital distance. In contrast, for a disk-like distribution, the inclination angles of the stellar orbits are constrained to $i = \pm 15^\circ$.

The setup summarized in Table~\ref{tab:distances} indicates that for the mean stellar distance of $r\sim 10\,{\rm mpc}$, the orbital velocity is $v=(GM_{\bullet}/r)^{1/2}\simeq 113.6(M_{\bullet}/3\times 10^4\,M_{\odot})^{1/2} (r/10\,{\rm mpc})^{-1/2}\,{\rm km\,s^{-1}}$, implying a stellar-wind velocity to orbital velocity ratio of $v_{\rm w}/v\sim 8.8$. For the stars bound to Sgr~A*, the orbital velocities within the clockwise disc are of the order of $v\sim 415\,(M_{\bullet}/4\times 10^6\,M_{\odot})^{1/2} (r/0.1\,{\rm pc})^{-1/2}\,{\rm km\,s^{-1}}$ and the ratio is $v_{\rm w}/v\sim 2.4$, which is $\sim 4$ times smaller than for the case of adopted orbits around the IMBH. The difference in $v_{\rm w}/v$ clearly motivates the exploration of the wind-fed accretion regime in compact stellar systems around the IMBH. 

Another motivation is the much higher mean stellar density than young massive stars around Sgr~A*. Around Sgr~A*, there are about $\sim 200$ young massive stars within $r\lesssim 0.5\,{\rm pc}$ \citep{2022ApJ...932L...6V}, which implies a mean stellar density of $n_{\star}\sim 382\,{\rm pc^{-3}}$ and a mean stellar distance of $d_{\star}\sim 2.8\times 10^4\,{\rm AU}$. In the stellar system analogous to IRS 13E that potentially hosts an IMBH, the mean stellar density within $20\,{\rm mpc}$ is $n_{\star}\sim 1.8\times 10^5\,{\rm pc^{-3}}$, which is three orders of magnitude larger than around Sgr~A*. The mean distance between stars is $d_{\star}\sim 3.7\times 10^3\,{\rm AU}$, one order of magnitude smaller than around Sgr~A*. This implies a much more frequent and vigorous wind-wind interaction in compact stellar systems around the IMBH compared to the Sgr~A* stellar environment. 

\subsection{Computational grid and boundary conditions}
Most simulations are performed on a 3D Cartesian grid with a $256^3$ grid cell resolution, covering a $(0.06 \, \text{pc})^3$ volume. We employ five levels of nested mesh refinement at points $r = r_0/2^n$, where $r_0$ is the box edge length and $n$ indicates the refinement level. This approach effectively doubles the simulation's resolution with each halving of the radial distance. Therefore, the edge length of the smallest grid cell is $\approx 7 \times 10^{-6}$ pc. The global character of the fluid variables is relatively independent of resolution, as we have tested. We ran simulations at $64^3$ and $128^3$ and found qualitatively similar results. However, we found $256^3$ necessary to resolve the cold clumps in some simulations.
For the simulations with the chemical abundance model I, where cold clumps do not form, we use a base resolution of 128$^3$. The inner boundary is equal to twice the edge of the smallest grid cell, $r_\text{in} \approx 1.4 \times 10^{-5}\, \text{pc} \approx 9765\, r_\text{g}$, to ensure proper variable reconstruction for cells bordering the inner boundary. We let the simulations evolve in time up to $t \approx 2$ kyr, or about 1.5 orbital periods for the most distant star, E7. 

The code evolves the conservative variables of mass density, momentum density, and total energy density. Consequently, $\rho$ and $P$ can reach unphysical (i.e., negative) values in regions with low mass. To prevent code failure in such an event, we utilize floors on density and pressure such that if $\rho < \rho_{\rm floor}$, we set $\rho = \rho_{\rm floor}$, and similarly if $P < P_{\rm floor}$ we set $P = P_{\rm floor}$. We adopt the values $\rho_{\rm floor} = 10^{-7}M_\odot\,{\rm pc}^{-3}$ and $P_{\rm floor} = 10^{-10}M_\odot\,{\rm pc}^{-1}{\rm kyr}^{-2}$. Furthermore, we impose a minimum temperature of $10^4$ K, providing an additional, density-dependent floor on pressure. The floors are activated sufficiently rarely and, therefore, do not affect our results.

We set the region within the inner boundary to have a floored density, pressure, and zero velocity. This approach allows the material to fall into the `black hole' while the unphysical boundary affects only a few cells outside of $r_\text{in}$. The outer boundary condition is set to outflow in all directions.

\subsection{Model Approximations}

Our study is the first to investigate wind-fed accretion onto an IMBH embedded in a compact stellar cluster using a 3D MHD framework. Given the lack of prior studies in this specific parameter regime, we carefully selected our model parameters to ensure a physically realistic, yet computationally feasible approach. We aim to isolate the dominant physical mechanisms governing the accretion process while minimizing additional complexities that could obscure fundamental trends. However, certain approximations are necessary to consider when interpreting our results.

First, we assume the stars follow fixed, circular orbits with zero eccentricity. In reality, stellar orbits in a dense cluster are expected to have a range of eccentricities due to dynamical interactions, which could introduce periodic variations in wind densities and collision velocities. 
Over longer timescales, eccentric orbits may enhance variability in the wind structure and introduce episodic increases in the accretion rate as stars approach the pericenter. Additionally, previous studies have shown that when winds originate from stars on eccentric orbits, the trailing portions of the wind can be captured by the black hole, leading to angular momentum accumulation and the possible formation of a rotationally supported accretion disk \citep{Cuadra2008}. By restricting our setup to circular orbits, we may underestimate how wind-fed accretion transitions from a quasi-spherical inflow to a more disk-like structure.
Furthermore, the orbits in IRS 13E are likely to evolve over time.
However, given that our simulations cover only a relatively short timescale of $\sim 2$ kyr, the effects of orbital evolution should be minimal.

Second, the WR stars in our model are not resolved as individual stellar bodies but are instead treated as mass, momentum, energy, and magnetic field source terms that continuously inject their winds into the computational domain. This approach captures the large-scale effects of wind-wind interactions and accretion but does not model the stars' physical presence. Additionally, all six WR stars are assumed to be identical, with the same wind velocity ($v_w \sim 1000$ km s$^{-1}$) and mass-loss rate ($\dot{M}_w \sim 10^{-5} M_{\odot} \,\mathrm{yr}^{-1}$). This symmetry results in a nearly uniform distribution of wind-wind collisions, which may underestimate the diversity in wind interaction structures found in realistic clusters, where stars have varying mass-loss rates, wind velocities, and magnetic field strengths. 

\section{Results}
\label{sec_results}

\subsection{Flow morphology}

\begin{figure*}[h]
    \centering
    \includegraphics[width=\textwidth]{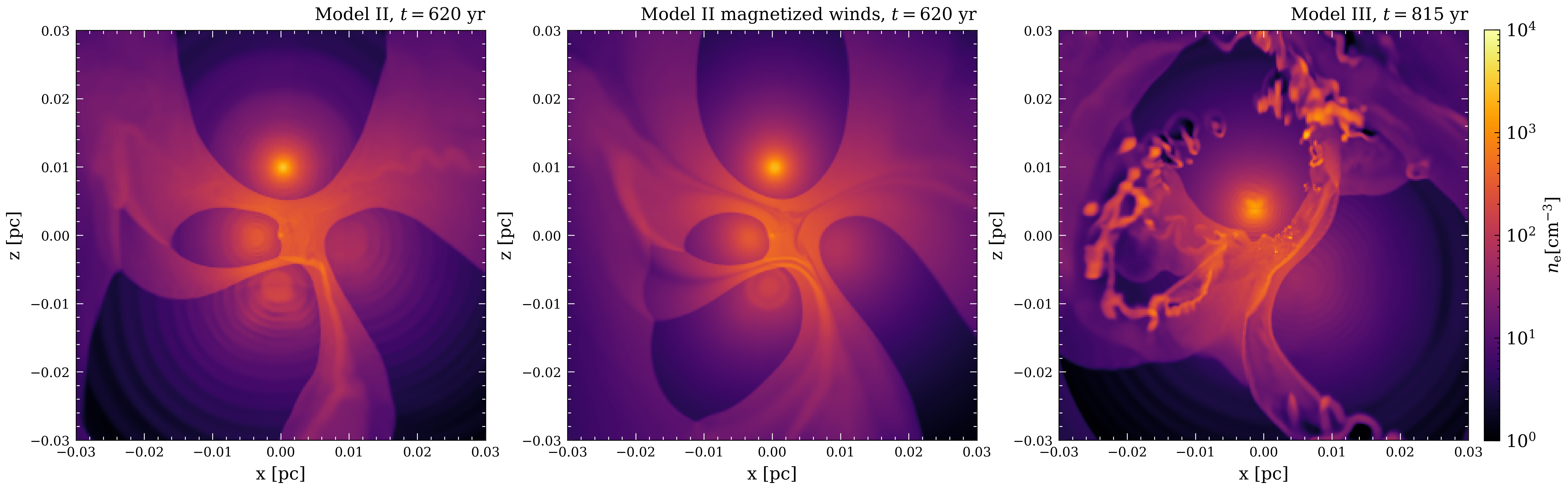}
    \includegraphics[width=1.02\textwidth]{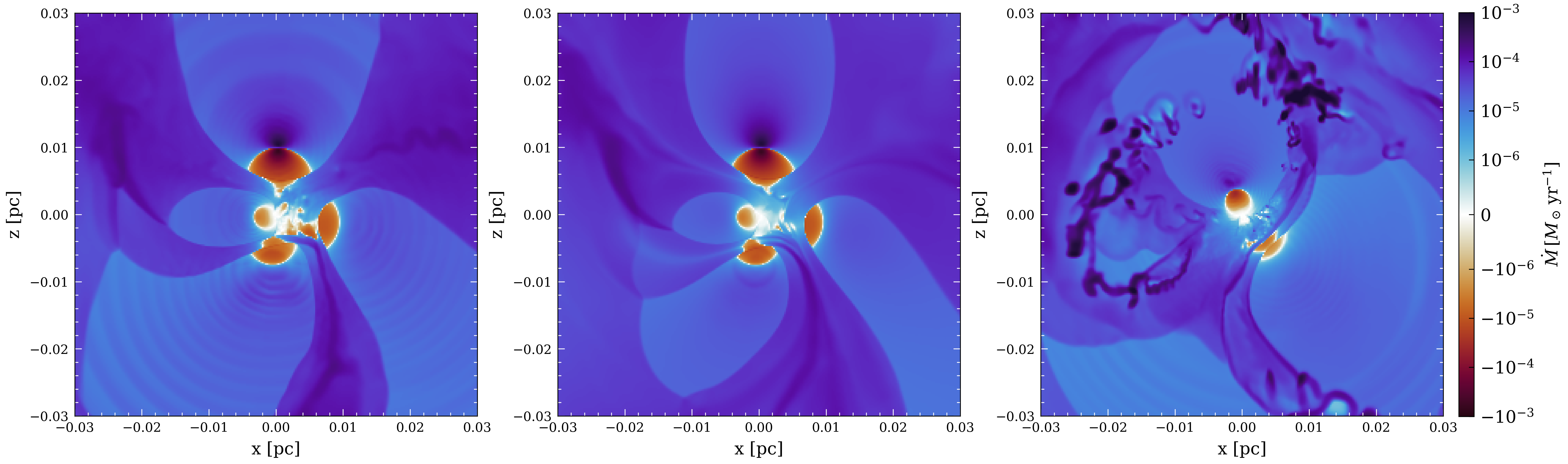}    \includegraphics[width=\textwidth]{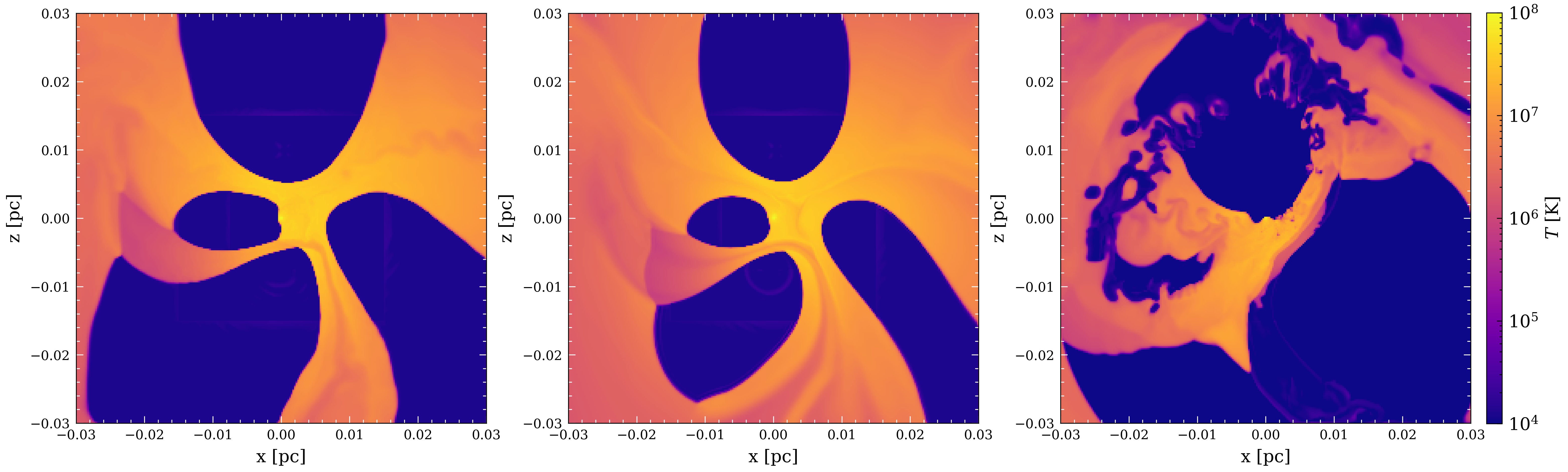}
    \caption{Plane-parallel cuts ($y=0$) of fluid quantities for different models. Left and middle column: Model II without ($\beta_\mathrm{w} = \infty$) and with magnetized winds ($\beta_\mathrm{w} =100$) 620 years into the simulation. Right column: Model III simulation 815 years into the simulation. Top row: electron number density $n_{\rm e}$; middle row: accretion rate $\dot{M}$; bottom row: temperature.}
    \label{fig:slices}
\end{figure*}

\begin{figure*}[h]
    \centering
    \includegraphics[width=1\linewidth]{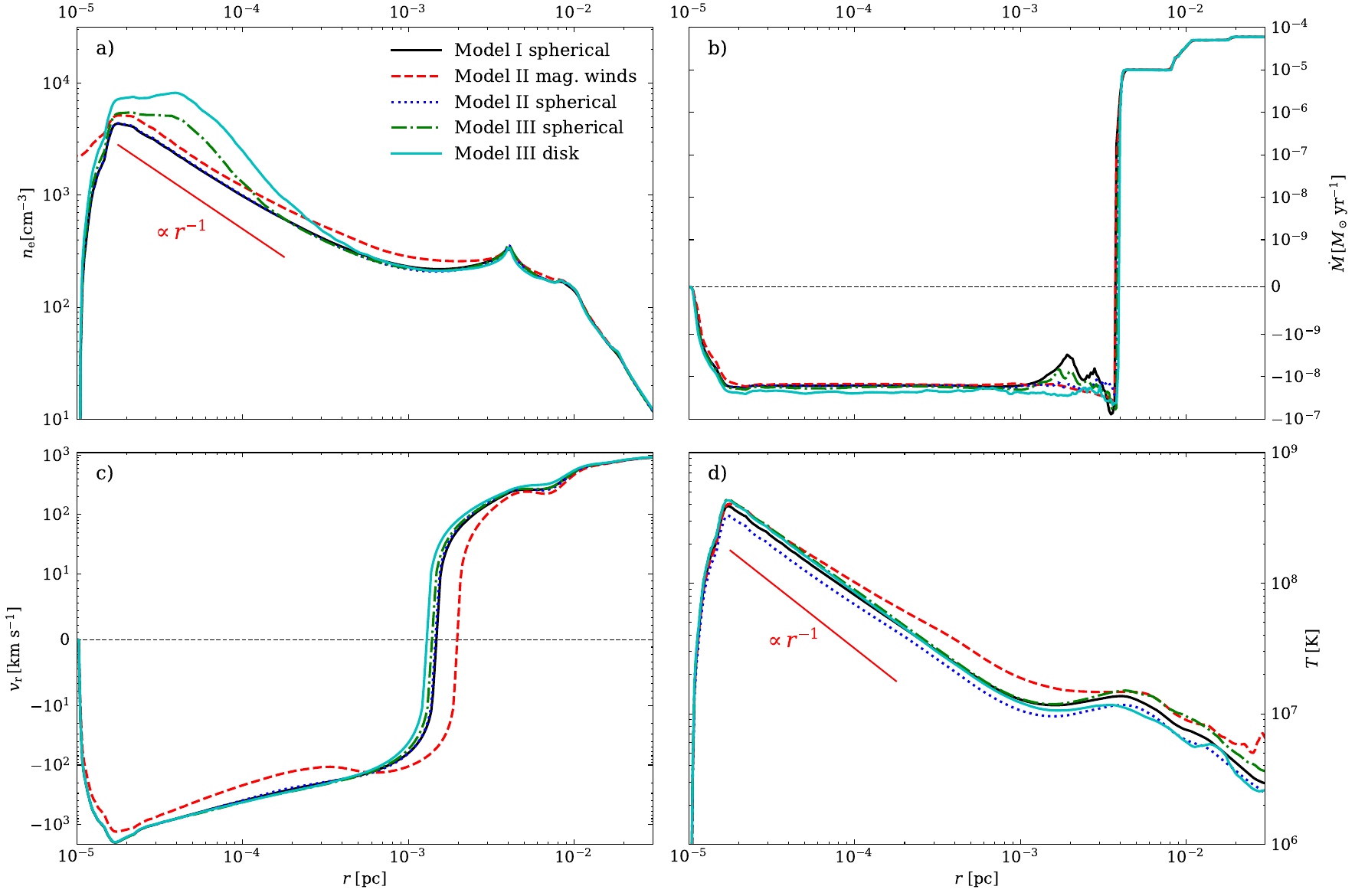}
    \caption{Time and angle-averaged quantities in our simulations as a function of distance from the black hole, $r$. Different lines denote different configurations of the simulation. \textit{a)} density in the units of electron number density. \textit{b)} accretion rate $\dot{M}$. \textit{c)} radial velocity of the gas $v_\mathrm{r}$. \textit{d)} Temperature of the gas. For the density and temperature radial profiles, a line representing $\propto r^{-1}$ is also plotted for reference, highlighting the approximate power law in both quantities in the inner regions of the simulations.}
    \label{fig:profiles}
\end{figure*}

Across multiple simulations, we surveyed a variety of physical parameters, such as different stellar distributions of the cluster, the magnetization of the winds, and different chemical abundance models of the stars, listed in Table \ref{tab:abundance}. Fig. \ref{fig:snap} presents 3D volume renderings of density for two distinct simulation setups featuring a spherically isotropic stellar distribution. The left panel depicts a snapshot taken 620 years into the Model II spherical simulation, whereas the right panel shows a snapshot captured 815 years into the Model III spherical simulation. The snapshots reveal the stellar winds of the 'stars' in our simulations, emphasizing their dynamic nature and intense interactions. By comparing the snapshots, it is clear that the behavior of the wind is significantly sensitive to the mass fractions within the stellar composition, given that Model II and Model III differ not only in hydrogen content but mainly in the metallicity, $Z$, with Model III having around six times higher $Z$ at the expense of no hydrogen and a lower helium content. 

The differences are also visible in Fig. \ref{fig:slices}, which shows plane-parallel cuts at $y=0$ of the simulation box (note that there is no axis of symmetry) of electron number density $n_\mathrm{e} = \rho/(\mu_\mathrm{e}m_\mathrm{p})$,
accretion rate, $\dot{M} = 4\pi r^2 \rho v_\mathrm{r}$, 
and temperature $T= \mu m_\mathrm{p}P/(\rho \mathrm{k_B})$,
where $m_{\rm p}$ is the proton mass, $v_{\rm r}$ radial velocity, and $\mathrm{k_B}$ is the Boltzmann constant, for the two discussed scenarios (left and right columns), as well as Model II with magnetized winds (middle column). In the lower-metallicity case (Fig. \ref{fig:slices}, left column), reduced radiative cooling efficiency prevents the formation of cool, dense clumps within the wind. Instead, this scenario facilitates the development of high-temperature ($\sim 10^7$ K) outflow channels that extend beyond the cluster and originate where the winds collide. 

In the simulation with higher metallicity of the stellar winds (Fig. \ref{fig:slices}, right column), the radiative cooling is more efficient, leading to the formation and ejection of cool, dense clumps. There is no sign of an accretion disk forming in our simulations; instead, accretion of matter happens primarily through close encounters of the stellar winds and the central black hole. This is true even for simulations with a disk-like distribution of stars (not shown). 

Compared to the pure HD case, the Model II simulation with magnetic fields (Fig. \ref{fig:slices}, middle column) shows minimal differences, as magnetic fields do not appear to alter the global dynamics significantly.

\subsection{Radial profiles}
We present radial profiles of fluid quantities averaged over angles and time to better understand accretion dynamics. 
To do this, we first interpolate the data onto a spherical grid ($r,\theta,\varphi$) logarithmically spaced in radius $r$, where the polar $\theta$-axis is aligned with the $z$-axis of the simulation. 
Then we define the time (from $t_\mathrm{min}$ to $t_\mathrm{max}$) and angle average of a quantity $A$ as
\begin{equation}
    \langle A \rangle \equiv \frac{1}{4\pi(t_\mathrm{max}-t_\mathrm{min})} \int\displaylimits_{t_{\rm min}}^{t_{\rm max}}\int\displaylimits_{0}^{2\pi}\int\displaylimits_{0}^{\pi} A\sin(\theta)\,\mathrm{d}\theta\,\mathrm{d}\varphi\,\mathrm{d}t.
\end{equation}

Fig. \ref{fig:profiles}a shows the radial dependence of the angle and time-averaged electron number density for our five simulations. In the outer region of the domain ($r > 10^{-3}$ pc), the profiles are nearly identical. Closer to the IMBH, Models I and II generally follow a density scaling of $n_{\rm e} \propto r^{-1}$, whereas Model III exhibits a steeper density increase toward the system's center. Fig. \ref{fig:profiles}b shows the radial dependence of the accretion rate,
$\dot{M} = 4\pi \langle r^2\rho v_{\rm r} \rangle$.
Positive values indicate outflow, while negative values correspond to inflow. The accretion rate profiles are similar across all scenarios, except in the region occupied by stellar winds due to their dynamic interaction. The system is predominantly outflow-driven, with the outflow rate ($\sim 10^{-5}M_\odot\,\mathrm{yr}^{-1}$) being approximately three orders of magnitude greater than the inflow rate at the simulation's inner boundary ($\sim 10^{-8}M_\odot\,\mathrm{yr}^{-1}$). Assuming the accretion rate scales with the radius of the inner boundary of the simulation as
\begin{equation}
\dot{M}_{\bullet \rm,new}=\dot{M}_{\rm in}\sqrt{\frac{r_{\rm in,new}}{r_{\rm in}}},
\label{eq_mdot_imbh}
\end{equation}
which is appropriate for an $r^{-1}$ scaling of density \citep{Ressler_Quataert_Stone_2018,Guo2023,SCAF}, extrapolating to the event horizon of a central black hole with mass $M_\bullet = 3 \times 10^{4}\,M_\odot$ yields an estimated accretion rate $\dot{M}_{\bullet} \approx 2.1 \times 10^{-10}M_\odot\,\mathrm{yr}^{-1}$. 

Figures \ref{fig:profiles}c and \ref{fig:profiles}d present the radial profiles of radial velocity, $v_{\rm r}$, and temperature, $T$, respectively, demonstrating that these quantities are largely unaffected by the stellar wind configuration (at least for our assumption of circular orbits). In the inner region ($r \lesssim 10^{-3}$ pc), the temperature radial profile is similar to that of the density, $T \propto r^{-1}$, as expected for an approximately virial temperature. The temperature gradually decreases to $\sim 10^6$ K at the outer domain boundary.
The magnetized stellar wind simulation with $\beta_{\rm w} = 100$ is a minor outlier from these trends. Magnetic fields partially slow the outflow, so the winds reach slightly lower radial velocities. Dissipation of magnetic fields also contributes additional heating, leading to marginally higher temperatures. 

\begin{figure*}[h]
    \centering
    \begin{minipage}{0.49\textwidth}
    \includegraphics[width=1\linewidth]{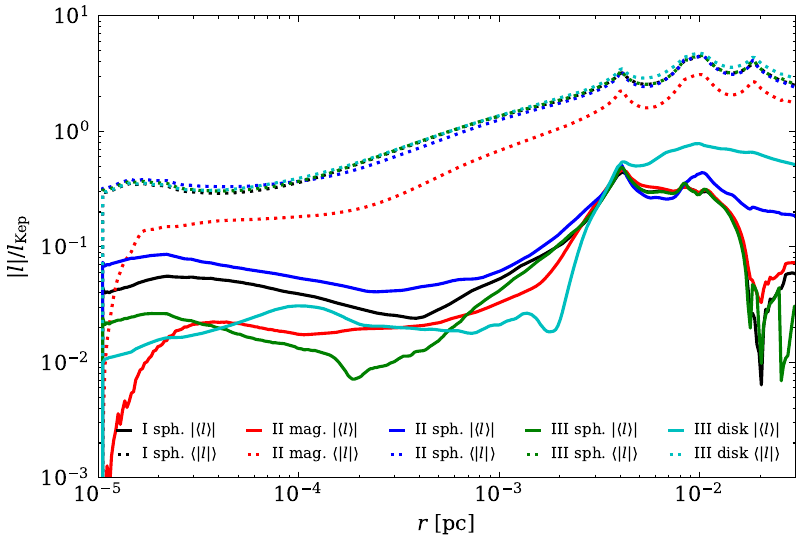}
    \caption{Time and angle-averaged absolute value of the specific angular momentum $\langle|l|\rangle$ (dashed lines) as well as the absolute value of the average $|\langle l\rangle|$ (solid lines) normalized by the Keplerian value, $l_\mathrm{Kep}$, as a function of distance from the black hole, $r$.
    \vspace{1em}}
    \label{fig:angular_momentum}
    \end{minipage}
    \hfill
    \centering
    \begin{minipage}{0.49\textwidth}
    \includegraphics[width=1\linewidth]{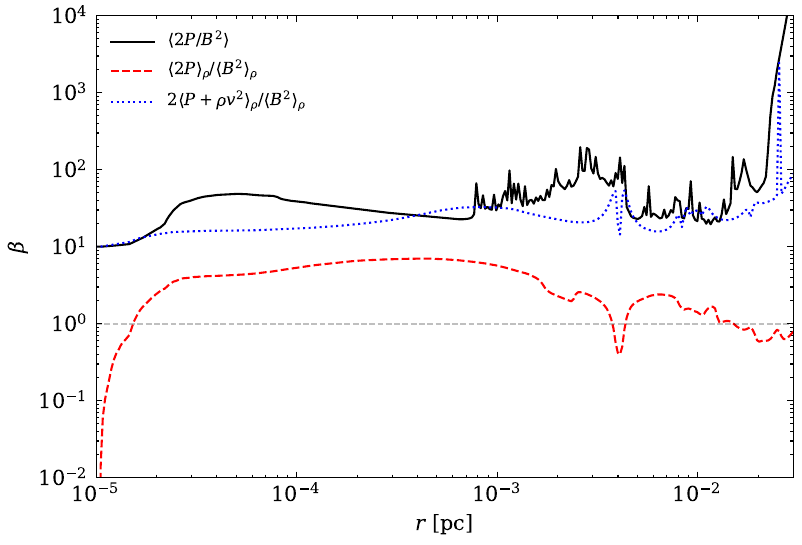}
    \caption{Time and angle-averaged value of plasma $\beta$ as a function of distance from the black hole, $r$, for the Model II simulation with magnetized winds using $\beta_{\rm w} = 100$. Different lines denote average $\langle \beta \rangle$ (solid line), density-weighted average $\langle \beta \rangle_\rho$ (dashed line), and $\langle \tilde \beta \rangle_\rho = 2\langle P +\rho v^2\rangle_\rho/\langle B^2\rangle_\rho$ (dotted line).}
    \label{fig:beta}
    \end{minipage}
\end{figure*}

\subsection{Angular momentum and magnetic effects}

Fig.~\ref{fig:angular_momentum} plots radial profiles of the time and angle-averaged specific angular momentum, $\mathbf{l} = \mathbf{r} \times \mathbf{v}$, where $\mathbf{r}$ and $\mathbf{v}$ are the position and velocity vectors, respectively, for both the magnitude of the average and the average of the magnitude scaled to the Keplerian specific angular momentum, $l_\mathrm{kep} = \sqrt{GM_\bullet r}$.
These curves show that individual gas elements injected into the domain by the stellar winds generally have a super-Keplerian amount of angular momentum, meaning they tend to move outwards in radius.  
On the other hand, the net angular momentum is relatively low in comparison because there is a significant amount of cancellation due to the lack of a consistent direction for the angular momentum.  
This is true even for the Model III disk-like stellar configuration, primarily because the wind speeds are higher than the orbital speeds.
For distances from the black hole within the feeding region, fluid elements tend to have sub-Keplerian angular momentum ($\langle |\mathbf{l}| \rangle \sim 0.3$--0.4 $l_{\rm kep}$ for the hydrodynamic simulations), implying that only the lowest angular momentum gas from the stellar winds falls inwards. 
Since the wind speeds are relatively large, this gas corresponds with either low impact parameter trajectories (i.e., gas blown almost directly at the black hole) or gas that loses its angular momentum through wind-wind shocks. 
Again, in this region, the magnitude of the net angular momentum is much lower than the average magnitude $|\langle \mathbf{l} \rangle | \lesssim 0.07$ $l_{\rm kep}$, implying the lack of coherent accretion structure. The simulation with magnetic fields has the lowest $\langle |\mathbf{l}|\rangle$, $\lesssim 0.2 l_{\rm kep}$, because magnetic forces tend to slow the gas.

Finally, for the magnetized wind simulation, Fig. \ref{fig:beta} presents angle and time-averaged radial profiles of plasma beta ($\beta = 2P/B^2$) for different averaging methods:  $\langle \beta \rangle$ (solid line), $\langle P \rangle_\rho/\langle B^2/2\rangle_\rho$ (dashed line), and $\langle \tilde \beta \rangle_\rho = \langle P +\rho v^2\rangle_\rho/\langle B^2/2\rangle_\rho$ (dotted line), where we include the gas ram pressure along with the thermal pressure. 
Of these three curves, $\langle \tilde \beta \rangle_\rho$ best represents the relative strength of magnetic energy/pressure relative to purely hydrodynamic energy/pressure.
We find that $\langle \tilde \beta \rangle_\rho \sim 10$ for all radii implies that hydrodynamic forces dominate over magnetic ones. 
This is consistent with the radial profiles shown in Fig. \ref{fig:profiles}, where the MHD simulation shows only moderate differences with the hydrodynamic simulations.  
$\langle P \rangle_\rho/\langle B^2/2\rangle_\rho$ is generally lower than $\langle \tilde \beta \rangle_\rho$ because the ram pressure is always significant when compared with the thermal pressure.  
In the region containing the stellar winds ($r \sim 10^{-3}\mathrm{pc}$--$10^{-2}\mathrm{pc}$), $\langle P \rangle_\rho/\langle B^2/2\rangle_\rho$ is particularly low ($\sim $1--3) compared to $\langle \tilde \beta \rangle_\rho$ because most of the hydrodynamical energy is contained in the kinetic energy of the winds, which have low thermal pressure. 

\section{Connection to observations} \label{sec:obs}
\subsection{X-ray Surface brightness density}
To evaluate the detectability of the IMBH embedded within the cluster, we calculate the X-ray surface brightness density, $S_\mathrm{X}$. We use Model II spherical as our fiducial model, since the wind chemical composition here best aligns with the realistic expectation, and we see from Fig.~\ref{fig:slices}-\ref{fig:angular_momentum} that all models are qualitatively similar. First, we use \textsc{spex} as described in Sect. \ref{subsec:cooling} to calculate the cooling function, $\Lambda$, except we consider only the contributions from frequencies corresponding to the Chandra merged hard-band ($2 - 7$ keV) \citep{2010ApJS..189...37E}, denoting this as $\Lambda_\mathrm{X}$. Then we transform our simulation data to cylindrical coordinates ($s, z, \varphi$, with $z$ being the line of sight) and integrate along the line of sight to obtain
\begin{equation} \label{eq:sx}
    S_{\rm X}=\int\displaylimits_{-z_{\rm max}}^{z_{\rm max}}\frac{\rho^2}{\mu_\mathrm{e}\mu_\mathrm{H,\odot}}\Lambda_\mathrm{X}\,\mathrm{d}z,
\end{equation}
where $z_\mathrm{max}$ is half of the box length of the simulation. Fig. \ref{fig:sx} presents two snapshots of $S_\mathrm{X}$. The left panel corresponds to 650 years into the Model II spherical simulation, while the right panel depicts 830 years into the Model III spherical simulation, shortly after the timeframes shown in Fig. \ref{fig:snap}. At specific moments, emission from near the black hole becomes visible in the X-ray band (Fig. \ref{fig:sx}, left panel); however, the emission is typically dominated by the colliding stellar winds, rendering the black hole undetectable for most of the simulation (Fig. \ref{fig:sx}, right panel). Even when visible, the X-ray surface brightness at the location of the black hole remains lower than that of the surrounding winds.

\begin{figure}[h]
    \centering 
    \includegraphics[width=1\linewidth]{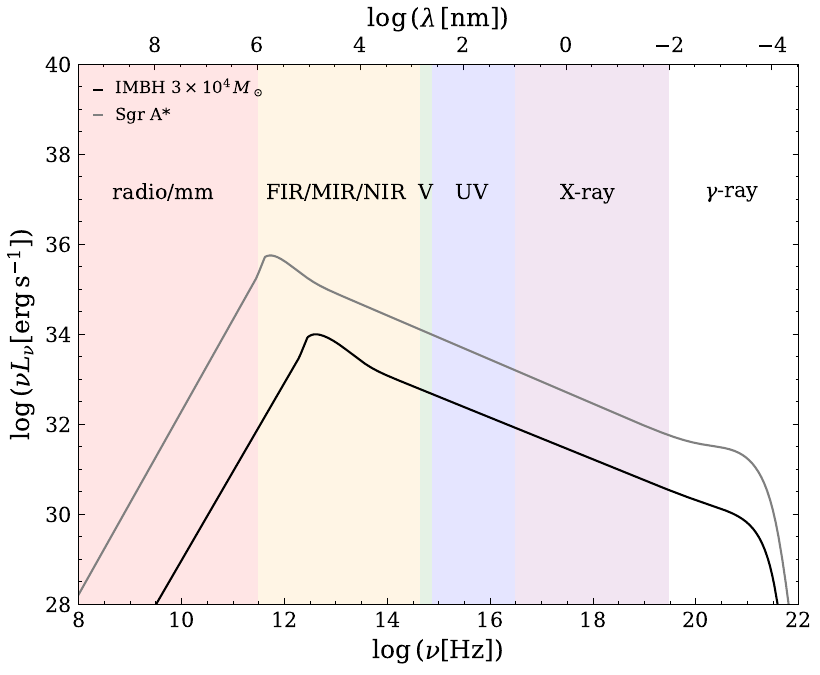}
    \caption{Spectral energy distribution (SED) for the accretion flow near the IMBH at the center of the stellar association calculated using a simple advection-dominated accretion flow (ADAF) model with a numerical prescription described in \citet{2021ApJ...923..260P}. The input for the model is the time-averaged accretion rate extrapolated to the event horizon of the IMBH ($\dot{M} = 2.1\times10^{-10}\,M_\odot\,\mathrm{yr}^{-1}$)  and the associated  Eddington ratio ($\lambda_\mathrm{Edd}=3.168\times 10^{-7}$).  For comparison, we also show an SED  for Sgr~A* calculated with the the same model using $\dot{M} = 2\times10^{-8}\,M_\odot\,\mathrm{yr}^{-1}$ and Eddington ratio $\lambda_\mathrm{Edd}=2.258\times 10^{-7}$.}
    \label{fig:sed}
\end{figure}

Additionally, we roughly estimate the quiescent, thermal spectral energy distribution (SED) of the near-horizon flow (i.e., the flow within the inner boundary not modeled by our simulation) using the numerical prescription for an advection-dominated accretion flow (ADAF) using the formalism of \citet{1997ApJ...477..585M} presented in \citet{2021ApJ...923..260P}. Based on the black hole mass $M_\bullet$, accretion rate $\dot{M}$, radiative efficiency $\eta$, plasma-$\beta$ parameter, viscosity parameter $\alpha$, power-law index for the mass accretion rate as a function of radius $s$, fraction of viscously dissipated energy that is advected $f$, and fraction of viscous heating that goes directly to the electrons $\delta$ \citep[for numerical values see][Table~3]{2021ApJ...923..260P}, this one-zone model accounts for synchrotron, bremsstrahlung, and inverse Compton radiation. Fig. \ref{fig:sed} displays this SED, calculated using the time-averaged accretion rate extrapolated to the event horizon in the Model II spherical simulation in terms of the Eddington ratio $\lambda_\mathrm{Edd}$, defined as 
\begin{equation}
\lambda_\mathrm{Edd} = \frac{\eta c \, \sigma_{\rm T}}{4\pi G m_{\rm p}}\,\frac{\dot{M_\bullet}}{M_\bullet},
\end{equation}
where $\eta$ is the radiative efficiency (assumed to be 0.1) and $\sigma_{\rm T}$ is the Thomson cross-section.
We highlight specific spectral bands and include an approximate Sgr A* SED calculated using the same model for a "typical" accretion rate of $ 2\times 10^{-8}\,M_\odot\,\mathrm{yr}^{-1}$, corresponding to an Eddington ratio of  $\lambda_\mathrm{Edd}=2.25\times 10^{-7}$ \citep{2007ApJ...654L..57M}. Compared to Sgr A*, the IMBH's radiative flux is roughly three orders of magnitude lower across the spectrum, with its peak slightly shifted from the far-infrared (FIR) to the mid-infrared (MIR).

\begin{figure*}[t]
    \centering
    \includegraphics[width=1\linewidth]{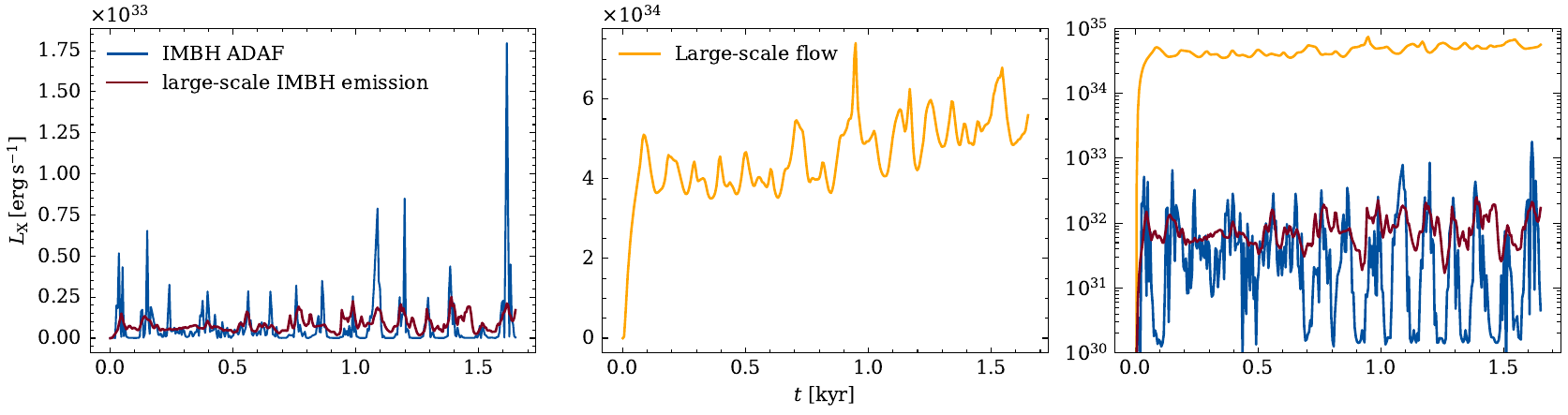}
    \caption{Time evolution of the integrated X-ray luminosity computed from the Model II simulation. \textit{Left:} X-ray luminosity of a near-horizon ADAF model for the IMBH as well as well as the large-scale emission from the vicinity of the IMBH. \textit{Middle:} X-ray luminosity of the large-scale flow. \textit{Right:} IMBH ADAF, IMBH large-scale flow, and total X-ray luminosity combined; note that the vertical scale is logarithmic.}
    \label{fig:l_comparison}
\end{figure*}

To estimate the relative contributions of the IMBH near-horizon emission and stellar wind collisions to the X-ray luminosity, we compute an IMBH ADAF spectrum every 5 years of the simulation time and integrate over the X-ray band. 
We then compare this to the total X-ray luminosity, $L_{\rm X}$, computed from the simulation by integrating
\begin{equation} \label{eq:sx}
    L_{\rm X}(s) = \int\displaylimits_{r_{\rm in}}^{s}\int\displaylimits_0^{2\pi}S_\mathrm{X}(s)\,s\mathrm{d}s\mathrm{d}\varphi=\int\displaylimits_0^{2\pi}\int\displaylimits_{-z_{\rm max}}^{z_{\rm max}}\int\displaylimits_{r_{\rm in}}^{s}\frac{\rho^2}{\mu_\mathrm{e}\mu_\mathrm{H,\odot}}\Lambda_\mathrm{X}\,s\mathrm{d}s\,\mathrm{d}z\,\mathrm{d}\varphi,
\end{equation}
where $r_\mathrm{in}$ is the radius of the inner boundary.
Fig. \ref{fig:l_comparison} illustrates the time evolution of the X-ray luminosity for the ADAF SED model associated with the IMBH and the large-scale emission from the IMBH, computed from Eq. \eqref{eq:sx} integrated over the the central $10^{-4}$ pc encompassing the black hole (left), the large-scale flow modeled by our simulations, computed from Eq. \eqref{eq:sx} integrated over the whole domain (middle), and all lines combined on the same plot (right). The luminosity produced by the large-scale flow is approximately two orders of magnitude greater than the ADAF luminosity and the large-scale emission associated with the IMBH, reaffirming that stellar wind collisions dominate the X-ray spectrum. 

\subsection{X-ray data analysis}
\label{sec_light_curves}

To compare the simulated X-ray signal with actual X-ray data of the IRS 13E source, we use archival observations obtained by the Chandra X-ray Observatory. We analyzed 128 non-grating ACIS-S and ACIS-I Chandra Observational IDs (OBSIDs) with Sgr~A* in the aim-point spanning almost 25 years. All observations are reprocessed using the \texttt{chandra\_repro} script (\textsc{ciao}~4.17; \citealp{Fruscione2006}) and the latest calibration files (\textsc{caldb}~4.11.6).

We perform relative astrometric corrections using the \texttt{fine\_astro} script to obtain the best possible match between observations. As a reference, we use the OBSID 11843, which provides long enough exposure (78 ks) with a minimum of bright transient sources close to the chip aim-point, which would complicate source-matching with other OBSIDs. Before merging all OBSIDs, we further omit observations containing very bright transient sources in the vicinity of IRS 13E that would contaminate its signal, and we end up with 112 observations with a total cleaned exposure time of $\approx 3.6$ Ms ($\approx 2$ Ms for ACIS-S and $\approx 1.6$ for ACIS-I).

When reprocessing the individual OBSIDs using the \texttt{chandra\_repro} script, we use the default value for the \texttt{pix\_adj} parameter (\texttt{pix\_adj=EDSER}), which allows us to leverage the sub-pixel Chandra resolution. We then produce a merged hard-band ($2-7$ keV) image binned to a pixel size of $\approx 0.125$ arcsec. This allows us to study the morphology of the X-ray source corresponding to IRS 13E.

\begin{figure*}[h]
    \sidecaption
    \includegraphics[width=12cm]{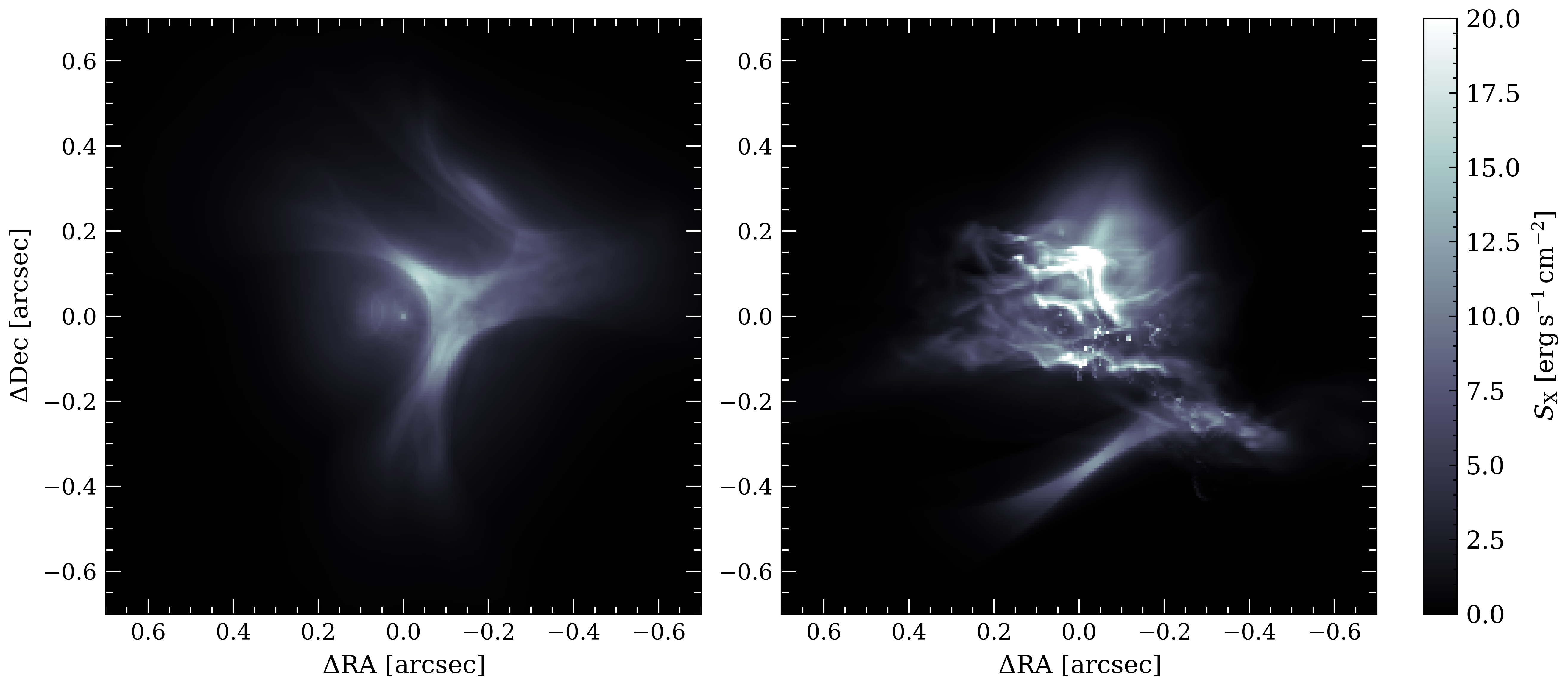}
    \caption{Predicted intrinsic X-ray surface brightness density $S_\mathrm{X}$ computed from our simulations scaled to Galactic center distance. \textit{Left:} 650 years into the Model II spherical simulation. \textit{Right:} 830 years into the Model III spherical simulation. The black hole is visible in the center of the left panel; however, the colliding winds dominate the X-ray emission and, most of the time, make the black hole undetectable (as seen in the right panel).}
    \label{fig:sx}
\end{figure*}

\begin{figure*}[h]
    \sidecaption
    \includegraphics[width=12cm]{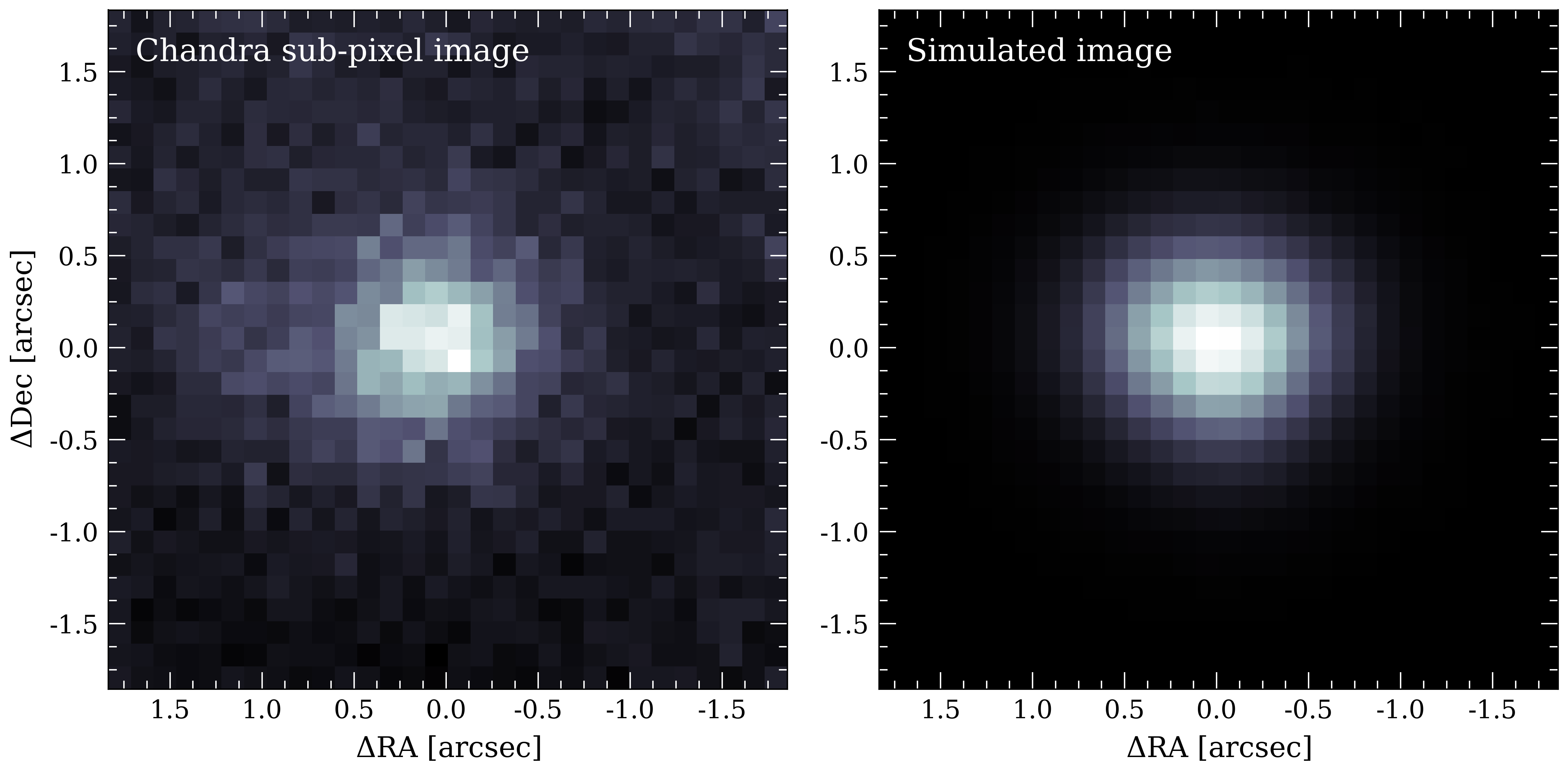}
    \caption{\textit{Left:} Merged sub-pixel Chandra image obtained by combining 112 non-grating ACIS-S and ACIS-I observations with a total cleaned exposure time of almost 3.6 Ms. \textit{Right:} Mock Chandra ACIS-S image obtained using the \textsc{SOXS} simulator based on a Model II spherical simulation snapshot at 50 years.}
    \label{fig:real_vs_sim}
\end{figure*}

The simulated mock Chandra image obtained by scaling the X-ray surface brightness map (Fig. \ref{fig:sx}) to a distance of $8$ kpc and then using the SOXS package \citep{2023ascl.soft01024Z} to simulate a mock Chandra ACIS-S\footnote{Effective area corresponding to Cycle 22 was assumed.} observation with 3.6 Ms exposure time. Based on the simulated event file, we then produce a hard-band image binned to match the pixel size of the real Chandra sub-pixel image ($\approx 0.125$ arcsec). We compare the sub-pixel Chandra image to the simulated image at $t = 50$ yr in Fig. \ref{fig:real_vs_sim}. We chose this time arbitrarily, since the simulated image does not change much throughout the simulation, with only the emission centroid moving slightly depending on where the winds collide the most.
Given the uncertainties in the stellar wind parameters, the qualitative agreement between the Chandra image and the mock image from our simulation is quite good except for minor differences. In Appendix~\ref{appendix_surf_brightness} and in particular in Fig.~\ref{fig_Xray_profiles}, we compare the surface-brightness profiles between the observed image and the simulated one.
The simulation produces a slightly more spherical distribution of X-ray luminosity, which we demonstrate by evaluating the asymmetry parameter ($A=0.13$ for the simulated image vs. $A=0.21$ for the observed one). This result is not surprising because we assumed that all the stars were on spherical orbits.  
More importantly, the Chandra image also contains an extended, diffuse emission outside the brighter core. In addition, there are additional background/foreground X-ray point sources in the IRS 13E field. Our simulations do not contain any more distant sources, and the simulated image is thus based entirely on the six stars listed in Table \ref{tab:distances} that orbit the IMBH. However, apart from the differences in the tails of the X-ray surface-brightness profile, both profiles are qualitatively consistent with the uncertainties. They closely follow and are slightly wider than the Chandra point-spread function (PSF). From a quantitative point of view, the total flux/luminosity of IRS 13E and the simulated cluster differ significantly, with the simulated cluster being more luminous by an order of magnitude. The agreement could be reached by fine-tuning, e.g., the WR stars' mass-loss rates and wind velocities within the uncertainties. Since we did not seek an exact model for IRS 13E at this point, we will explore the parameter space of the cluster in relation to IRS 13E in the upcoming study in more detail. 

\subsection{Accretion rate and X-ray Variability}
\begin{figure*}[h]
    \centering
    \includegraphics[width=\textwidth]{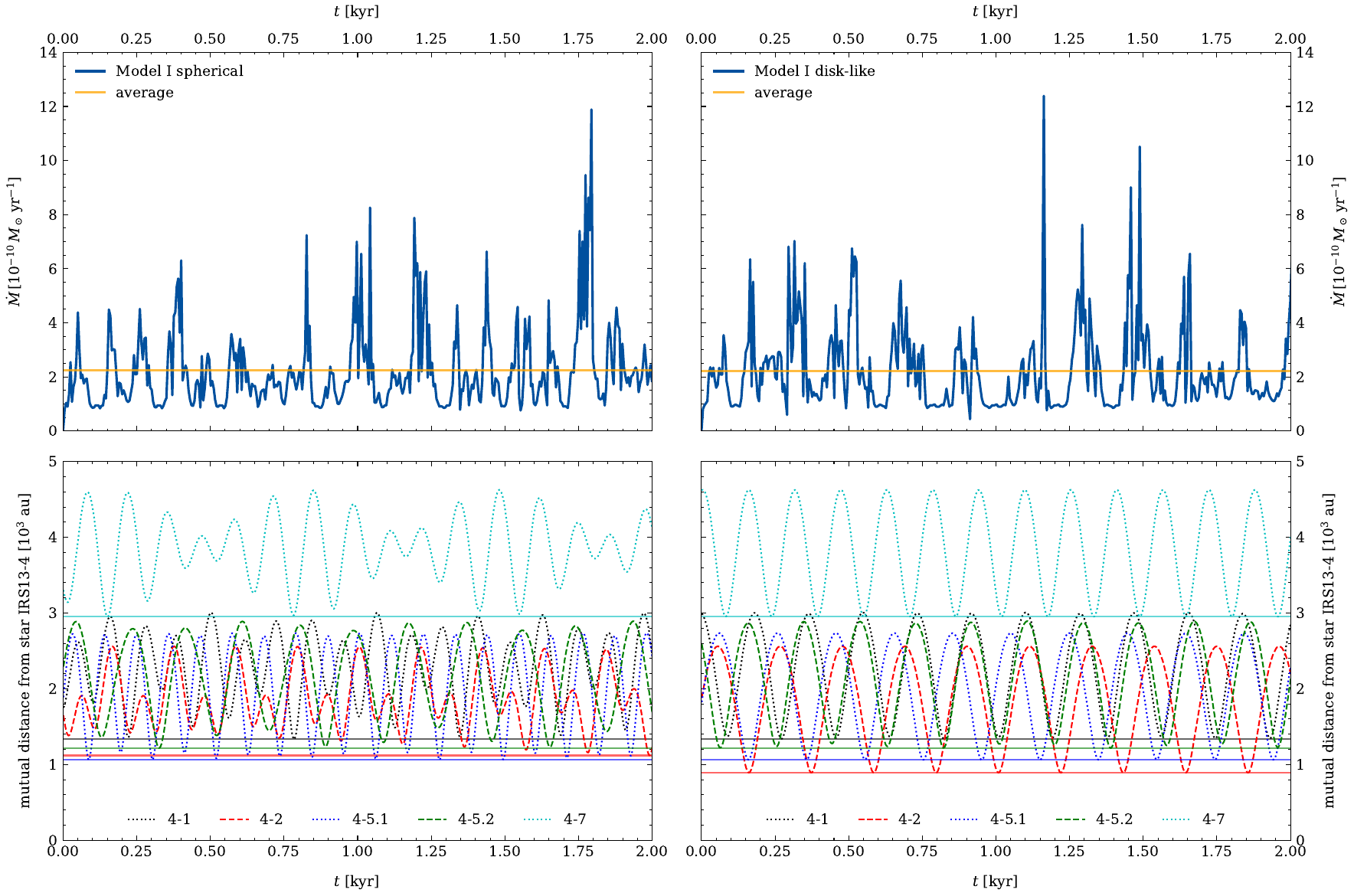}
    \caption{Temporal evolution of the mass inflow rate $\dot{M}$ (top row) and mutual distances between the shortest-period star E4 and other stars expressed in au (bottom row) for the isotropic spherical (left column) and disk-like (right column) stellar configurations. The minimum distances for each pair are represented via solid horizontal lines of corresponding color. In both configurations, periodic approaches between the closest star, E4, and all other stars are responsible for the periodic spikes in the inflow rate, with the shortest separations reaching $\sim 1000\,{\rm au}$.}
    \label{fig:mdot_mutual_distance}
\end{figure*}

In addition to the time-averaged and angle-integrated radial profiles of the accretion rate (Fig.~\ref{fig:profiles}), we also analyze the time variability. 
In particular, we focus on the accretion rate at the inner boundary of the computational domain extrapolated to the event horizon using Eq.~\eqref{eq_mdot_imbh}, shown in the top left and top right panels of Fig~\ref{fig:mdot_mutual_distance} for the spherical and disk-like stellar distributions. 
The variability is quite pronounced, with spikes in accretion rate up to an order of magnitude, i.e. from $\dot{M}_{\bullet}\sim 10^{-10}\,{\rm M_{\odot}\,yr^{-1}}$ up to $\dot{M_{\bullet}}\sim 10^{-9}\,{\rm M_{\odot}\,yr^{-1}}$ on a timescale of $\sim 20$ years.

In order to isolate candidate periodicities in the accretion rate, we compute Lomb-Scargle periodograms \citep{1976Ap&SS..39..447L,1982ApJ...263..835S}, as detailed in Appendix~\ref{appendix_periodicities}. Our simulations typically span 2340 years, and we sample every 5 years, meaning that we can identify periodic signals in the time range from 10 to 1000 years. For the accretion rate in spherically distributed stellar systems, we detect a prominent periodicity peak at $\sim 110$ years for both low and high metallicity. In addition, peaks at $\sim 200$ years, $\sim 55$ years, and $\sim 70$ years are also pronounced, as shown in Figs.~\ref{fig_mdot1_ls_psd}-\ref{fig_mdot3_ls_psd} (left panels). For the disk-like stellar configuration, the most prominent peak is at $\sim 186$ years, followed by peaks at $\sim 217$ and $\sim 100$ years, respectively, as shown in Fig.~\ref{fig_mdot_d_ls_psd} (left panel). 

These detected periodicities induced close encounters between stars, during which their strong stellar winds are compressed and shocked. The most shocked, denser gas flows out of the system while a small fraction accretes onto the black hole. In Appendix~\ref{appendix_mutual_dist}, there is a comprehensive overview of the approaches of the stellar pairs, including minimum, mean, and maximum distances (in AU) and the associated recurrence timescales (in years). The most pronounced periods are associated with the star E4, which has the most compact orbit (at a distance of $4.06$ mpc). For the adopted spherical setup of stars, the closest approaches between stars E4 and E5.1 take place every 108 years, which is consistent with the most prominent periodicity peak in the inflow rate, compare Fig.~\ref{fig:mdot_mutual_distance} (left column) and Figs.~\ref{fig_mdot1_ls_psd}-\ref{fig_mdot3_ls_psd}. On the other hand, for the disk-like setup, the approaches between the same stars (E4 and E5.1) take place every 198 years, which is again consistent with the broader peak around 186-217 years in the periodogram (Fig.~\ref{fig_mdot_d_ls_psd}; compare with Fig.~\ref{fig:mdot_mutual_distance}, right column). The closest approaches between star E4 and other stars take place with comparable periods of $\sim 100-200$ years (see Tables~\ref{tab:rel_distances_isot} and \ref{tab:rel_distances_disk}), which further affects the mass inflow rate; specific inflow peaks are strengthened or prolonged when the stellar approaches take place almost simultaneously. In addition, occasional rare approaches on the scale of $\sim 100-200\,{\rm AU}$ among other stars can further significantly modulate the inflow rate at some epochs. Altogether, this results in multiple periodicity peaks for all the runs in the periodograms.    

In Appendix \ref{appendix_periodicities}, we also construct power spectral densities (PSDs) of the accretion rates (shown in the right panels of  Figs.~\ref{fig_mdot1_ls_psd} and \ref{fig_mdot_d_ls_psd}). 
Overall, except for the presence of peaks due to the periodocities already discussed (at $\sim 100$ and $200$ years), the PSD profiles are consistent with a simple power law function with the mean slope of $\sim -0.60$, which implies the presence of underlying stochastic variability.  

\begin{figure*}[t]
    \centering
    \includegraphics[width=0.49\textwidth]{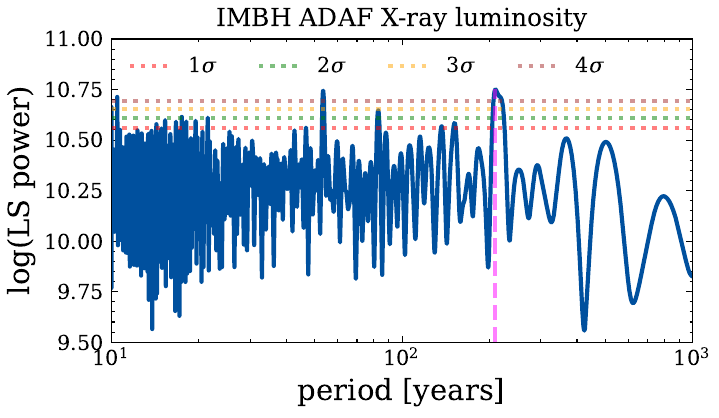}
     \includegraphics[width=0.49\textwidth]{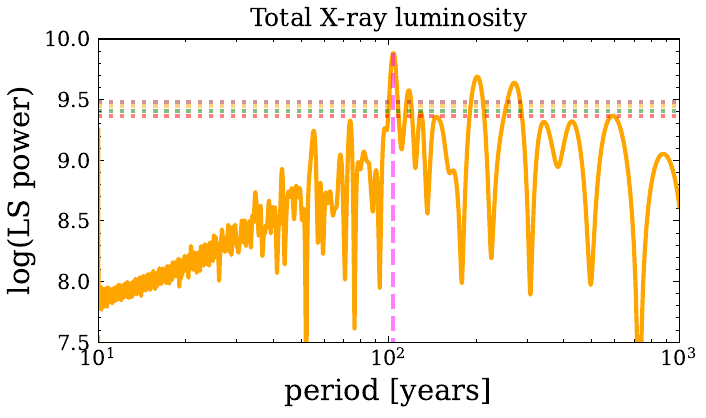}
    \caption{Lomb-Scargle (LS) periodograms of the X-ray light curves computed from the simulations. \textit{Left:} The LS periodogram of the X-ray light curve produced by the near-horizon flow around the IMBH as calculated from an ADAF model.  The highest peak is at $\sim 210$ years, while other prominent peaks are at $\sim 54$ and $\sim 83$ years. Dotted horizontal lines mark inferred confidence levels based on 10000 bootstrap realizations. \textit{Right:} The LS periodogram of the total X-ray luminosity exhibits the highest peak at $\sim 104$ years. Other prominent peaks are at $\sim 202$, $\sim 271$, $\sim 55$, and $\sim 74$ years. Dotted horizontal lines stand for the confidence levels, as in the left panel.}
    \label{fig:ls_periodogram_Xray}
\end{figure*}

\begin{figure*}[t]
    \centering
    \includegraphics[width=0.49\textwidth]{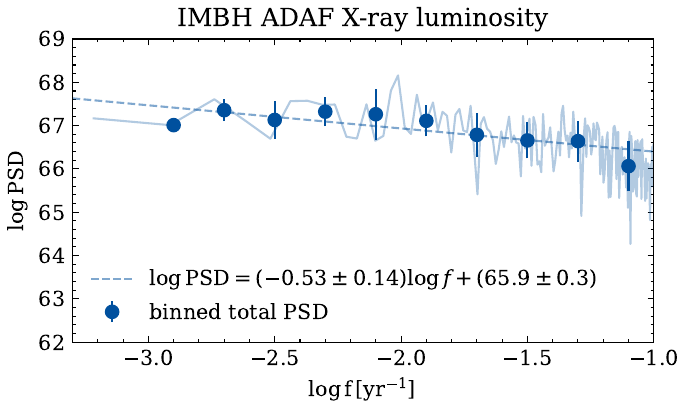}
     \includegraphics[width=0.49\textwidth]{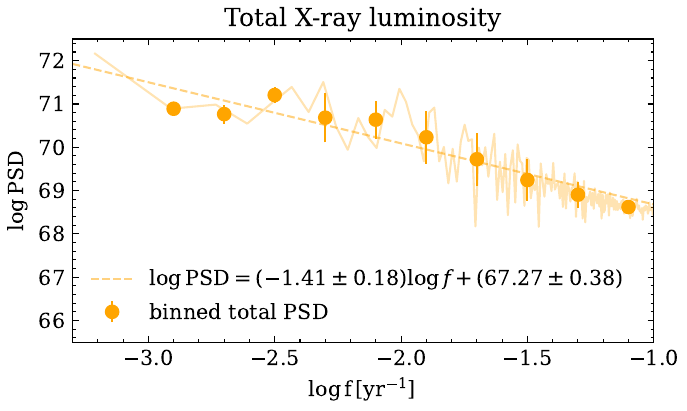}
    \caption{Power spectral densities (PSDs) of the X-ray variability of the compact stellar cluster with an IMBH. \textit{Left:} The PSD of the X-ray light curve produced by the near-horizon flow around the IMBH as calculated from an ADAF model, which has a flat power-law slope of $\sim -0.53$. \textit{Right:} The PSD of the total X-ray light curve, which exhibits a steeper power-law slope of $\sim -1.41$. The points represent mean PSD values within the ten regular logarithmic bins in frequency, while the error bars represent standard deviations in these bins. We fit the power-law functions to these binned PSD values for consistency. In both panels, the extents of the logarithms of PSDs and frequencies are kept the same for consistency and easier comparison.}
    \label{fig:psd_Xray}
\end{figure*}

Finally, we compute Lomb-Scargle periodograms (Fig.~\ref{fig:ls_periodogram_Xray}) and PSDs (Fig.~\ref{fig:psd_Xray}) for the X-ray light curves presented in Sect. \ref{sec_light_curves}.
Because the magnitude and the spectral shape of the X-ray portion of the ADAF model's SED (Fig.~\ref{fig:sed}) depend on the variable accretion rate, the time-variability is not one-to-one with the accretion rate. 
The Lomb-Scargle periodograms of the ADAF models display significant peaks at $\sim 210$ years and $\sim 54$ years. In contrast, the peak around $\sim 110$ years appears to be suppressed compared to the periodogram of the accretion rate in  Fig.~\ref{fig_mdot1_ls_psd}. 
On the other hand, the total X-ray luminosity, which dominated the shocked stellar winds, has a significant periodic component at $\sim 104$ years, corresponding to the periods of close approaches among the stellar pairs E4-E1 and E4-E5.1. 
In terms of power, the PSD for the ADAF model is significantly flatter than the PSD for the whole cluster (with power-law slopes of $\sim -0.53$ vs. $\sim -1.41$). 
This is consistent with the fact that there is more short-term variability on timescales of $<100$ years in the region closer to the inner boundary of the simulation than on larger scales of the cluster. 

We also characterize the intrinsic IMBH ADAF variability in terms of the fractional variability \citep{1997ApJ...476...70N,1997ApJS..110....9R,2019ApJ...883..170M},
\begin{equation}
    F_{\rm var}=\frac{(\sigma^2-\Delta^2)^{1/2}}{\overline{L}}
    \label{eq_fractional_variability}
\end{equation}
where $\sigma^2$ is the variance of the luminosity, $\Delta^2$ is the mean square uncertainty associated with individual luminosities $L_i$, and $\overline{L}$ is the mean luminosity. Assuming one percent relative uncertainty for both the IMBH ADAF and the total (cluster) luminosities (see Fig.~\ref{fig:l_comparison} left and middle panels), we obtain $F_{\rm var}\simeq 213\%$ for the IMBH's ADAF X-ray luminosity, while we infer $F_{\rm var}\simeq 20\%$ for the X-ray luminosity of the whole cluster (for a luminosity uncertainty of $10\%$, $F_{\rm var}\simeq 212\%$ and $F_{\rm var}\simeq 17\%$ for the IMBH's ADAF and the total luminosities, respectively). Hence, the cluster X-ray emission, whose variability is mild, masks a much more vigorous X-ray emission of the IMBH ADAF. This is also consistent with the flatter power spectral density of the ADAF variability ($\propto f^{-0.5}$) in comparison with the PSD of the total luminosity ($\propto f^{-1.4}$), see Fig.~\ref{fig:psd_Xray}. The quasiperiodic X-ray spikes associated with the IMBH with the typical timescale of $\sim 100$ years are reminiscent of order-of-magnitude quasiperiodic X-ray eruptions (QPEs) detected for several both active and quiescent galactic nuclei \citep{2019Natur.573..381M,2021Natur.592..704A,2024arXiv241104592S}. However, in contrast, QPEs take place on timescales five orders of magnitude smaller, and the QPE X-ray properties are consistent with being associated with the soft thermal X-ray emission. Still, in future work, it is worth considering a compact association of wind-blowing stars orbiting around the massive black hole in a galactic nucleus as a potential model to address some of the QPE sources. Such a simulation would have to be a scaled-down version of the MHD simulations presented in this work with the characteristic stellar orbit radius of $r_{\rm QPE}=r_{\rm IRS13}(P_{\rm QPE}/P_{\rm IRS13})^{2/3}\sim 2 \times 10^{-6}\,(r_{\rm IRS13}/4\times 10^{-3}\,{\rm pc}) (P_{\rm QPE}/10\,\text{hours})^{2/3} (P_{\rm IRS13}/100\,\text{years})^{-2/3}  {\rm pc}$ for $M=3\times 10^4\,M_{\odot}$ (IMBH), which corresponds to $r_{\rm QPE}\sim 5.4 \times 10^{-6}\,{\rm pc}\sim 111 (P_{\rm QPE}/10\,\text{hours})^{2/3}(M/10^6\,M_{\odot})^{-2/3}$ gravitational radii for the $M=10^6\,M_{\odot}$ supermassive black hole. 

\section{Discussion}
\label{sec_discussion}
\subsection{IMBH accretion fueled by stellar winds}

Understanding accretion onto IMBHs in wind-fed systems is crucial for understanding their growth and observational signatures. 
In the compact WR stellar cluster scenario studied here, the winds create an unstructured inflow that affects accretion efficiency and detectability in X-rays and other wavelengths. As shown in Fig. \ref{fig:snap}, the interaction of these winds results in a turbulent, inhomogeneous flow with dense filaments and outflow channels. Furthermore, only about $10^{-5}$ of the injected mass reaches the black hole. This stark disparity highlights that only a tiny fraction of the available wind material is captured. Consequently, the black hole is unlikely to experience significant growth in mass from stellar winds alone. The accretion rates in our models are several orders of magnitude smaller than would be expected if the IMBH fed a dense gas disk or a continuous gas inflow \citep[e.g.,][]{2013MNRAS.432..506P}. This implies that other processes—such as external gas infall from the larger environment, tidal disruptions of stars, or mergers with compact objects—would be needed for the IMBH to increase its mass substantially. 

The accretion process is highly intermittent and variable, primarily driven by the orbital motions of the stars: periodic close WR-WR and WR-IMBH passages induce spikes in the inflow rate (see Fig.~\ref{fig:mdot_mutual_distance}). These findings illustrate that wind-wind interactions dominate the gas dynamics, which is comprised almost entirely of unbound material. Instead, the resulting flow structure is largely outflow-dominated, which can further limit the amount of infalling gas. In summary, wind-fed accretion in this environment is a fundamentally chaotic process markedly different from the orderly disk accretion scenarios often considered for black hole growth.

Our simulations identify several physical factors that collectively limit the accretion efficiency in this wind-fed system. First, the high velocity of the WR winds (on the order of $v_w \sim 1000$~km~s$^{-1}$; \citealt{1995ApJ...447L..95K}) means that much of the ejected gas moves faster than the escape speed defined with respect to the IMBH (the escape speed at $r = 0.01$ pc is 165 km s$^{-1}$). The radial profile of accretion rate (Fig.~\ref{fig:profiles}) confirms that outflows dominate, indicating that most of the wind material is expelled before it can be captured. Shock heating from wind–wind collisions keeps the gas hot and turbulent, further impeding any steady inflow. 

The IMBH’s radiative output would be pretty low with such a meager feeding rate. Assuming an advection-dominated (radiatively inefficient) accretion flow, we predict a quiescent X-ray luminosity of only $L_X \sim 10^{32}$~erg~s$^{-1}$, with short-lived bursts up to $\sim 10^{33}$~erg~s$^{-1}$ during quasi-periodic accretion spikes, many orders of magnitude below the Eddington luminosity for a $3\times10^4\,M_\odot$ black hole ($L_{\rm Edd} \sim 10^{42}\text{--}10^{43}$~erg~s$^{-1}$).

The efficiency of radiative cooling---which depends on the wind’s metal content---strongly influences the morphology of the flow. In the high-metallicity case ($Z=0.4$, representative of WC8–9 type WR stars; \citealt{2007ARA&A..45..177C}), cooling instabilities induce the formation of dense clumps via thin-shell shocks at wind–wind collision interfaces \citep[e.g.,][]{1992ApJ...386..265S,2025A&A...693A.180C}. Nevertheless, as noted above, these clumps do not substantially boost accretion. By contrast, in the lower-metallicity case, the winds stay hotter and smoother, and the added thermal pressure further inhibits gas from falling inward.

The specific angular momentum of the inflowing gas (Fig.~\ref{fig:angular_momentum}) gives further insight into the structure and behavior of the accretion flow. 
Because the wind speeds are so high relative to the orbital speeds of the stars, most gas is injected into the domain with super-Keplerian angular momentum pointing in quasi-random directions.  
As a result, only the gas that is either directed very close to the black hole or gas that loses its angular momentum through wind-wind shocks can accrete. 
In the inner domain, this leads to a low angular momentum (in terms of individual fluid elements and the net average) accretion flow without a coherent structure.

Magnetic fields present in the stellar winds introduce additional physical effects. In the model with moderately magnetized winds ($\beta_{\rm w} = 100$), we observe that magnetic pressure slows the inflow and enhances turbulence. Magnetic dissipation converts some field energy into heat, raising the gas pressure and counteracting a portion of the radiative cooling. Although including magnetic fields does not dramatically change the global accretion rate in our simulations, it alters the geometry of outflows. It suggests that even stronger wind magnetization could further inhibit accretion. $\beta$ profiles shown in Fig. \ref{fig:beta} demonstrate that the combined thermal and ram pressure of the gas dominates the balance – evidenced by $\beta \gg 1$ at virtually all radii – meaning that the overall inflow–outflow structure is shaped primarily by hydrodynamic forces. 

Beyond accretion, the IMBH’s presence could influence the cluster’s overall dynamics. The kinetic energy imparted by the stellar winds and their interaction with the IMBH drives turbulence, potentially limiting the retention of cold gas in the cluster. Similar effects have been seen in AGN feedback studies involving observations, models, or both (e.g., \citealt{2012ApJ...745L..34Z,2015ARA&A..53..115K,2016MNRAS.461.3724W}), where accretion flow-driven outflows can heat and expel surrounding material. This turbulence and gas expulsion may impact the long-term evolution of the cluster, affecting star formation and modifying the density profile. However, these effects are not solely dependent on the presence of IMBH. Future studies incorporating stellar evolution and external gas supply mechanisms will be crucial for understanding the full impact of IMBHs in such environments. 

In summary, wind-fed accretion onto an IMBH in a dense stellar cluster is inefficient. The high-velocity winds effectively starve the black hole, keeping its accretion luminosity low. Consequently, the IMBH would remain practically invisible against the bright background of colliding-wind X-ray emission from the cluster (as discussed later), even though its presence can still perturb the surrounding medium through gravity and outflows.

\subsection{Comparison with other wind-fed accretion regimes}
Wind-fed accretion has been most extensively studied in the context of Sgr A* – the supermassive black hole at the Galactic center – which feeds the winds of nearby massive stars (e.g., \citealt{Cuadra2008,Ressler_Quataert_Stone_2018,Ressler_Quataert_Stone_2019a,2020ApJ...888L...2C,2025A&A...693A.180C}). Our work is the first to explore a similar wind-fed accretion scenario for an IMBH embedded in a compact star cluster. Comparing the two cases provides insight into how differences in black hole mass, stellar cluster size, wind properties, and cooling processes affect accretion dynamics and efficiency.

The physical scale and environment are key distinctions between Sgr A* and our IMBH scenario. Sgr A* ($M_\bullet \approx 4\times10^6\,M_\odot$) lies in the Galactic nucleus, surrounded by a relatively extended cluster of OB-spectral type and WR stars spread over a region of a few tenths of a parsec. In contrast, our simulations consider a much smaller black hole ($M_\bullet = 3\times10^4\,M_\odot$) embedded in a very compact cluster (with radial extent on the order of  0.1~pc) of only a handful of WR stars. Therefore, the stellar density in the IMBH's immediate vicinity is much higher in our scenario. On the other hand, the ratio between the gravitational capture radius of the IMBH and the scale of the cluster is comparable to that of the SMBH in the Galactic center.

Sgr A* and the simulated IMBH exhibit extremely low accretion efficiencies in their wind-fed modes. HD simulations of Sgr A*'s environment have shown that only a small fraction of the stellar wind material (of order $10^{-3}$) captured the SMBH, while the rest is blown away by outflows \citep{Cuadra2008,Ressler_Quataert_Stone_2018}. In our IMBH case, the captured fraction is even smaller, on the order of $10^{-5}$. In absolute terms, the time-averaged accretion rate for Sgr A* is estimated to be $\sim 10^{-9}\text{--}10^{-7}\,M_\odot\,\mathrm{yr}^{-1}$, whereas in our simulations it is $\sim 10^{-10}\,M_\odot$~yr$^{-1}$. The outcome is qualitatively similar despite the two-order-of-magnitude difference in black hole mass. In both systems, most available wind mass is expelled, and only a small fraction feeds the black hole.

In both cases, the primary factor driving accretion inefficiency is the high wind velocity relative to the gravitational capture threshold. For an IMBH of mass $M_{\bullet} = 3 \times 10^4 M_{\odot}$, the Bondi radius, which represents the scale at which gravitational influence becomes significant, is
\begin{equation}
    R_{\rm B} = \frac{2 G M_{\bullet}}{c_{\rm s}^2},
\end{equation}
where $c_{\rm s} = \sqrt{\gamma {\rm k_B}T/\mu m_{\rm p}}$ is the sound speed of the gas. Assuming an adiabatic index of  $\gamma = 5/3$ and a resulting $\mathrm{k_B}T = 3/16\,m_{\rm p}v_{\rm w}^2$ for strongly shocked gas, we can rewrite the Bondi radius in terms of wind speed $v_{\rm w}$ as 
\begin{align}
    R_{\rm B} &\simeq \frac{32}{5} \frac{\mu G M_{\bullet}}{v_{\rm w}^2} \notag\\
    &\sim   8.27 \times 10^{-4} \left(\frac{M_{\bullet}}{3\times 10^4\,M_{\odot}} \right) \left(\frac{v_{\rm w}}{1000\,{\rm km\,s^{-1}}} \right)^{-1} \, \text{pc}.
\end{align}
 This is one order of magnitude smaller than the cluster scale and consistent with the inflow-outflow boundary in the velocity and accretion rate profiles shown in Fig. \ref{fig:profiles}. Most wind material escapes before experiencing significant gravitational influence. In comparison, even though Sgr A*'s deeper gravitational potential results in a larger $R_{\rm B} \sim 0.05$ pc, stars in the inner parsec of the Galactic center are generally further out ($\sim 0.2$ pc). The ratio between the Bondi radius and star distances in both systems is comparable, and the high wind speeds relative to the escape velocity lead to substantial mass loss in outflows rather than sustained accretion.

In both systems, radiative cooling can lead to the formation of clumpy, filamentary structures in the accretion flow, depending on the gas composition.
In the Galactic center, these clumps can be bound and can therefore collect into an accretion disk and substantially raise the long-term accretion rate \citep[e.g.,][]{2020ApJ...888L...2C,2025A&A...693A.180C}. 
In contrast, in our IMBH simulations, the dense clumps only momentarily increase local densities but are quickly disrupted or ejected, yielding no significant enhancement of accretion or disk formation. 

Finally, while both observations of Sgr A* and our IMBH model display relatively low quiescent X-ray luminosities ($L_X \sim 10^{33}$~erg~s$^{-1}$ for Sgr A* and $\sim 10^{32}$~erg~s$^{-1}$ for the IMBH), they also show evidence of intermittent, episodic accretion events. Sgr A* exhibits frequent, daily flares in X-rays and near-infrared domains as well as at longer wavelengths \citep{2001Natur.413...45B,2003A&A...407L..17P,2006ApJ...644..198Y,2010RvMP...82.3121G,2012A&A...537A..52E,2017FoPh...47..553E,2021ApJ...917...73W,vonFellenberg2025}, thought to be triggered by sudden increases in accretion or magnetic reconnection events in the near-horizon flow that accelerate electrons to nonthermal energies \citep[e.g.,][]{Dexter2020b,Porth2021,Ripperda2022}. Our simulations also observe short-lived accretion spikes when individual stars pass particularly close to each other and their wind collisions funnel gas inward. However, these bursts are much weaker relative to the "quiescent" X-ray emisison ($L_{\rm X} \lesssim 10^{32}\text{--}10^{33}$~erg~s$^{-1}$) than the observed Sgr A* flares (largest being $L_{\rm X} \lesssim 3-4\times10^{35}$~erg~s$^{-1}$, 160 times the quiescent level \citealt{2003A&A...407L..17P}), since our simulations do not capture the relevant scales or non-ideal physics important for such larger flares, we cannot rule them out. 
In fact, since both systems are in the radiatively inefficient regime with many qualitatively similar features in their larger-scale flow, we posit that high-magnitude flares originating from the near-horizon flow around the IMBH are likely (though on comparatively shorter timescales).

\subsection{Observability}
Our findings suggest that detecting an embedded IMBH in a WR stellar cluster poses a major observational challenge because the radiation from the accreting IMBH overwhelmed the emission from the surrounding stellar winds (as shown by our synthetic X-ray images in Fig.~\ref{fig:sx}). Throughout our simulations, most high-energy photons originate from wind–wind shock heating. This is true even when we account for the analytically predicted ADAF spectrum produced by the near-horizon flow (Figures~\ref{fig:sed} and \ref{fig:l_comparison}), which results in an X-ray luminosity that is a factor of about $10^2\text{--}10^3$ smaller than that resulting from the wind-wind collisions. In absolute terms, the IMBH's X-ray luminosity in our models (of order $10^{32}$~erg~s$^{-1}$) is extremely low – consistent with the non-detections and upper limits for IMBH candidates in similar environments \citep[e.g.,][]{2018ApJ...863....1C,2020MNRAS.492.2481W} – whereas the total diffuse X-ray luminosity of the cluster (from the stellar winds) is about $10^{34}\text{--}10^{35}$~erg~s$^{-1}$.

Therefore, the primary difficulty for observers is one of contrast: disentangling the IMBH's fainter radiative signal from the luminous stellar wind environment. In our scenario, the black hole would be effectively "lost in the glare" of the cluster's collective emission, emphasizing the need for sensitive, high-resolution observations to find an IMBH's steady glow amid the cluster's background. 
On the other hand, our simulations do not capture the physics and scales required for Sgr A*-like X-ray and infrared flares, meaning that there could still be strong transient signals. Evaluating this possibility is an important task for future work that includes such physics. 

Future X-ray and radio missions will provide enhanced sensitivity for detecting these objects. The proposed \textit{Lynx X-ray Observatory} \citep{2019JATIS...5b1001G} will feature an angular resolution of $\sim 0.5$ arcseconds, comparable to \textit{Chandra X-ray Observatory}, but with significantly improved sensitivity. This could allow for detecting faint, persistent X-ray sources in dense stellar environments where IMBHs are suspected to reside. Given that the primary observational challenge is distinguishing the IMBH's weak, point-like accretion signature from the surrounding diffuse stellar-wind emission, Lynx's superior imaging and spectroscopic capabilities could be critical for resolving such sources. Similarly, the \textit{Advanced X-ray Imaging Satellite (AXIS)} \citep{2023SPIE12678E..1ER} is designed to provide high angular resolution X-ray imaging with a resolution of about 1.5 arcseconds across a wide field of view. Scheduled for launch in 2032, \textit{AXIS} aims to study high-energy phenomena in the universe, including detecting faint X-ray sources in dense stellar environments. Its rapid response capabilities to transient events and improved sensitivity make it a promising tool for identifying IMBHs. Looking further ahead, X-ray interferometry holds exciting potential for the future of high-resolution X-ray astronomy. By overcoming the diffraction limits of traditional telescopes, space-based X-ray interferometers could achieve microarcsecond angular resolution, making it possible to directly image IMBH environments and distinguish them from surrounding stellar activity \citep{2021ExA....51.1081U}. Although still in an early conceptual phase, continued advancements in this technology could one day provide an unprecedented tool for confirming and characterizing IMBHs in dense stellar clusters.

Beyond X-rays, radio observations offer complementary constraints. IMBHs accreting in a radiatively inefficient mode may exhibit faint synchrotron radio emission, analogous to low-luminosity AGN \citep{2018MNRAS.478.2576M}. Upcoming facilities such as the \textit{Square Kilometer Array (SKA)} \citep{2009IEEEP..97.1482D} could help probe this parameter space, particularly in nearby dense stellar clusters where IMBH candidates are suspected to reside. Furthermore, from the SED shown in Fig.~\ref{fig:sed} and after applying the extinction curve from \citet{2011ApJ...737...73F}, IRS 13E's putative IMBH should exhibit radiative flux of 0.5 – 1.2 mJy at 19 $\mu$m. Assuming that the 25-80\% FIR variability of Sgr A*'s peak thermal emission \citep{2018ApJ...862..129V,Wielgus2022} is typical for low-luminosity accretion flows, this would result in $\sim$ 0.1 – 1 mJy mid-IR variability for the IRS 13E IMBH's peak thermal emission, which should be detectable with JWST/MIRI \citep{2023PASP..135d8003W}.

Given the limitations of the current observational missions, future searches for IMBHs in stellar clusters should focus on:
\begin{enumerate}
    \item High-sensitivity X-ray imaging with sub-arcsecond resolution to detect weak, persistent accretion signatures.
    \item Time-domain surveys to search for transient accretion events or variability in cluster-integrated X-ray emission.
    \item Deep radio observations to constrain possible synchrotron emission from IMBHs accreting in a radiatively inefficient mode.
    \item Multi-wavelength approaches that combine NIR, X-ray, and radio data to build a comprehensive observational profile.
\end{enumerate}

Overall, while our simulations reinforce that an IMBH in a compact stellar cluster would be extremely difficult to detect with current instruments, the next generation of observatories --- coupled with coordinated multi-wavelength efforts --- may provide the necessary sensitivity and resolution to finally uncover these elusive objects.

\section{Conclusions}
\label{sec_conclusions}

We conducted 3D MHD simulations to study wind-fed accretion onto an IMBH of mass \( M_{\bullet} = 3 \times 10^4 \, M_{\odot} \) embedded in a compact stellar cluster of six WR stars. Using the grid-based code \textsc{athena++}, we modeled the stellar winds (with wind velocities of \( v_w \sim\,1000\  \mathrm{km\,s^{-1}} \)) and their interactions with the IMBH’s gravitational potential while varying the cluster’s stellar distribution, wind magnetization, and chemical composition. 

From the simulations, we found that only $10^{-5}$ of the injected wind mass reaches the black hole. Wind-wind interactions dominate the dynamics, and the system is outflow-dominated. The gas follows a virial temperature profile, reaching \( T \sim 10^8 \) K near the boundary, while high wind velocities exceed the escape speed, ensuring most gas remains unbound. In higher-metallicity environments, enhanced radiative cooling forms cool, dense clumps, but their transient nature and high velocities prevent efficient accretion. The inflowing gas remains well below the Keplerian angular momentum, favoring a quasi-Bondi-like, radial accretion mode. Magnetic fields slightly suppress outflows but do not significantly affect the net accretion rate.

Observationally, the system’s X-ray emission dominated wind-wind shocks, with total luminosities of \( L_X \sim 10^{34} - 10^{35} \,\mathrm{erg\,s}^{-1} \), while the extrapolated near-horizon flow around the IMBH  is much fainter at \( L_X \sim 10^{32} \,\mathrm{erg\,s}^{-1} \). X-ray emission is wind-dominated, so IMBH signatures remain below current detection thresholds. Although weak, the X-ray emission of the near-horizon flow around the IMBH is highly variable with quasiperiodic flares recurring every $\sim 100-200$ years due to the close passages of the nearest star with the surrounding stars. 

These results suggest that IMBHs in dense stellar clusters grow slowly if fed solely by stellar winds. In this quiescent regime, they do not affect their surroundings significantly through feedback.  
Instead, wind-wind collisions due to the cluster stars influence the closest environment through kinetic feedback.
Feedback from the IMBH could be enhanced by, e.g., the tidal disruption of one of the cluster stars or the accretion of an additional external gas reservoir. Future work should investigate these possibilities and study the long-term cluster evolution to better understand the detectability of IMBHs and their role in stellar dynamics.

\begin{acknowledgements}
 We thank S. von Fellenberg for useful discussions.
 ML and TP have received support from the GA\v{C}R EXPRO grant no. GX21-13491X. MZ acknowledges the financial support of the GA\v{C}R Junior Star grant no. GM24-10599M. BR and SMR supported the Natural Sciences \& Engineering Research Council of Canada (NSERC). BR supported the Canadian Space Agency (23JWGO2A01), and by a grant from the Simons Foundation (MP-SCMPS-00001470). BR acknowledges a guest researcher position at the Flatiron Institute, supported by the Simons Foundation. F.P. gratefully acknowledges the Collaborative Research Center 1601 funded by the Deutsche Forschungsgemeinschaft (DFG, German Research Foundation) -- SFB 1601 [sub-project A3] -- 500700252.
 The computational resources and services used in this work were partially provided by facilities supported by the VSC (Flemish Supercomputer Center), funded by the Research Foundation Flanders (FWO) and the Flemish Government – department EWI, and by Compute Ontario and the Digital Research Alliance of Canada (alliancecan.ca) compute allocation rrg-ripperda.
\end{acknowledgements}

\bibliographystyle{aa}
\bibliography{references}

\begin{thebibliography}{154}
\expandafter\ifx\csname natexlab\endcsname\relax\def\natexlab#1{#1}\fi

\bibitem[{{Abbott} {et~al.}(2020){Abbott}, {Abbott}, {Abraham}, {Acernese},
  {Ackley}, {Adams}, {Adhikari}, {Adya}, {Affeldt}, {Agathos}, {Agatsuma},
  {Aggarwal}, {Aguiar}, {Aich}, {Aiello}, {Ain}, {Ajith}, {Akcay}, {Allen},
  {Allocca}, {Altin}, {Amato}, {Anand}, {Ananyeva}, {Anderson}, {Anderson},
  {Angelova}, {Ansoldi}, {Antier}, {Appert}, {Arai}, {Araya}, {Areeda},
  {Ar{\`e}ne}, {Arnaud}, {Aronson}, {Arun}, {Asali}, {Ascenzi}, {Ashton},
  {Aston}, {Astone}, {Aubin}, {Aufmuth}, {AultONeal}, {Austin}, {Avendano},
  {Babak}, {Bacon}, {Badaracco}, {Bader}, {Bae}, {Baer}, {Baird}, {Baldaccini},
  {Ballardin}, {Ballmer}, {Bals}, {Balsamo}, {Baltus}, {Banagiri}, {Bankar},
  {Bankar}, {Barayoga}, {Barbieri}, {Barish}, {Barker}, {Barkett}, {Barneo},
  {Barone}, {Barr}, {Barsotti}, {Barsuglia}, {Barta}, {Bartlett}, {Bartos},
  {Bassiri}, {Basti}, {Bawaj}, {Bayley}, {Bazzan}, {B{\'e}csy}, {Bejger},
  {Belahcene}, {Bell}, {Beniwal}, {Benjamin}, {Bentley}, {Bergamin}, {Berger},
  {Bergmann}, {Bernuzzi}, {Berry}, {Bersanetti}, {Bertolini}, {Betzwieser},
  {Bhandare}, {Bhandari}, {Bidler}, {Biggs}, {Bilenko}, {Billingsley},
  {Birney}, {Birnholtz}, {Biscans}, {Bischi}, {Biscoveanu}, {Bisht},
  {Bissenbayeva}, {Bitossi}, {Bizouard}, {Blackburn}, {Blackman}, {Blair},
  {Blair}, {Blair}, {Bobba}, {Bode}, {Boer}, {Boetzel}, {Bogaert}, {Bondu},
  {Bonilla}, {Bonnand}, {Booker}, {Boom}, {Bork}, {Boschi}, {Bose},
  {Bossilkov}, {Bosveld}, {Bouffanais}, {Bozzi}, {Bradaschia}, {Brady},
  {Bramley}, {Branchesi}, {Brau}, {Breschi}, {Briant}, {Briggs}, {Brighenti},
  {Brillet}, {Brinkmann}, {Brockill}, {Brooks}, {Brooks}, {Brown}, {Brunett},
  {Bruno}, {Bruntz}, {Buikema}, {Bulik}, {Bulten}, {Buonanno}, {Buscicchio},
  {Buskulic}, {Byer}, {Cabero}, {Cadonati}, {Cagnoli}, {Cahillane}, {Bustillo},
  {Callaghan}, {Callister}, {Calloni}, {Camp}, {Canepa}, {Cannon}, {Cao},
  {Cao}, {Carapella}, {Carbognani}, {Caride}, {Carney}, {Carullo}, {Diaz},
  {Casentini}, {Casta{\~n}eda}, {Caudill}, {Cavagli{\`a}}, {Cavalier},
  {Cavalieri}, {Cella}, {Cerd{\'a}-Dur{\'a}n}, {Cesarini}, {Chaibi},
  {Chakravarti}, {Chan}, {Chan}, {Chao}, {Charlton}, {Chase},
  {Chassande-Mottin}, {Chatterjee}, {Chaturvedi}, {Chatziioannou}, {Chen},
  {Chen}, \& {Chen}}]{2020ApJ...900L..13A}
{Abbott}, R., {Abbott}, T.~D., {Abraham}, S., {et~al.} 2020, \apjl, 900, L13

\bibitem[{{Akiba} {et~al.}(2025){Akiba}, {Naoz}, \&
  {Madigan}}]{2025ApJ...987L..27A}
{Akiba}, T., {Naoz}, S., \& {Madigan}, A.-M. 2025, \apjl, 987, L27

\bibitem[{{Arcodia} {et~al.}(2021){Arcodia}, {Merloni}, {Nandra}, {Buchner},
  {Salvato}, {Pasham}, {Remillard}, {Comparat}, {Lamer}, {Ponti}, {Malyali},
  {Wolf}, {Arzoumanian}, {Bogensberger}, {Buckley}, {Gendreau}, {Gromadzki},
  {Kara}, {Krumpe}, {Markwardt}, {Ramos-Ceja}, {Rau}, {Schramm}, \&
  {Schwope}}]{2021Natur.592..704A}
{Arcodia}, R., {Merloni}, A., {Nandra}, K., {et~al.} 2021, \nat, 592, 704

\bibitem[{{Ba{\~n}ares-Hern{\'a}ndez}
  {et~al.}(2025){Ba{\~n}ares-Hern{\'a}ndez}, {Calore}, {Martin Camalich}, \&
  {Read}}]{2025A&A...693A.104B}
{Ba{\~n}ares-Hern{\'a}ndez}, A., {Calore}, F., {Martin Camalich}, J., \&
  {Read}, J.~I. 2025, \aap, 693, A104

\bibitem[{{Baganoff} {et~al.}(2001){Baganoff}, {Bautz}, {Brandt}, {Chartas},
  {Feigelson}, {Garmire}, {Maeda}, {Morris}, {Ricker}, {Townsley}, \&
  {Walter}}]{2001Natur.413...45B}
{Baganoff}, F.~K., {Bautz}, M.~W., {Brandt}, W.~N., {et~al.} 2001, \nat, 413,
  45

\bibitem[{{Banerjee} \& {Kroupa}(2011)}]{2011ApJ...741L..12B}
{Banerjee}, S. \& {Kroupa}, P. 2011, \apjl, 741, L12

\bibitem[{{Bartko} {et~al.}(2009){Bartko}, {Martins}, {Fritz}, {Genzel},
  {Levin}, {Perets}, {Paumard}, {Nayakshin}, {Gerhard}, {Alexander},
  {Dodds-Eden}, {Eisenhauer}, {Gillessen}, {Mascetti}, {Ott}, {Perrin},
  {Pfuhl}, {Reid}, {Rouan}, {Sternberg}, \& {Trippe}}]{2009ApJ...697.1741B}
{Bartko}, H., {Martins}, F., {Fritz}, T.~K., {et~al.} 2009, \apj, 697, 1741

\bibitem[{{Begelman} {et~al.}(2006){Begelman}, {Volonteri}, \&
  {Rees}}]{2006MNRAS.370..289B}
{Begelman}, M.~C., {Volonteri}, M., \& {Rees}, M.~J. 2006, \mnras, 370, 289

\bibitem[{{Britzen} {et~al.}(2023){Britzen}, {Zaja{\v{c}}ek}, {Gopal-Krishna},
  {Fendt}, {Kun}, {Jaron}, {Sillanp{\"a}{\"a}}, \&
  {Eckart}}]{2023ApJ...951..106B}
{Britzen}, S., {Zaja{\v{c}}ek}, M., {Gopal-Krishna}, {et~al.} 2023, \apj, 951,
  106

\bibitem[{{Calder{\'o}n} {et~al.}(2025){Calder{\'o}n}, {Cuadra}, {Russell},
  {Burkert}, {Rosswog}, \& {Balakrishnan}}]{2025A&A...693A.180C}
{Calder{\'o}n}, D., {Cuadra}, J., {Russell}, C. M.~P., {et~al.} 2025, \aap,
  693, A180

\bibitem[{Calder\'on {et~al.}(2016)Calder\'on, Cuadra, Russell, \&
  Wang}]{Calderon2016}
Calder\'on, D., Cuadra, J., Russell, C. M.~P., \& Wang, Q.~D. 2016, MNRAS, 455,
  4388

\bibitem[{{Calder{\'o}n} {et~al.}(2020{\natexlab{a}}){Calder{\'o}n}, {Cuadra},
  {Schartmann}, {Burkert}, {Prieto}, \& {Russell}}]{2020MNRAS.493..447C}
{Calder{\'o}n}, D., {Cuadra}, J., {Schartmann}, M., {et~al.}
  2020{\natexlab{a}}, \mnras, 493, 447

\bibitem[{{Calder{\'o}n} {et~al.}(2020{\natexlab{b}}){Calder{\'o}n}, {Cuadra},
  {Schartmann}, {Burkert}, \& {Russell}}]{2020ApJ...888L...2C}
{Calder{\'o}n}, D., {Cuadra}, J., {Schartmann}, M., {Burkert}, A., \&
  {Russell}, C. M.~P. 2020{\natexlab{b}}, \apjl, 888, L2

\bibitem[{{Cao} {et~al.}(2025){Cao}, {Liu}, {Li}, {Chen}, \&
  {Wang}}]{2025ApJ...982L..37C}
{Cao}, C., {Liu}, F.~K., {Li}, S., {Chen}, X., \& {Wang}, K. 2025, \apjl, 982,
  L37

\bibitem[{{Chatterjee} {et~al.}(2002{\natexlab{a}}){Chatterjee}, {Hernquist},
  \& {Loeb}}]{2002PhRvL..88l1103C}
{Chatterjee}, P., {Hernquist}, L., \& {Loeb}, A. 2002{\natexlab{a}}, \prl, 88,
  121103

\bibitem[{{Chatterjee} {et~al.}(2002{\natexlab{b}}){Chatterjee}, {Hernquist},
  \& {Loeb}}]{2002ApJ...572..371C}
{Chatterjee}, P., {Hernquist}, L., \& {Loeb}, A. 2002{\natexlab{b}}, \apj, 572,
  371

\bibitem[{{Chilingarian} {et~al.}(2018){Chilingarian}, {Katkov}, {Zolotukhin},
  {Grishin}, {Beletsky}, {Boutsia}, \& {Osip}}]{2018ApJ...863....1C}
{Chilingarian}, I.~V., {Katkov}, I.~Y., {Zolotukhin}, I.~Y., {et~al.} 2018,
  \apj, 863, 1

\bibitem[{{Crowther}(2007)}]{2007ARA&A..45..177C}
{Crowther}, P.~A. 2007, \araa, 45, 177

\bibitem[{Cuadra {et~al.}(2008)Cuadra, Nayakshin, Martins, Genzel, Gillessen,
  \& Alexander}]{Cuadra2008}
Cuadra, J., Nayakshin, S., Martins, F., {et~al.} 2008, MNRAS, 383, 458

\bibitem[{{Cuadra} {et~al.}(2005){Cuadra}, {Nayakshin}, {Springel}, \& {Di
  Matteo}}]{Cuadra2005}
{Cuadra}, J., {Nayakshin}, S., {Springel}, V., \& {Di Matteo}, T. 2005, \mnras,
  360, L55

\bibitem[{{Cuadra} {et~al.}(2006){Cuadra}, {Nayakshin}, {Springel}, \& {Di
  Matteo}}]{Cuadra2006}
{Cuadra}, J., {Nayakshin}, S., {Springel}, V., \& {Di Matteo}, T. 2006, \mnras,
  366, 358

\bibitem[{{Cuadra} {et~al.}(2015){Cuadra}, {Nayakshin}, \& {Wang}}]{cuadra2015}
{Cuadra}, J., {Nayakshin}, S., \& {Wang}, Q.~D. 2015, \mnras, 450, 277

\bibitem[{{Dewdney} {et~al.}(2009){Dewdney}, {Hall}, {Schilizzi}, \&
  {Lazio}}]{2009IEEEP..97.1482D}
{Dewdney}, P.~E., {Hall}, P.~J., {Schilizzi}, R.~T., \& {Lazio}, T.~J.~L.~W.
  2009, IEEE Proceedings, 97, 1482

\bibitem[{{Dexter} {et~al.}(2020){Dexter}, {Tchekhovskoy},
  {Jim{\'e}nez-Rosales}, {Ressler}, {Baub{\"o}ck}, {Dallilar}, {de Zeeuw},
  {Eisenhauer}, {von Fellenberg}, {Gao}, {Genzel}, {Gillessen}, {Habibi},
  {Ott}, {Stadler}, {Straub}, \& {Widmann}}]{Dexter2020b}
{Dexter}, J., {Tchekhovskoy}, A., {Jim{\'e}nez-Rosales}, A., {et~al.} 2020,
  MNRAS, 497, 4999

\bibitem[{{Do} {et~al.}(2013){Do}, {Lu}, {Ghez}, {Morris}, {Yelda}, {Martinez},
  {Wright}, \& {Matthews}}]{2013ApJ...764..154D}
{Do}, T., {Lu}, J.~R., {Ghez}, A.~M., {et~al.} 2013, \apj, 764, 154

\bibitem[{{Eckart} {et~al.}(2012){Eckart}, {Garc{\'\i}a-Mar{\'\i}n}, {Vogel},
  {Teuben}, {Morris}, {Baganoff}, {Dexter}, {Sch{\"o}del}, {Witzel},
  {Valencia-S.}, {Karas}, {Kunneriath}, {Straubmeier}, {Moser}, {Sabha},
  {Buchholz}, {Zamaninasab}, {Mu{\v{z}}i{\'c}}, {Moultaka}, \&
  {Zensus}}]{2012A&A...537A..52E}
{Eckart}, A., {Garc{\'\i}a-Mar{\'\i}n}, M., {Vogel}, S.~N., {et~al.} 2012,
  \aap, 537, A52

\bibitem[{{Eckart} {et~al.}(2017){Eckart}, {H{\"u}ttemann}, {Kiefer},
  {Britzen}, {Zaja{\v{c}}ek}, {L{\"a}mmerzahl}, {St{\"o}ckler}, {Valencia-S},
  {Karas}, \& {Garc{\'\i}a-Mar{\'\i}n}}]{2017FoPh...47..553E}
{Eckart}, A., {H{\"u}ttemann}, A., {Kiefer}, C., {et~al.} 2017, Foundations of
  Physics, 47, 553

\bibitem[{{Eisenstein} \& {Loeb}(1995)}]{1995ApJ...443...11E}
{Eisenstein}, D.~J. \& {Loeb}, A. 1995, \apj, 443, 11

\bibitem[{{Evans} {et~al.}(2010){Evans}, {Primini}, {Glotfelty}, {Anderson},
  {Bonaventura}, {Chen}, {Davis}, {Doe}, {Evans}, {Fabbiano}, {Galle}, {Gibbs},
  {Grier}, {Hain}, {Hall}, {Harbo}, {He}, {Houck}, {Karovska}, {Kashyap},
  {Lauer}, {McCollough}, {McDowell}, {Miller}, {Mitschang}, {Morgan},
  {Mossman}, {Nichols}, {Nowak}, {Plummer}, {Refsdal}, {Rots}, {Siemiginowska},
  {Sundheim}, {Tibbetts}, {Van Stone}, {Winkelman}, \&
  {Zografou}}]{2010ApJS..189...37E}
{Evans}, I.~N., {Primini}, F.~A., {Glotfelty}, K.~J., {et~al.} 2010, \apjs,
  189, 37

\bibitem[{{Ferrarese} \& {Merritt}(2000)}]{2000ApJ...539L...9F}
{Ferrarese}, L. \& {Merritt}, D. 2000, \apjl, 539, L9

\bibitem[{{Filippenko} \& {Ho}(2003)}]{2003ApJ...588L..13F}
{Filippenko}, A.~V. \& {Ho}, L.~C. 2003, \apjl, 588, L13

\bibitem[{{Fragione} {et~al.}(2022){Fragione}, {Kocsis}, {Rasio}, \&
  {Silk}}]{2022ApJ...927..231F}
{Fragione}, G., {Kocsis}, B., {Rasio}, F.~A., \& {Silk}, J. 2022, \apj, 927,
  231

\bibitem[{{Freitag} {et~al.}(2006){Freitag}, {G{\"u}rkan}, \&
  {Rasio}}]{2006MNRAS.368..141F}
{Freitag}, M., {G{\"u}rkan}, M.~A., \& {Rasio}, F.~A. 2006, \mnras, 368, 141

\bibitem[{{Fritz} {et~al.}(2011){Fritz}, {Gillessen}, {Dodds-Eden}, {Lutz},
  {Genzel}, {Raab}, {Ott}, {Pfuhl}, {Eisenhauer}, \&
  {Yusef-Zadeh}}]{2011ApJ...737...73F}
{Fritz}, T.~K., {Gillessen}, S., {Dodds-Eden}, K., {et~al.} 2011, \apj, 737, 73

\bibitem[{{Fritz} {et~al.}(2010){Fritz}, {Gillessen}, {Dodds-Eden}, {Martins},
  {Bartko}, {Genzel}, {Paumard}, {Ott}, {Pfuhl}, {Trippe}, {Eisenhauer}, \&
  {Gratadour}}]{2010ApJ...721..395F}
{Fritz}, T.~K., {Gillessen}, S., {Dodds-Eden}, K., {et~al.} 2010, \apj, 721,
  395

\bibitem[{{Fruscione} {et~al.}(2006){Fruscione}, {McDowell}, {Allen},
  {Brickhouse}, {Burke}, {Davis}, {Durham}, {Elvis}, {Galle}, {Harris},
  {Huenemoerder}, {Houck}, {Ishibashi}, {Karovska}, {Nicastro}, {Noble},
  {Nowak}, {Primini}, {Siemiginowska}, {Smith}, \& {Wise}}]{Fruscione2006}
{Fruscione}, A., {McDowell}, J.~C., {Allen}, G.~E., {et~al.} 2006, in Society
  of Photo-Optical Instrumentation Engineers (SPIE) Conference Series, Vol.
  6270, Society of Photo-Optical Instrumentation Engineers (SPIE) Conference
  Series, ed. D.~R. {Silva} \& R.~E. {Doxsey}, 62701V

\bibitem[{{Fujii} {et~al.}(2024){Fujii}, {Wang}, {Tanikawa}, {Hirai}, \&
  {Saitoh}}]{2024Sci...384.1488F}
{Fujii}, M.~S., {Wang}, L., {Tanikawa}, A., {Hirai}, Y., \& {Saitoh}, T.~R.
  2024, Science, 384, 1488

\bibitem[{{Gaskin} {et~al.}(2019){Gaskin}, {Swartz}, {Vikhlinin}, {{\"O}zel},
  {Gelmis}, {Arenberg}, {Bandler}, {Bautz}, {Civitani}, {Dominguez}, {Eckart},
  {Falcone}, {Figueroa-Feliciano}, {Freeman}, {G{\"u}nther}, {Havey},
  {Heilmann}, {Kilaru}, {Kraft}, {McCarley}, {McEntaffer}, {Pareschi},
  {Purcell}, {Reid}, {Schattenburg}, {Schwartz}, {Schwartz}, {Tananbaum},
  {Tremblay}, {Zhang}, \& {Zuhone}}]{2019JATIS...5b1001G}
{Gaskin}, J.~A., {Swartz}, D.~A., {Vikhlinin}, A., {et~al.} 2019, JATIS, 5,
  021001

\bibitem[{{Gebhardt} {et~al.}(2000){Gebhardt}, {Bender}, {Bower}, {Dressler},
  {Faber}, {Filippenko}, {Green}, {Grillmair}, {Ho}, {Kormendy}, {Lauer},
  {Magorrian}, {Pinkney}, {Richstone}, \& {Tremaine}}]{2000ApJ...539L..13G}
{Gebhardt}, K., {Bender}, R., {Bower}, G., {et~al.} 2000, \apjl, 539, L13

\bibitem[{{Genzel} {et~al.}(2010){Genzel}, {Eisenhauer}, \&
  {Gillessen}}]{2010RvMP...82.3121G}
{Genzel}, R., {Eisenhauer}, F., \& {Gillessen}, S. 2010, Reviews of Modern
  Physics, 82, 3121

\bibitem[{{Giersz} {et~al.}(2015){Giersz}, {Leigh}, {Hypki}, {L{\"u}tzgendorf},
  \& {Askar}}]{2015MNRAS.454.3150G}
{Giersz}, M., {Leigh}, N., {Hypki}, A., {L{\"u}tzgendorf}, N., \& {Askar}, A.
  2015, \mnras, 454, 3150

\bibitem[{{Greene} {et~al.}(2020){Greene}, {Strader}, \&
  {Ho}}]{2020ARA&A..58..257G}
{Greene}, J.~E., {Strader}, J., \& {Ho}, L.~C. 2020, \araa, 58, 257

\bibitem[{{G{\"u}ltekin} {et~al.}(2004){G{\"u}ltekin}, {Miller}, \&
  {Hamilton}}]{2004ApJ...616..221G}
{G{\"u}ltekin}, K., {Miller}, M.~C., \& {Hamilton}, D.~P. 2004, \apj, 616, 221

\bibitem[{{Guo} {et~al.}(2023){Guo}, {Stone}, {Kim}, \& {Quataert}}]{Guo2023}
{Guo}, M., {Stone}, J.~M., {Kim}, C.-G., \& {Quataert}, E. 2023, \apj, 946, 26

\bibitem[{{Haas} {et~al.}(2025){Haas}, {Kroupa}, {{\v{S}}ubr}, \&
  {Singhal}}]{2025arXiv250315598H}
{Haas}, J., {Kroupa}, P., {{\v{S}}ubr}, L., \& {Singhal}, M. 2025, \aap, 695,
  L19

\bibitem[{{H{\"a}berle} {et~al.}(2024){H{\"a}berle}, {Neumayer}, {Seth},
  {Bellini}, {Libralato}, {Baumgardt}, {Whitaker}, {Dumont}, {Alfaro-Cuello},
  {Anderson}, {Clontz}, {Kacharov}, {Kamann}, {Feldmeier-Krause}, {Milone},
  {Nitschai}, {Pechetti}, \& {van de Ven}}]{2024Natur.631..285H}
{H{\"a}berle}, M., {Neumayer}, N., {Seth}, A., {et~al.} 2024, \nat, 631, 285

\bibitem[{{Hadrava} \& {{\v{C}}echura}(2012)}]{2012A&A...542A..42H}
{Hadrava}, P. \& {{\v{C}}echura}, J. 2012, \aap, 542, A42

\bibitem[{{Hansen} \& {Milosavljevi{\'c}}(2003)}]{2003ApJ...593L..77H}
{Hansen}, B. M.~S. \& {Milosavljevi{\'c}}, M. 2003, \apjl, 593, L77

\bibitem[{{Hendrix} {et~al.}(2016){Hendrix}, {Keppens}, {van Marle}, {Camps},
  {Baes}, \& {Meliani}}]{2016MNRAS.460.3975H}
{Hendrix}, T., {Keppens}, R., {van Marle}, A.~J., {et~al.} 2016, \mnras, 460,
  3975

\bibitem[{Herald {et~al.}(2001)Herald, Hillier, \&
  Schulte-Ladbeck}]{Herald2001}
Herald, J.~E., Hillier, D.~J., \& Schulte-Ladbeck, R.~E. 2001, ApJ, 548, 932

\bibitem[{{Hosseini} {et~al.}(2024){Hosseini}, {Eckart}, {Zaja{\v{c}}ek},
  {Britzen}, {Bhat}, \& {Karas}}]{2024ApJ...975..261H}
{Hosseini}, S.~E., {Eckart}, A., {Zaja{\v{c}}ek}, M., {et~al.} 2024, \apj, 975,
  261

\bibitem[{{Hu} {et~al.}(2024){Hu}, {Zaja{\v{c}}ek}, {Werner}, {Grossov{\'a}},
  {J{\'a}chym}, {Roberts}, {Ignesti}, {Kenney}, {Pl{\v{s}}ek}, {Breuer},
  {Shimwell}, {Tasse}, {Zhu}, \& {Wu}}]{2024MNRAS.527.1062H}
{Hu}, D., {Zaja{\v{c}}ek}, M., {Werner}, N., {et~al.} 2024, \mnras, 527, 1062

\bibitem[{{Jin} {et~al.}(2025){Jin}, {Li}, {Jiang}, {Dai}, {Cheng}, {Zhu},
  {Yang}, {Rau}, {Baldini}, {Wang}, {Zhou}, {Yuan}, {Zhang}, {Shu}, {Shen},
  {Wang}, {Wen}, {Wu}, {Wang}, {Thomsen}, {Zhang}, {Zhang}, {Coleiro},
  {Eyles-Ferris}, {Fang}, {Ho}, {Hu}, {Jin}, {Li}, {Liu}, {Liu}, {Liu}, {Liu},
  {Lu}, {Merloni}, {Qiao}, {Saxton}, {Soria}, {Wang}, {Xue}, {Yang}, {Zhang},
  {Zhang}, {Cai}, {Chen}, {Chen}, {Chen}, {Chen}, {Chen}, {Chen}, {Chen},
  {Cordier}, {Cui}, {Cui}, {Dai}, {Ding}, {Fan}, {Fan}, {Feng}, {Garcia},
  {Guan}, {Han}, {Hou}, {Hu}, {Huang}, {Huo}, {Jia}, {Jia}, {Jiang}, {Jin},
  {Kong}, {Kuulkers}, {Lei}, {Li}, {Li}, {Li}, {Li}, {Li}, {Li}, {Lian},
  {Ling}, {Liu}, {Liu}, {Liu}, {Liu}, {Liu}, {Lu}, {Luo}, {Ma}, {Mao}, {Mu},
  {Nandra}, {O'Brien}, {Pan}, {Pan}, {Qin}, {Rea}, {Sanders}, {Song}, {Sun},
  {Sun}, {Sun}, {Tan}, {Tang}, {Tao}, {Wang}, {Wang}, {Wang}, {Wang}, {Wang},
  {Wang}, {Wang}, {Wu}, {Wu}, {Xu}, {Xu}, {Xu}, {Xu}, {Xu}, {Xue}, {Xue},
  {Xue}, {Yan}, {Yang}, {Yang}, {Zhang}, {Zhang}, {Zhang}, {Zhang}, {Zhang},
  {Zhang}, {Zhang}, {Zhao}, {Zhao}, {Zhao}, {Zhao}, {Zheng}, {Zhu}, {Zhu},
  {Zhu}, \& {Zou}}]{2025arXiv250109580J}
{Jin}, C.~C., {Li}, D.~Y., {Jiang}, N., {et~al.} 2025, arXiv e-prints,
  arXiv:2501.09580

\bibitem[{Kaastra {et~al.}(1996)Kaastra, Mewe, Nieuwenhuijzen, Yamashita,
  Watanabe, {et~al.}}]{kaastra1996uv}
Kaastra, J., Mewe, R., Nieuwenhuijzen, H., {et~al.} 1996, UV and X-ray
  Spectroscopy of Astrophysical and Laboratory Plasmas

\bibitem[{{Kee} {et~al.}(2014){Kee}, {Owocki}, \& {ud-Doula}}]{Kee2014}
{Kee}, N.~D., {Owocki}, S., \& {ud-Doula}, A. 2014, \mnras, 438, 3557

\bibitem[{{King} \& {Pounds}(2015)}]{2015ARA&A..53..115K}
{King}, A. \& {Pounds}, K. 2015, \araa, 53, 115

\bibitem[{{K{\i}z{\i}ltan} {et~al.}(2017){K{\i}z{\i}ltan}, {Baumgardt}, \&
  {Loeb}}]{2017Natur.542..203K}
{K{\i}z{\i}ltan}, B., {Baumgardt}, H., \& {Loeb}, A. 2017, \nat, 542, 203

\bibitem[{{Krabbe} {et~al.}(1995){Krabbe}, {Genzel}, {Eckart}, {Najarro},
  {Lutz}, {Cameron}, {Kroker}, {Tacconi-Garman}, {Thatte}, {Weitzel},
  {Drapatz}, {Geballe}, {Sternberg}, \& {Kudritzki}}]{1995ApJ...447L..95K}
{Krabbe}, A., {Genzel}, R., {Eckart}, A., {et~al.} 1995, \apjl, 447, L95

\bibitem[{{Lamberts} {et~al.}(2012){Lamberts}, {Dubus}, {Lesur}, \&
  {Fromang}}]{2012A&A...546A..60L}
{Lamberts}, A., {Dubus}, G., {Lesur}, G., \& {Fromang}, S. 2012, \aap, 546, A60

\bibitem[{{Lamberts} {et~al.}(2011){Lamberts}, {Fromang}, \&
  {Dubus}}]{Lamberts2011}
{Lamberts}, A., {Fromang}, S., \& {Dubus}, G. 2011, \mnras, 418, 2618

\bibitem[{{Latif} {et~al.}(2013){Latif}, {Schleicher}, {Schmidt}, \&
  {Niemeyer}}]{2013MNRAS.436.2989L}
{Latif}, M.~A., {Schleicher}, D.~R.~G., {Schmidt}, W., \& {Niemeyer}, J.~C.
  2013, \mnras, 436, 2989

\bibitem[{{Lin} {et~al.}(2018){Lin}, {Strader}, {Carrasco}, {Page},
  {Romanowsky}, {Homan}, {Irwin}, {Remillard}, {Godet}, {Webb}, {Baumgardt},
  {Wijnands}, {Barret}, {Duc}, {Brodie}, \& {Gwyn}}]{2018NatAs...2..656L}
{Lin}, D., {Strader}, J., {Carrasco}, E.~R., {et~al.} 2018, NatAs, 2, 656

\bibitem[{Lodders(2003)}]{Lodders_2003}
Lodders, K. 2003, ApJ, 591, 1220

\bibitem[{{Loeb}(2004)}]{Loeb2004}
{Loeb}, A. 2004, \mnras, 350, 725

\bibitem[{{Loeb} \& {Rasio}(1994)}]{1994ApJ...432...52L}
{Loeb}, A. \& {Rasio}, F.~A. 1994, \apj, 432, 52

\bibitem[{{Lomb}(1976)}]{1976Ap&SS..39..447L}
{Lomb}, N.~R. 1976, \apss, 39, 447

\bibitem[{{Lu} {et~al.}(2009){Lu}, {Ghez}, {Hornstein}, {Morris}, {Becklin}, \&
  {Matthews}}]{Lu2009}
{Lu}, J.~R., {Ghez}, A.~M., {Hornstein}, S.~D., {et~al.} 2009, \apj, 690, 1463

\bibitem[{{Madau} \& {Rees}(2001)}]{2001ApJ...551L..27M}
{Madau}, P. \& {Rees}, M.~J. 2001, \apjl, 551, L27

\bibitem[{{Magorrian} {et~al.}(1998){Magorrian}, {Tremaine}, {Richstone},
  {Bender}, {Bower}, {Dressler}, {Faber}, {Gebhardt}, {Green}, {Grillmair},
  {Kormendy}, \& {Lauer}}]{1998AJ....115.2285M}
{Magorrian}, J., {Tremaine}, S., {Richstone}, D., {et~al.} 1998, \aj, 115, 2285

\bibitem[{{Mahadevan}(1997)}]{1997ApJ...477..585M}
{Mahadevan}, R. 1997, \apj, 477, 585

\bibitem[{{Maillard} {et~al.}(2004){Maillard}, {Paumard}, {Stolovy}, \&
  {Rigaut}}]{2004A&A...423..155M}
{Maillard}, J.~P., {Paumard}, T., {Stolovy}, S.~R., \& {Rigaut}, F. 2004, \aap,
  423, 155

\bibitem[{{Marconi} \& {Hunt}(2003)}]{2003ApJ...589L..21M}
{Marconi}, A. \& {Hunt}, L.~K. 2003, \apjl, 589, L21

\bibitem[{{Marrone} {et~al.}(2007){Marrone}, {Moran}, {Zhao}, \&
  {Rao}}]{2007ApJ...654L..57M}
{Marrone}, D.~P., {Moran}, J.~M., {Zhao}, J.-H., \& {Rao}, R. 2007, \apjl, 654,
  L57

\bibitem[{{Mart{\'\i}nez-Aldama} {et~al.}(2019){Mart{\'\i}nez-Aldama},
  {Czerny}, {Kawka}, {Karas}, {Panda}, {Zaja{\v{c}}ek}, \&
  {{\.Z}ycki}}]{2019ApJ...883..170M}
{Mart{\'\i}nez-Aldama}, M.~L., {Czerny}, B., {Kawka}, D., {et~al.} 2019, \apj,
  883, 170

\bibitem[{{Martins} {et~al.}(2007){Martins}, {Genzel}, {Hillier}, {Eisenhauer},
  {Paumard}, {Gillessen}, {Ott}, \& {Trippe}}]{Martins2007}
{Martins}, F., {Genzel}, R., {Hillier}, D.~J., {et~al.} 2007, \aap, 468, 233

\bibitem[{{Mezcua} {et~al.}(2018){Mezcua}, {Civano}, {Marchesi}, {Suh},
  {Fabbiano}, \& {Volonteri}}]{2018MNRAS.478.2576M}
{Mezcua}, M., {Civano}, F., {Marchesi}, S., {et~al.} 2018, \mnras, 478, 2576

\bibitem[{{Miller} \& {Hamilton}(2002)}]{2002MNRAS.330..232C}
{Miller}, M.~C. \& {Hamilton}, D.~P. 2002, \mnras, 330, 232

\bibitem[{{Miniutti} {et~al.}(2019){Miniutti}, {Saxton}, {Giustini},
  {Alexander}, {Fender}, {Heywood}, {Monageng}, {Coriat}, {Tzioumis}, {Read},
  {Knigge}, {Gandhi}, {Pretorius}, \&
  {Ag{\'\i}s-Gonz{\'a}lez}}]{2019Natur.573..381M}
{Miniutti}, G., {Saxton}, R.~D., {Giustini}, M., {et~al.} 2019, \nat, 573, 381

\bibitem[{{Moultaka} {et~al.}(2005){Moultaka}, {Eckart}, {Sch{\"o}del},
  {Viehmann}, \& {Najarro}}]{2005A&A...443..163M}
{Moultaka}, J., {Eckart}, A., {Sch{\"o}del}, R., {Viehmann}, T., \& {Najarro},
  F. 2005, \aap, 443, 163

\bibitem[{{Murchikova} {et~al.}(2019){Murchikova}, {Phinney}, {Pancoast}, \&
  {Blandford}}]{Murchikova2019}
{Murchikova}, E.~M., {Phinney}, E.~S., {Pancoast}, A., \& {Blandford}, R.~D.
  2019, \nat, 570, 83

\bibitem[{{Murchikova} {et~al.}(2022){Murchikova}, {White}, \&
  {Ressler}}]{2022ApJ...932L..21M}
{Murchikova}, L., {White}, C.~J., \& {Ressler}, S.~M. 2022, \apjl, 932, L21

\bibitem[{{Nandra} {et~al.}(1997){Nandra}, {George}, {Mushotzky}, {Turner}, \&
  {Yaqoob}}]{1997ApJ...476...70N}
{Nandra}, K., {George}, I.~M., {Mushotzky}, R.~F., {Turner}, T.~J., \&
  {Yaqoob}, T. 1997, \apj, 476, 70

\bibitem[{{Netzer}(2019)}]{2019MNRAS.488.5185N}
{Netzer}, H. 2019, \mnras, 488, 5185

\bibitem[{{Neumayer} {et~al.}(2020){Neumayer}, {Seth}, \&
  {B{\"o}ker}}]{2020A&ARv..28....4N}
{Neumayer}, N., {Seth}, A., \& {B{\"o}ker}, T. 2020, \aapr, 28, 4

\bibitem[{{Pandey} {et~al.}(2024){Pandey}, {Rakshit}, {Chand}, {Stalin}, {Cho},
  {Woo}, {Jalan}, {Mandal}, {Omar}, {Jose}, \& {Gupta}}]{2024ApJ...976..116P}
{Pandey}, S., {Rakshit}, S., {Chand}, K., {et~al.} 2024, \apj, 976, 116

\bibitem[{{Parkin} {et~al.}(2011){Parkin}, {Broos}, {Townsley}, {Pittard},
  {Moffat}, {Naz{\'e}}, {Rauw}, {Oskinova}, \& {Waldron}}]{2011ApJS..194....8P}
{Parkin}, E.~R., {Broos}, P.~S., {Townsley}, L.~K., {et~al.} 2011, \apjs, 194,
  8

\bibitem[{{Parkin} \& {Gosset}(2011)}]{2011A&A...530A.119P}
{Parkin}, E.~R. \& {Gosset}, E. 2011, \aap, 530, A119

\bibitem[{{Parkin} \& {Pittard}(2008)}]{2008MNRAS.388.1047P}
{Parkin}, E.~R. \& {Pittard}, J.~M. 2008, \mnras, 388, 1047

\bibitem[{{Pasham} {et~al.}(2024){Pasham}, {Tombesi}, {Sukov{\'a}},
  {Zaja{\v{c}}ek}, {Rakshit}, {Coughlin}, {Kosec}, {Karas}, {Masterson},
  {Mummery}, {Holoien}, {Guolo}, {Hinkle}, {Ripperda}, {Witzany}, {Shappee},
  {Kara}, {Horesh}, {van Velzen}, {Sfaradi}, {Kaplan}, {Burger}, {Murphy},
  {Remillard}, {Steiner}, {Wevers}, {Arcodia}, {Buchner}, {Merloni}, {Malyali},
  {Fabian}, {Fausnaugh}, {Daylan}, {Altamirano}, {Payne}, \&
  {Ferraraa}}]{2024SciA...10J8898P}
{Pasham}, D.~R., {Tombesi}, F., {Sukov{\'a}}, P., {et~al.} 2024, SciA, 10,
  eadj8898

\bibitem[{{Paumard} {et~al.}(2006){Paumard}, {Genzel}, {Martins}, {Nayakshin},
  {Beloborodov}, {Levin}, {Trippe}, {Eisenhauer}, {Ott}, {Gillessen}, {Abuter},
  {Cuadra}, {Alexander}, \& {Sternberg}}]{2006ApJ...643.1011P}
{Paumard}, T., {Genzel}, R., {Martins}, F., {et~al.} 2006, \apj, 643, 1011

\bibitem[{{Pavl{\'\i}k} {et~al.}(2024){Pavl{\'\i}k}, {Karas}, {Bhat},
  {Pei{\ss}ker}, \& {Eckart}}]{2024A&A...692A.104P}
{Pavl{\'\i}k}, V., {Karas}, V., {Bhat}, B., {Pei{\ss}ker}, F., \& {Eckart}, A.
  2024, \aap, 692, A104

\bibitem[{{Pei{\ss}ker} {et~al.}(2024){Pei{\ss}ker}, {Zaja{\v{c}}ek}, {Labaj},
  {Thomkins}, {Elbe}, {Eckart}, {Labadie}, {Karas}, {Sabha}, {Steiniger}, \&
  {Melamed}}]{2024ApJ...970...74P}
{Pei{\ss}ker}, F., {Zaja{\v{c}}ek}, M., {Labaj}, M., {et~al.} 2024, \apj, 970,
  74

\bibitem[{{Pei{\ss}ker} {et~al.}(2023){Pei{\ss}ker}, {Zaja{\v{c}}ek},
  {Thomkins}, {Eckart}, {Labadie}, {Karas}, {Sabha}, {Steiniger}, \&
  {Melamed}}]{2023ApJ...956...70P}
{Pei{\ss}ker}, F., {Zaja{\v{c}}ek}, M., {Thomkins}, L., {et~al.} 2023, \apj,
  956, 70

\bibitem[{{Pesce} {et~al.}(2021){Pesce}, {Palumbo}, {Narayan}, {Blackburn},
  {Doeleman}, {Johnson}, {Ma}, {Nagar}, {Natarajan}, \&
  {Ricarte}}]{2021ApJ...923..260P}
{Pesce}, D.~W., {Palumbo}, D. C.~M., {Narayan}, R., {et~al.} 2021, \apj, 923,
  260

\bibitem[{{Peterson} {et~al.}(2005){Peterson}, {Bentz}, {Desroches},
  {Filippenko}, {Ho}, {Kaspi}, {Laor}, {Maoz}, {Moran}, {Pogge}, \&
  {Quillen}}]{2005ApJ...632..799P}
{Peterson}, B.~M., {Bentz}, M.~C., {Desroches}, L.-B., {et~al.} 2005, \apj,
  632, 799

\bibitem[{{Pittard}(2009)}]{Pittard2009}
{Pittard}, J.~M. 2009, \mnras, 396, 1743

\bibitem[{{Porquet} {et~al.}(2003){Porquet}, {Predehl}, {Aschenbach}, {Grosso},
  {Goldwurm}, {Goldoni}, {Warwick}, \& {Decourchelle}}]{2003A&A...407L..17P}
{Porquet}, D., {Predehl}, P., {Aschenbach}, B., {et~al.} 2003, \aap, 407, L17

\bibitem[{{Portegies Zwart} \& {McMillan}(2002)}]{2002ApJ...576..899P}
{Portegies Zwart}, S.~F. \& {McMillan}, S. L.~W. 2002, \apj, 576, 899

\bibitem[{{Porth} {et~al.}(2021){Porth}, {Mizuno}, {Younsi}, \&
  {Fromm}}]{Porth2021}
{Porth}, O., {Mizuno}, Y., {Younsi}, Z., \& {Fromm}, C.~M. 2021, MNRAS, 502,
  2023

\bibitem[{{Poutanen} {et~al.}(2013){Poutanen}, {Fabrika}, {Valeev},
  {Sholukhova}, \& {Greiner}}]{2013MNRAS.432..506P}
{Poutanen}, J., {Fabrika}, S., {Valeev}, A.~F., {Sholukhova}, O., \& {Greiner},
  J. 2013, \mnras, 432, 506

\bibitem[{{Quataert}(2004)}]{2004ApJ...613..322Q}
{Quataert}, E. 2004, \apj, 613, 322

\bibitem[{{Ressler} {et~al.}(2024){Ressler}, {Combi}, {Li}, {Ripperda}, \&
  {Yang}}]{Ressler2024}
{Ressler}, S.~M., {Combi}, L., {Li}, X., {Ripperda}, B., \& {Yang}, H. 2024,
  \apj, 967, 70

\bibitem[{Ressler {et~al.}(2018)Ressler, Quataert, \&
  Stone}]{Ressler_Quataert_Stone_2018}
Ressler, S.~M., Quataert, E., \& Stone, J.~M. 2018, MNRAS, 478, 3544–3563

\bibitem[{{Ressler} {et~al.}(2019a){Ressler}, {Quataert}, \&
  {Stone}}]{Ressler_Quataert_Stone_2019a}
{Ressler}, S.~M., {Quataert}, E., \& {Stone}, J.~M. 2019a, \mnras, 482, L123

\bibitem[{Ressler {et~al.}(2019b)Ressler, Quataert, \&
  Stone}]{Ressler_Quataert_Stone_2019}
Ressler, S.~M., Quataert, E., \& Stone, J.~M. 2019b, MNRAS, 492, 3272–3293

\bibitem[{{Ressler} {et~al.}(2023){Ressler}, {White}, \&
  {Quataert}}]{2023MNRAS.521.4277R}
{Ressler}, S.~M., {White}, C.~J., \& {Quataert}, E. 2023, \mnras, 521, 4277

\bibitem[{{Ressler} {et~al.}(2020){Ressler}, {White}, {Quataert}, \&
  {Stone}}]{2020ApJ...896L...6R}
{Ressler}, S.~M., {White}, C.~J., {Quataert}, E., \& {Stone}, J.~M. 2020,
  \apjl, 896, L6

\bibitem[{{Reynolds} {et~al.}(2023){Reynolds}, {Kara}, {Mushotzky}, {Ptak},
  {Koss}, {Williams}, {Allen}, {Bauer}, {Bautz}, {Bogadhee}, {Burdge},
  {Cappelluti}, {Cenko}, {Chartas}, {Chan}, {Corrales}, {Daylan}, {Falcone},
  {Foord}, {Grant}, {Habouzit}, {Haggard}, {Herrmann}, {Hodges-Kluck},
  {Kargaltsev}, {King}, {Kounkel}, {Lopez}, {Marchesi}, {McDonald}, {Meyer},
  {Miller}, {Nynka}, {Okajima}, {Pacucci}, {Russell}, {Safi-Harb}, {Strassun},
  {Trindade Falc{\~a}o}, {Walker}, {Wilms}, {Yukita}, \&
  {Zhang}}]{2023SPIE12678E..1ER}
{Reynolds}, C.~S., {Kara}, E.~A., {Mushotzky}, R.~F., {et~al.} 2023, in SPIE,
  Vol. 12678, UV, X-Ray, and Gamma-Ray Space Instrumentation for Astronomy
  XXIII, ed. O.~H. {Siegmund} \& K.~{Hoadley}, 126781E

\bibitem[{{Richards} {et~al.}(2006){Richards}, {Lacy}, {Storrie-Lombardi},
  {Hall}, {Gallagher}, {Hines}, {Fan}, {Papovich}, {Vanden Berk}, {Trammell},
  {Schneider}, {Vestergaard}, {York}, {Jester}, {Anderson}, {Budav{\'a}ri}, \&
  {Szalay}}]{2006ApJS..166..470R}
{Richards}, G.~T., {Lacy}, M., {Storrie-Lombardi}, L.~J., {et~al.} 2006, \apjs,
  166, 470

\bibitem[{{Ripperda} {et~al.}(2022){Ripperda}, {Liska}, {Chatterjee}, {Musoke},
  {Philippov}, {Markoff}, {Tchekhovskoy}, \& {Younsi}}]{Ripperda2022}
{Ripperda}, B., {Liska}, M., {Chatterjee}, K., {et~al.} 2022, \apjl, 924, L32

\bibitem[{{Rockefeller} {et~al.}(2004){Rockefeller}, {Fryer}, {Melia}, \&
  {Warren}}]{Rockefeller2004}
{Rockefeller}, G., {Fryer}, C.~L., {Melia}, F., \& {Warren}, M.~S. 2004, \apj,
  604, 662

\bibitem[{{Rodr{\'\i}guez-Pascual} {et~al.}(1997){Rodr{\'\i}guez-Pascual},
  {Alloin}, {Clavel}, {Crenshaw}, {Horne}, {Kriss}, {Krolik}, {Malkan},
  {Netzer}, {O'Brien}, {Peterson}, {Reichert}, {Wamsteker}, {Alexander},
  {Barr}, {Blandford}, {Bregman}, {Carone}, {Clements}, {Courvoisier}, {De
  Robertis}, {Dietrich}, {Dottori}, {Edelson}, {Filippenko}, {Gaskell},
  {Huchra}, {Hutchings}, {Kollatschny}, {Koratkar}, {Korista}, {Laor},
  {MacAlpine}, {Martin}, {Maoz}, {McCollum}, {Morris}, {Perola}, {Pogge},
  {Ptak}, {Recondo-Gonz{\'a}lez}, {Rodr{\'\i}guez-Espinoza}, {Rokaki},
  {Santos-Lle{\'o}}, {Sekiguchi}, {Shull}, {Snijders}, {Sparke}, {Stirpe},
  {Stoner}, {Sun}, {Wagner}, {Wanders}, {Wilkes}, {Winge}, \&
  {Zheng}}]{1997ApJS..110....9R}
{Rodr{\'\i}guez-Pascual}, P.~M., {Alloin}, D., {Clavel}, J., {et~al.} 1997,
  \apjs, 110, 9

\bibitem[{{Rose} {et~al.}(2022){Rose}, {Naoz}, {Sari}, \&
  {Linial}}]{2022ApJ...929L..22R}
{Rose}, S.~C., {Naoz}, S., {Sari}, R., \& {Linial}, I. 2022, \apjl, 929, L22

\bibitem[{Russell {et~al.}(2017)Russell, Wang, \& Cuadra}]{Russell2017}
Russell, C. M.~P., Wang, Q.~D., \& Cuadra, J. 2017, MNRAS, 467, 1410

\bibitem[{{Scargle}(1982)}]{1982ApJ...263..835S}
{Scargle}, J.~D. 1982, \apj, 263, 835

\bibitem[{{Sch{\"o}del} {et~al.}(2005){Sch{\"o}del}, {Eckart}, {Iserlohe},
  {Genzel}, \& {Ott}}]{2005ApJ...625L.111S}
{Sch{\"o}del}, R., {Eckart}, A., {Iserlohe}, C., {Genzel}, R., \& {Ott}, T.
  2005, \apjl, 625, L111

\bibitem[{{Schure} {et~al.}(2009){Schure}, {Kosenko}, {Kaastra}, {Keppens}, \&
  {Vink}}]{2009A&A...508..751S}
{Schure}, K.~M., {Kosenko}, D., {Kaastra}, J.~S., {Keppens}, R., \& {Vink}, J.
  2009, \aap, 508, 751

\bibitem[{{Seepaul} {et~al.}(2022){Seepaul}, {Pacucci}, \&
  {Narayan}}]{2022MNRAS.515.2110S}
{Seepaul}, B.~S., {Pacucci}, F., \& {Narayan}, R. 2022, \mnras, 515, 2110

\bibitem[{{Shcherbakov} \& {Baganoff}(2010)}]{Shcherbakov2010}
{Shcherbakov}, R.~V. \& {Baganoff}, F.~K. 2010, ApJ, 716, 504

\bibitem[{{Solanki} {et~al.}(2023){Solanki}, {Ressler}, {Murchikova}, {Stone},
  \& {Morris}}]{2023ApJ...953...22S}
{Solanki}, S., {Ressler}, S.~M., {Murchikova}, L., {Stone}, J.~M., \& {Morris},
  M.~R. 2023, \apj, 953, 22

\bibitem[{{Stevens} {et~al.}(1992){Stevens}, {Blondin}, \&
  {Pollock}}]{1992ApJ...386..265S}
{Stevens}, I.~R., {Blondin}, J.~M., \& {Pollock}, A.~M.~T. 1992, \apj, 386, 265

\bibitem[{Stone {et~al.}(2020)Stone, Tomida, White, \& Felker}]{Stone_2020}
Stone, J.~M., Tomida, K., White, C.~J., \& Felker, K.~G. 2020, ApJS, 249, 4

\bibitem[{{Sukov{\'a}} {et~al.}(2024){Sukov{\'a}}, {Tombesi}, {Pasham},
  {Zaja{\v{c}}ek}, {Wevers}, {Ryu}, {Linial}, \&
  {Franchini}}]{2024arXiv241104592S}
{Sukov{\'a}}, P., {Tombesi}, F., {Pasham}, D.~R., {et~al.} 2024, arXiv
  e-prints, arXiv:2411.04592

\bibitem[{{Sukov{\'a}} {et~al.}(2021){Sukov{\'a}}, {Zaja{\v{c}}ek}, {Witzany},
  \& {Karas}}]{2021ApJ...917...43S}
{Sukov{\'a}}, P., {Zaja{\v{c}}ek}, M., {Witzany}, V., \& {Karas}, V. 2021,
  \apj, 917, 43

\bibitem[{{Takekawa} {et~al.}(2017){Takekawa}, {Oka}, {Iwata}, {Tokuyama}, \&
  {Nomura}}]{2017ApJ...843L..11T}
{Takekawa}, S., {Oka}, T., {Iwata}, Y., {Tokuyama}, S., \& {Nomura}, M. 2017,
  \apjl, 843, L11

\bibitem[{{The LIGO Scientific Collaboration} {et~al.}(2025){The LIGO
  Scientific Collaboration}, {the Virgo Collaboration}, {the KAGRA
  Collaboration}, {Abac}, {Abouelfettouh}, {Acernese}, {Ackley}, {Adamcewicz},
  {Adhicary}, {Adhikari}, {Adhikari}, {Adhikari}, {Adkins}, {Afroz}, {Agapito},
  {Agarwal}, {Agathos}, {Aggarwal}, {Aggarwal}, {Aguiar}, {Ahrend}, {Aiello},
  {Ain}, {Ajith}, {Akutsu}, {Albanesi}, {Ali}, {Al-Kershi}, {All{\'e}n{\'e}},
  {Allocca}, {Al-Shammari}, {Altin}, {Alvarez-Lopez}, {Amar}, {Amarasinghe},
  {Amato}, {Amicucci}, {Amra}, {Ananyeva}, {Anderson}, {Anderson}, {Andia},
  {Ando}, {Andr{\'e}s-Carcasona}, {Andri{\'c}}, {Anglin}, {Ansoldi}, {Antelis},
  {Antier}, {Aoumi}, {Appavuravther}, {Appert}, {Apple}, {Arai}, {Araujo
  Alvarez}, {Araya}, {Araya}, {Arca Sedda}, {Areeda}, {Aritomi}, {Armato},
  {Armstrong}, {Arnaud}, {Arogeti}, {Aronson}, {Arun}, {Ashton}, {Aso},
  {Asprea}, {Assiduo}, {Assis de Souza Melo}, {Aston}, {Astone}, {Attadio},
  {Aubin}, {AultONeal}, {Avallone}, {Avila}, {Babak}, {Badger}, {Bae},
  {Bagnasco}, {Baiotti}, {Bajpai}, {Baka}, {Baker}, {Baker}, {Baker}, {Baldi},
  {Baldicchi}, {Ball}, {Ballardin}, {Ballmer}, {Banagiri}, {Banerjee},
  {Bankar}, {Baptiste}, {Baral}, {Baratti}, {Barayoga}, {Barish}, {Barker},
  {Barman}, {Barneo}, {Barone}, {Barr}, {Barsotti}, {Barsuglia}, {Barta},
  {Bartoletti}, {Barton}, {Bartos}, {Basalaev}, {Bassiri}, {Basti}, {Bawaj},
  {Baxi}, {Bayley}, {Baylor}, {Baynard}, {Bazzan}, {Bedakihale}, {Beirnaert},
  {Bejger}, {Belardinelli}, {Bell}, {Bellie}, {Bellizzi}, {Benoit}, {Bentara},
  {Bentley}, {Ben Yaala}, {Bera}, {Bergamin}, {Berger}, {Bernuzzi}, {Beroiz},
  {Berry}, {Bersanetti}, {Bertheas}, {Bertolini}, {Betzwieser}, {Beveridge},
  {Bevilacqua}, {Bevins}, {Bhandare}, {Bhatt}, {Bhattacharjee},
  {Bhattacharyya}, {Bhaumik}, {Bhagwat}, {Biancalana}, {Bianchi}, {Bilenko},
  {Billingsley}, {Binetti}, {Bini}, {Binu}, {Biot}, {Birnholtz}, {Biscoveanu},
  {Bisht}, {Bitossi}, {Bizouard}, {Blaber}, {Blackburn}, {Blagg}, {Blair},
  {Blair}, {Bode}, {Boettner}, {Boileau}, {Boldrini}, {Bolingbroke},
  {Bolliand}, {Bonavena}, {Bondarescu}, {Bondu}, {Bonilla}, {Bonilla},
  {Bonino}, {Bonnand}, {Borchers}, {Borhanian}, {Boschi}, {Bose}, {Bossilkov},
  {Bothra}, {Boudon}, {Bourg}, {Bouyer}, {Boyle}, {Bozzi}, {Bradaschia},
  {Brady}, {Branch}, {Branchesi}, {Braun}, {Briant}, \&
  {Brillet}}]{2025arXiv250708219T}
{The LIGO Scientific Collaboration}, {the Virgo Collaboration}, {the KAGRA
  Collaboration}, {et~al.} 2025, arXiv e-prints, arXiv:2507.08219

\bibitem[{{Townsend}(2009)}]{2009ApJS..181..391T}
{Townsend}, R.~H.~D. 2009, \apjs, 181, 391

\bibitem[{{Tsuboi} {et~al.}(2019){Tsuboi}, {Kitamura}, {Tsutsumi}, {Miyawaki},
  {Miyoshi}, \& {Miyazaki}}]{2019PASJ...71..105T}
{Tsuboi}, M., {Kitamura}, Y., {Tsutsumi}, T., {et~al.} 2019, \pasj, 71, 105

\bibitem[{{Tsuboi} {et~al.}(2017){Tsuboi}, {Kitamura}, {Tsutsumi}, {Uehara},
  {Miyoshi}, {Miyawaki}, \& {Miyazaki}}]{2017ApJ...850L...5T}
{Tsuboi}, M., {Kitamura}, Y., {Tsutsumi}, T., {et~al.} 2017, \apjl, 850, L5

\bibitem[{{Uttley} {et~al.}(2021){Uttley}, {Hartog}, {Bambi}, {Barret},
  {Bianchi}, {Bursa}, {Cappi}, {Casella}, {Cash}, {Costantini}, {Dauser},
  {Trigo}, {Gendreau}, {Grinberg}, {Herder}, {Ingram}, {Kara}, {Markoff},
  {Mingo}, {Panessa}, {Poppenh{\"a}ger}, {R{\'o}{\.Z}a{\'n}ska}, {Svoboda},
  {Wijers}, {Willingale}, {Wilms}, \& {Wise}}]{2021ExA....51.1081U}
{Uttley}, P., {Hartog}, R.~d., {Bambi}, C., {et~al.} 2021, ExA, 51, 1081

\bibitem[{{van Marle} {et~al.}(2011){van Marle}, {Keppens}, \&
  {Meliani}}]{vanMarle2011}
{van Marle}, A.~J., {Keppens}, R., \& {Meliani}, Z. 2011, \aap, 527, A3

\bibitem[{{{\v{C}}echura} \& {Hadrava}(2015)}]{2015A&A...575A...5C}
{{\v{C}}echura}, J. \& {Hadrava}, P. 2015, \aap, 575, A5

\bibitem[{{{\v{C}}echura} {et~al.}(2015){{\v{C}}echura}, {Vrtilek}, \&
  {Hadrava}}]{2015MNRAS.450.2410C}
{{\v{C}}echura}, J., {Vrtilek}, S.~D., \& {Hadrava}, P. 2015, \mnras, 450, 2410

\bibitem[{{von Fellenberg} {et~al.}(2018){von Fellenberg}, {Gillessen},
  {Graci{\'a}-Carpio}, {Fritz}, {Dexter}, {Baub{\"o}ck}, {Ponti}, {Gao},
  {Habibi}, {Plewa}, {Pfuhl}, {Jimenez-Rosales}, {Waisberg}, {Widmann}, {Ott},
  {Eisenhauer}, \& {Genzel}}]{2018ApJ...862..129V}
{von Fellenberg}, S.~D., {Gillessen}, S., {Graci{\'a}-Carpio}, J., {et~al.}
  2018, \apj, 862, 129

\bibitem[{{von Fellenberg} {et~al.}(2022){von Fellenberg}, {Gillessen},
  {Stadler}, {Baub{\"o}ck}, {Genzel}, {de Zeeuw}, {Pfuhl}, {Amaro Seoane},
  {Drescher}, {Eisenhauer}, {Habibi}, {Ott}, {Widmann}, \&
  {Young}}]{2022ApJ...932L...6V}
{von Fellenberg}, S.~D., {Gillessen}, S., {Stadler}, J., {et~al.} 2022, \apjl,
  932, L6

\bibitem[{{von Fellenberg} {et~al.}(2023){von Fellenberg}, {Janssen},
  {Davelaar}, {Zaja{\v{c}}ek}, {Britzen}, {Falcke}, {K{\"o}rding}, \&
  {Ros}}]{2023A&A...672L...5V}
{von Fellenberg}, S.~D., {Janssen}, M., {Davelaar}, J., {et~al.} 2023, \aap,
  672, L5

\bibitem[{{von Fellenberg} {et~al.}(2025){von Fellenberg}, {Roychowdhury},
  {Michail}, {Sumners}, {Sanger-Johnson}, {Fazio}, {Haggard}, {Hora},
  {Philippov}, {Ripperda}, {Smith}, {Willner}, {Witzel}, {Zhang}, {Becklin},
  {Bower}, {Chandra}, {Do}, {Garcia Marin}, {Gurwell}, {Ford}, {Hada},
  {Markoff}, {Morris}, {Neilsen}, {Sabha}, \&
  {Seefeldt-Gail}}]{vonFellenberg2025}
{von Fellenberg}, S.~D., {Roychowdhury}, T., {Michail}, J.~M., {et~al.} 2025,
  \apjl, 979, L20

\bibitem[{{{\v{S}}ubr} {et~al.}(2019){{\v{S}}ubr}, {Fragione}, \&
  {Dabringhausen}}]{2019MNRAS.484.2974S}
{{\v{S}}ubr}, L., {Fragione}, G., \& {Dabringhausen}, J. 2019, \mnras, 484,
  2974

\bibitem[{{Wang} {et~al.}(2020){Wang}, {Li}, {Russell}, \&
  {Cuadra}}]{2020MNRAS.492.2481W}
{Wang}, Q.~D., {Li}, J., {Russell}, C. M.~P., \& {Cuadra}, J. 2020, \mnras,
  492, 2481

\bibitem[{{Wielgus} {et~al.}(2022){Wielgus}, {Marchili}, {Mart{\'\i}-Vidal},
  {Keating}, {Ramakrishnan}, {Tiede}, {Fomalont}, {Issaoun}, {Neilsen},
  {Nowak}, {Blackburn}, {Gammie}, {Goddi}, {Haggard}, {Lee}, {Moscibrodzka},
  {Tetarenko}, {Bower}, {Chan}, {Chatterjee}, {Chesler}, {Dexter}, {Doeleman},
  {Georgiev}, {Gurwell}, {Johnson}, {Marrone}, {Mus}, {Psaltis}, {Ripperda},
  {Witzel}, {Akiyama}, {Alberdi}, {Alef}, {Algaba}, {Anantua}, {Asada},
  {Azulay}, {Bach}, {Baczko}, {Ball}, {Balokovi{\'c}}, {Barrett},
  {Baub{\"o}ck}, {Benson}, {Bintley}, {Blundell}, {Boland}, {Bouman}, {Boyce},
  {Bremer}, {Brinkerink}, {Brissenden}, {Britzen}, {Broderick}, {Broguiere},
  {Bronzwaer}, {Bustamante}, {Byun}, {Carlstrom}, {Ceccobello}, {Chael},
  {Chatterjee}, {Chen}, {Chen}, {Cho}, {Christian}, {Conroy}, {Conway},
  {Cordes}, {Crawford}, {Crew}, {Cruz-Osorio}, {Cui}, {Davelaar}, {De
  Laurentis}, {Deane}, {Dempsey}, {Desvignes}, {Dhruv}, {Dzib}, {Eatough},
  {Emami}, {Falcke}, {Farah}, {Fish}, {Ford}, {Fraga-Encinas}, {Freeman},
  {Friberg}, {Fromm}, {Fuentes}, {Galison}, {Garc{\'\i}a}, {Gentaz}, {Gold},
  {G{\'o}mez-Ruiz}, {G{\'o}mez}, {Gu}, {Hada}, {Haworth}, {Hecht}, {Hesper},
  {Ho}, {Ho}, {Honma}, {Huang}, {Huang}, {Hughes}, {Ikeda}, {Impellizzeri},
  {Inoue}, {James}, {Jannuzi}, {Janssen}, {Jeter}, {Jiang},
  {Jim{\'e}nez-Rosales}, {Jorstad}, {Joshi}, {Jung}, {Karami}, {Karuppusamy},
  {Kawashima}, {Kettenis}, {Kim}, {Kim}, {Kim}, {Kim}, {Kino}, {Koay},
  {Kocherlakota}, {Kofuji}, {Koch}, {Koyama}, {Kramer}, {Kramer}, {Krichbaum},
  {Kuo}, {La Bella}, {Lauer}, {Lee}, {Leung}, {Levis}, {Li}, {Lico}, {Lindahl},
  {Lindqvist}, {Lisakov}, {Liu}, {Liu}, {Liuzzo}, {Lo}, {Lobanov}, {Loinard},
  {Lonsdale}, {Lu}, {Mao}, {Markoff}, {Marscher}, {Matsushita}, {Matthews},
  {Medeiros}, {Menten}, {Michalik}, {Mizuno}, {Mizuno}, {Moran}, {Moriyama},
  {M{\"u}ller}, {Musoke}, {Myserlis}, {Nadolski}, {Nagai}, {Nagar}, {Nakamura},
  {Narayan}, {Narayanan}, {Natarajan}, {Nathanail}, {Navarro Fuentes}, {Neri},
  {Ni}, {Noutsos}, {Oh}, {Okino}, {Olivares}, {Ortiz-Le{\'o}n}, {Oyama},
  {{\"O}zel}, {Palumbo}, {Paraschos}, {Park}, {Parsons}, {Patel}, {Pen},
  {Pesce}, {Pi{\'e}tu}, {Plambeck}, \& {PopStefanija}}]{Wielgus2022}
{Wielgus}, M., {Marchili}, N., {Mart{\'\i}-Vidal}, I., {et~al.} 2022, ApJL,
  930, L19

\bibitem[{{Witzel} {et~al.}(2021){Witzel}, {Martinez}, {Willner}, {Becklin},
  {Boyce}, {Do}, {Eckart}, {Fazio}, {Ghez}, {Gurwell}, {Haggard},
  {Herrero-Illana}, {Hora}, {Li}, {Liu}, {Marchili}, {Morris}, {Smith},
  {Subroweit}, \& {Zensus}}]{2021ApJ...917...73W}
{Witzel}, G., {Martinez}, G., {Willner}, S.~P., {et~al.} 2021, \apj, 917, 73

\bibitem[{{Woosley} \& {Heger}(2021)}]{2021ApJ...912L..31W}
{Woosley}, S.~E. \& {Heger}, A. 2021, \apjl, 912, L31

\bibitem[{{Wright} {et~al.}(2023){Wright}, {Rieke}, {Glasse}, {Ressler},
  {Garc{\'\i}a Mar{\'\i}n}, {Aguilar}, {Alberts}, {{\'A}lvarez-M{\'a}rquez},
  {Argyriou}, {Banks}, {Baudoz}, {Boccaletti}, {Bouchet}, {Bouwman}, {Brandl},
  {Breda}, {Bright}, {Cale}, {Colina}, {Cossou}, {Coulais}, {Cracraft}, {De
  Meester}, {Dicken}, {Engesser}, {Etxaluze}, {Fox}, {Friedman}, {Fu},
  {Gasman}, {G{\'a}sp{\'a}r}, {Gastaud}, {Geers}, {Glauser}, {Gordon},
  {Greene}, {Greve}, {Grundy}, {G{\"u}del}, {Guillard}, {Haderlein},
  {Hashimoto}, {Henning}, {Hines}, {Holler}, {Detre}, {Jahromi}, {James},
  {Jones}, {Justtanont}, {Kavanagh}, {Kendrew}, {Klaassen}, {Krause},
  {Labiano}, {Lagage}, {Lambros}, {Larson}, {Law}, {Lee}, {Libralato}, {Lorenzo
  Alverez}, {Meixner}, {Morrison}, {Mueller}, {Murray}, {Mycroft}, {Myers},
  {Nayak}, {Naylor}, {Nickson}, {Noriega-Crespo}, {{\"O}stlin}, {O'Sullivan},
  {Ottens}, {Patapis}, {Penanen}, {Pietraszkiewicz}, {Ray}, {Regan},
  {Roteliuk}, {Royer}, {Samara-Ratna}, {Samuelson}, {Sargent}, {Scheithauer},
  {Schneider}, {Schreiber}, {Shaughnessy}, {Sheehan}, {Shivaei}, {Sloan},
  {Tamas}, {Teague}, {Temim}, {Tikkanen}, {Tustain}, {van Dishoeck},
  {Vandenbussche}, {Weilert}, {Whitehouse}, \& {Wolff}}]{2023PASP..135d8003W}
{Wright}, G.~S., {Rieke}, G.~H., {Glasse}, A., {et~al.} 2023, \pasp, 135,
  048003

\bibitem[{{Wylezalek} \& {Zakamska}(2016)}]{2016MNRAS.461.3724W}
{Wylezalek}, D. \& {Zakamska}, N.~L. 2016, \mnras, 461, 3724

\bibitem[{{Xu}(2023)}]{SCAF}
{Xu}, W. 2023, ApJ, 954, 180

\bibitem[{{Yusef-Zadeh} {et~al.}(2006){Yusef-Zadeh}, {Bushouse}, {Dowell},
  {Wardle}, {Roberts}, {Heinke}, {Bower}, {Vila-Vilar{\'o}}, {Shapiro},
  {Goldwurm}, \& {B{\'e}langer}}]{2006ApJ...644..198Y}
{Yusef-Zadeh}, F., {Bushouse}, H., {Dowell}, C.~D., {et~al.} 2006, \apj, 644,
  198

\bibitem[{{Yusef-Zadeh} {et~al.}(2015){Yusef-Zadeh}, {Bushouse}, {Sch{\"o}del},
  {Wardle}, {Cotton}, {Roberts}, {Nogueras-Lara}, \&
  {Gallego-Cano}}]{YusefZadeh2015}
{Yusef-Zadeh}, F., {Bushouse}, H., {Sch{\"o}del}, R., {et~al.} 2015, \apj, 809,
  10

\bibitem[{{Zaja{\v{c}}ek} {et~al.}(2024){Zaja{\v{c}}ek}, {Sukov{\'a}}, {Karas},
  {Pasham}, {Tombesi}, {Kurf{\"u}rst}, {Best}, {Garland}, {Labaj}, \&
  {Pikhartov{\'a}}}]{2024arXiv241012090Z}
{Zaja{\v{c}}ek}, M., {Sukov{\'a}}, P., {Karas}, V., {et~al.} 2024, arXiv
  e-prints, arXiv:2410.12090

\bibitem[{{Zhang} {et~al.}(2025){Zhang}, {Shu}, {Sun}, {Shen}, {Dou}, {Jiang},
  \& {Wang}}]{2025NatAs...Zhang}
{Zhang}, W., {Shu}, X., {Sun}, L., {et~al.} 2025, Nature Astronomy, 9, 702

\bibitem[{{Zhu} {et~al.}(2020){Zhu}, {Li}, {Ciurlo}, {Morris}, {Zhang}, {Do},
  \& {Ghez}}]{2020ApJ...897..135Z}
{Zhu}, Z., {Li}, Z., {Ciurlo}, A., {et~al.} 2020, \apj, 897, 135

\bibitem[{{Zocchi} {et~al.}(2017){Zocchi}, {Gieles}, \&
  {H{\'e}nault-Brunet}}]{2017MNRAS.468.4429Z}
{Zocchi}, A., {Gieles}, M., \& {H{\'e}nault-Brunet}, V. 2017, \mnras, 468, 4429

\bibitem[{{Zocchi} {et~al.}(2019){Zocchi}, {Gieles}, \&
  {H{\'e}nault-Brunet}}]{2019MNRAS.482.4713Z}
{Zocchi}, A., {Gieles}, M., \& {H{\'e}nault-Brunet}, V. 2019, \mnras, 482, 4713

\bibitem[{{Zubovas} \& {King}(2012)}]{2012ApJ...745L..34Z}
{Zubovas}, K. \& {King}, A. 2012, \apjl, 745, L34

\bibitem[{{ZuHone} {et~al.}(2023){ZuHone}, {Vikhlinin}, {Tremblay}, {Randall},
  {Andrade-Santos}, \& {Bourdin}}]{2023ascl.soft01024Z}
{ZuHone}, J.~A., {Vikhlinin}, A., {Tremblay}, G.~R., {et~al.} 2023, {SOXS:
  Simulated Observations of X-ray Sources}, Astrophysics Source Code Library,
  record ascl:2301.024

\end{thebibliography}

\begin{appendix}

\section{Mutual distances of stars}
\label{appendix_mutual_dist}

In this Appendix, we calculate the temporal evolution of the separations of stars in our simulations, adopting Keplerian orbits based on the randomly generated orbital elements that yield the projected orbital radii consistent with observations of early-type stars in the Galactic center IRS 13 E association. We consider two orbital distributions of 6 early-type stars (see Table~\ref{tab:distances}): isotropic and disk-like. In Table~\ref{tab:rel_distances_isot}, we list the basic statistics of the separations among all the pairs of stars for the isotropic orbital distribution, including the minimum, mean, and maximum distances (expressed in astronomical units) as well as the characteristic recurrence timescale (expressed in years) based on the most prominent peaks in the Lomb-Scargle periodogram of the mutual separation. In Table~\ref{tab:rel_distances_disk}, we list the same information for the disk-like distribution.  

\begin{table}[h!]
\caption{Minimum, mean, and maximum distances of all the pairs of six stars (in AU) in the compact cluster and their recurrence timescales corresponding to the most prominent peaks in the Lomb-Scargle periodogram (in years) in the isotropic distribution around the IMBH.}
    \centering
    \begin{threeparttable}
    \begin{tabular}{lcc}
    \toprule
    \midrule
    stars  & min, mean, max [au] & recurrence timescale [yr] \\
    \midrule
    E1-E2       &    448, 2579, 3893   &      1429            \\
    E1-E4       &    1332, 2284, 3009  &      113          \\
    E1-E5.1     &    288, 2762, 4045  &      2667 (292)             \\
    E1-E5.2     &   117, 2838, 4223  &     280 (6645)         \\
    E1-E7       &   1620, 4155, 5962 &    408              \\
    E2-E4       &  884, 1872, 2562  &    212    \\
    E2-E5.1     &  189, 2404, 3622  &    3077 (221)  \\
    E2-E5.2     &  332, 2565, 3778 &    233 (1775)     \\
    E2-E7       &   2068, 4026, 5513   &    316 \\
    E4-E5.1    &   1060, 2000, 2738  &    108\\
    E4-E5.2    &   1215, 2158, 2892 &   189    \\
    E4-E7      &   2952, 3858, 4630  &   127     \\
    E5.1-E5.2  &   171, 2601, 3953 &    253 (4281)        \\
    E5.1-E7    &  1890, 4141, 5688  &   353        \\
    E5.2-E7    &  1742, 4196, 5845  &  385 (894)           \\
    \bottomrule
    \end{tabular}   
    \end{threeparttable}
     \label{tab:rel_distances_isot}
\end{table}

\begin{table}[h!]
\caption{Minimum, mean, and maximum distances of all the pairs of six stars (in AU) in the compact cluster and their recurrence timescales corresponding to the most prominent peaks in the Lomb-Scargle periodogram (in years) in the disk-like distribution around the IMBH.}
    \centering
    \begin{threeparttable}
    \begin{tabular}{lcc}
    \toprule
    \midrule
    stars  & min, mean, max [au] & recurrence timescale [yr] \\
    \midrule
    E1-E2       & 450, 2563, 3895     &   1412           \\
    E1-E4       & 1332, 2253, 3009      &  184          \\
    E1-E5.1     & 295, 2664, 4062     &  2609         \\
    E1-E5.2     & 130, 2706, 4226    & 6667        \\
    E1-E7       & 1619, 4106, 5961  &  1035          \\
    E2-E4       & 884, 1829, 2562   &  212       \\
    E2-E5.1     & 175, 2315, 3623   &  3077    \\
    E2-E5.2     & 330, 2490, 3778  &  1791      \\
    E2-E7       & 330, 2490, 3778     &  594   \\
    E4-E5.1    & 1059, 1992, 2737    & 198 \\
    E4-E5.2    & 1215, 2142, 2893  &  189  \\
    E4-E7      & 2951, 3837, 4629    &  156     \\
    E5.1-E5.2  & 217, 2509, 3950  &  4286          \\
    E5.1-E7    &  1891, 4035, 5690  &  741       \\
    E5.2-E7    & 1736, 4068, 5844  & 896         \\
    \bottomrule
    \end{tabular}      
    \end{threeparttable}
    \label{tab:rel_distances_disk}
\end{table}

\section{Periodicities and power spectral densities of the inflow rate}
\label{appendix_periodicities}

\begin{figure*}[h!]
    \includegraphics[width=\columnwidth]{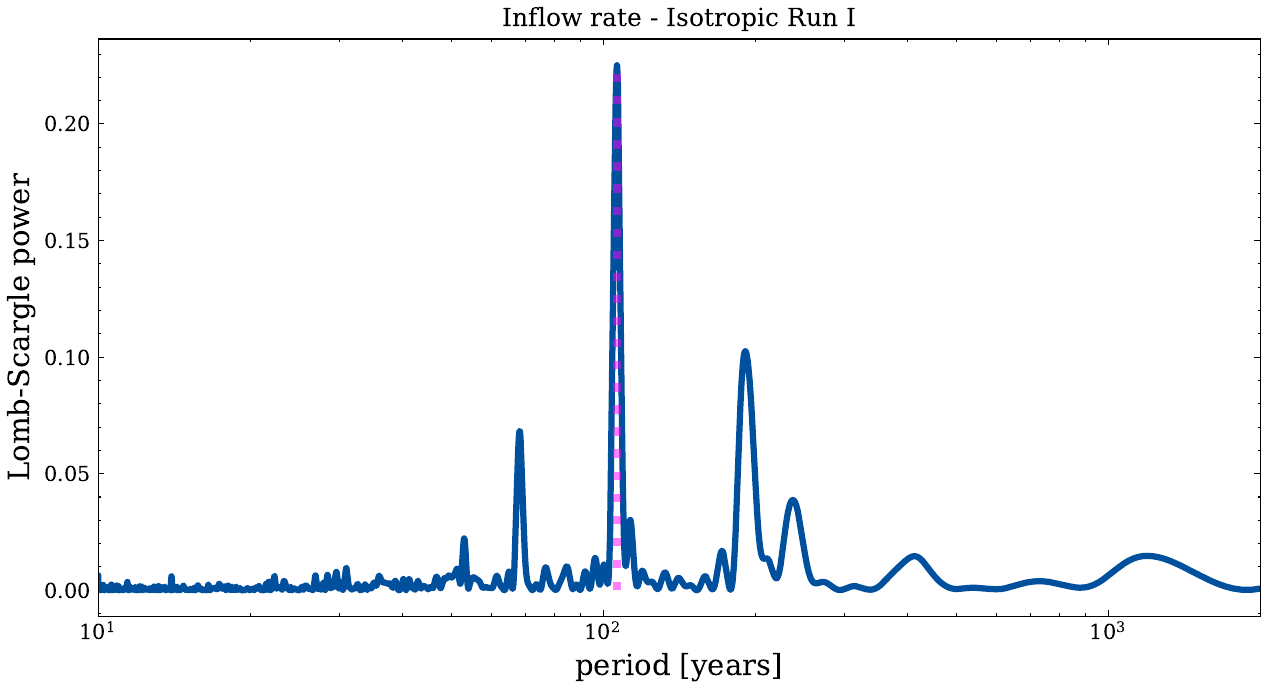}
    \includegraphics[width=\columnwidth]{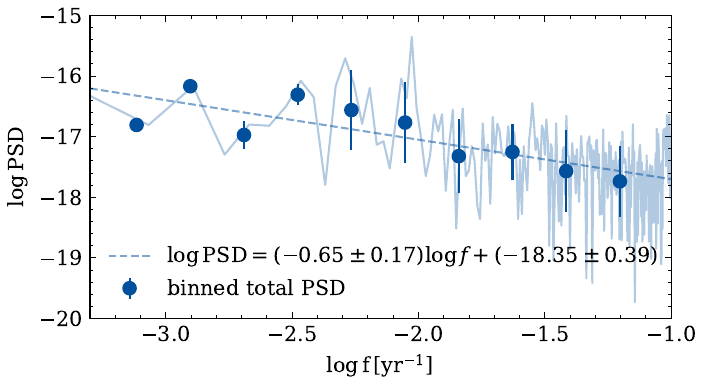}
    \caption{Periodicities and the power spectral density for the accretion rate in the isotropic Run I. \textit{Left:} The Lomb-Scargle periodogram (power vs. period in years) with the most prominent peak at $\sim 106$ years. Other prominent peaks are at $\sim 191$ and $\sim 68$ years. \textit{Right:} The PSD with the best-fit power-law function with the slope of $-0.65 \pm 0.17$.}
    \label{fig_mdot1_ls_psd}
\end{figure*}

\begin{figure*}[h!]
    \includegraphics[width=\columnwidth]{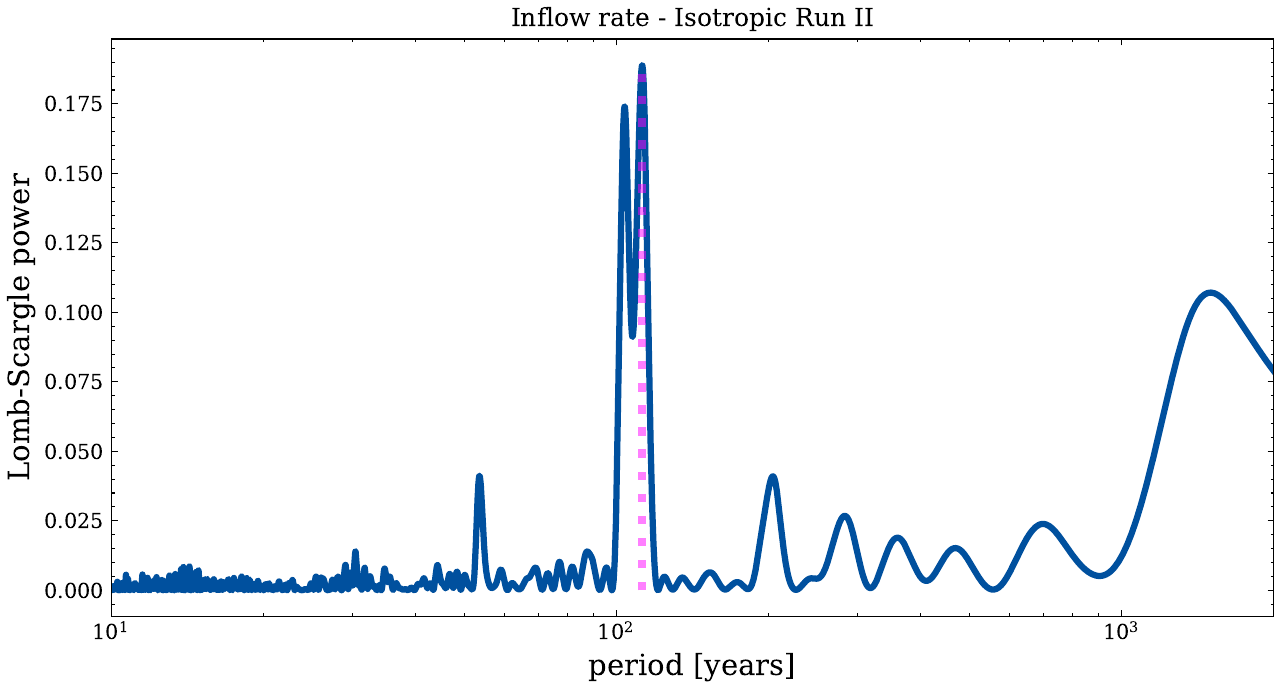}
    \includegraphics[width=\columnwidth]{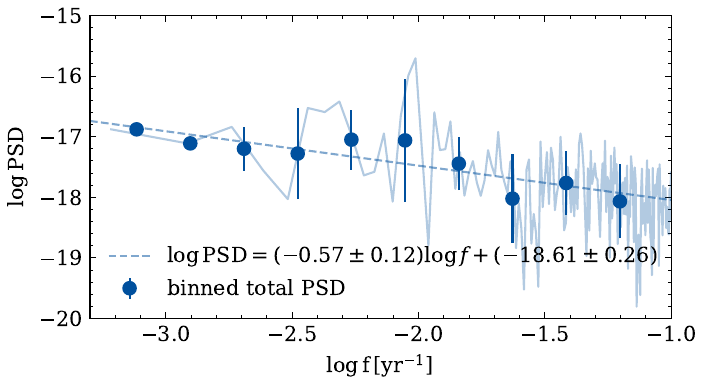}
    \caption{Periodicities and the power spectral density for the accretion rate in the isotropic Run II. \textit{Left:} The Lomb-Scargle periodogram (power vs. period in years) with the most prominent peak at $\sim 112$ years. Other prominent peaks are at $\sim 204$ and $\sim 54$ years. \textit{Right:} The PSD with the best-fit power-law function with the slope of $-0.57 \pm 0.12$.}
    \label{fig_mdot2_ls_psd}
\end{figure*}

\begin{figure*}[h!]
    \includegraphics[width=\columnwidth]{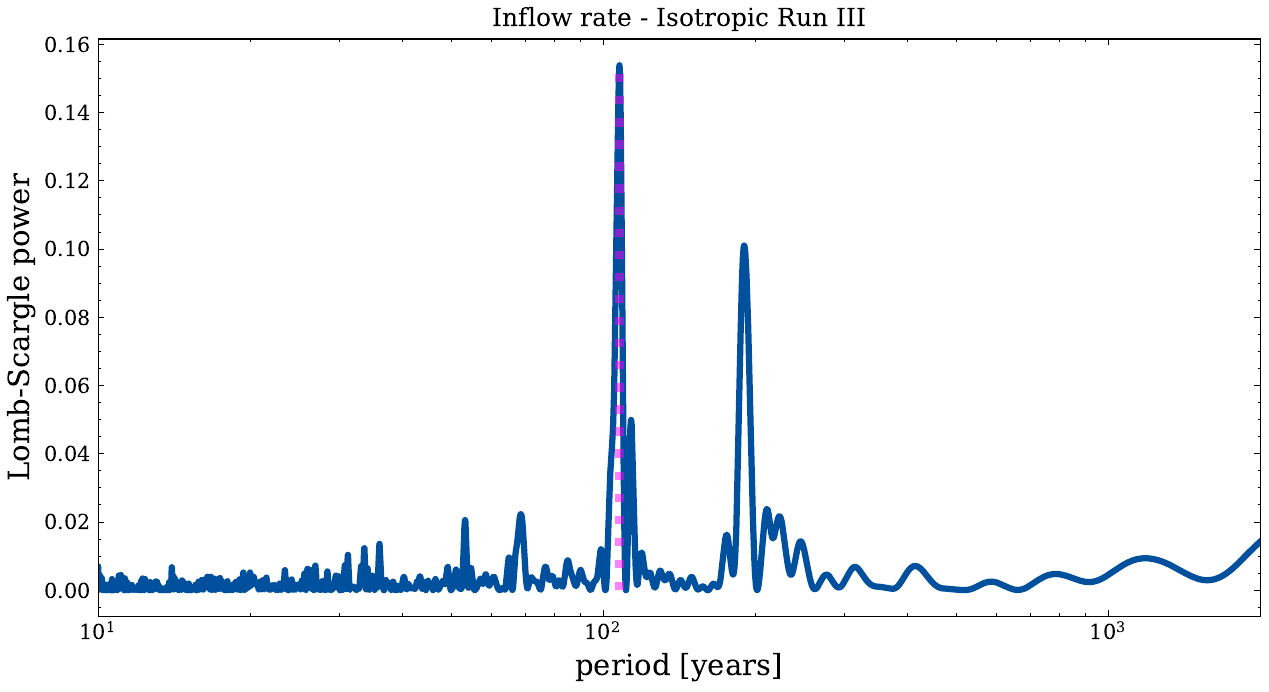}
    \includegraphics[width=\columnwidth]{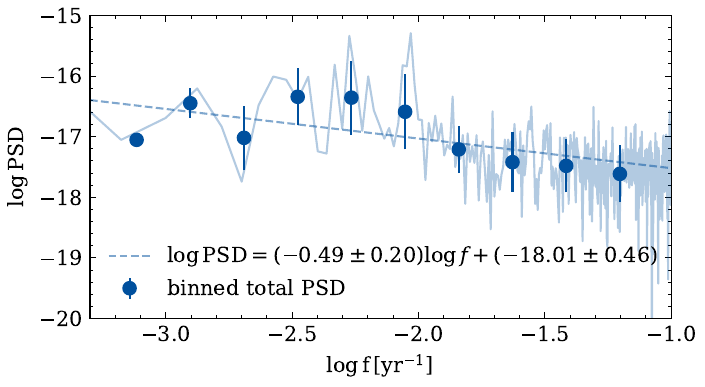}
    \caption{Periodicities and the power spectral density for the accretion rate in isotropic Run III. \textit{Left:} The Lomb-Scargle periodogram (power vs. period in years) with the most prominent peak at $\sim 108$ years. Other prominent peaks are at $\sim 190$, $\sim 69$, and $\sim 55$ years. \textit{Right:} The PSD with the best-fit power-law function with the slope of $-0.49 \pm 0.20$.}
    \label{fig_mdot3_ls_psd}
\end{figure*}

\begin{figure*}[h!]
    \includegraphics[width=\columnwidth]{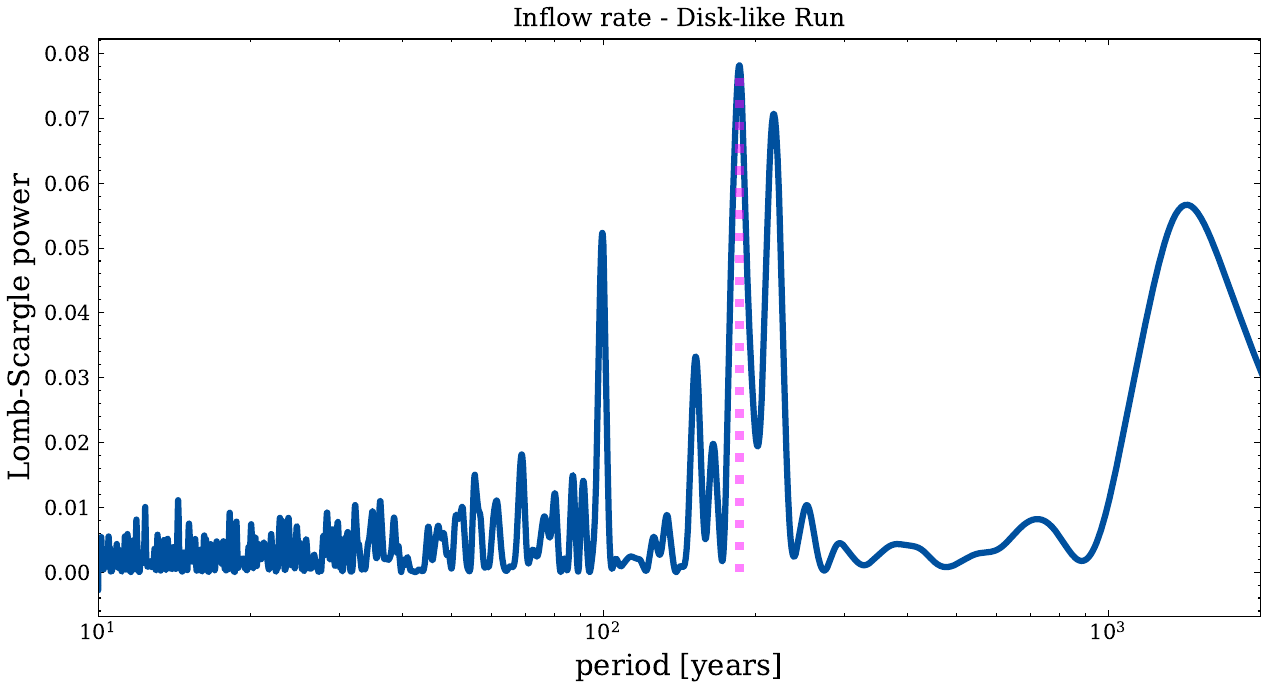}
    \includegraphics[width=\columnwidth]{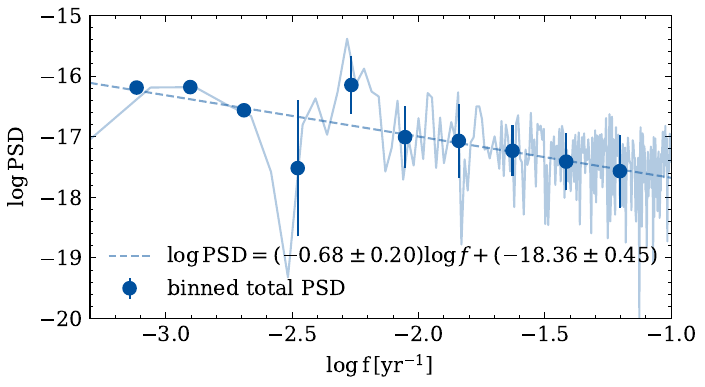}
    \caption{Periodicities and the power spectral density for the accretion rate in the disk-like Run. \textit{Left:} The Lomb-Scargle periodogram (power vs. period in years) with the most prominent peak at $\sim 186$ years. Other prominent peaks are at $\sim 217$, and $\sim 100$ years. \textit{Right:} The PSD with the best-fit power-law function with the slope of $-0.68 \pm 0.20$.}
    \label{fig_mdot_d_ls_psd}
\end{figure*}

In this Appendix, we include the most prominent periodicities present in the accretion rate of different isotropic and disk-like models. In addition, we derive power spectral densities (PSDs) as a function of the frequency (expressed in ${\rm yr^{-1}}$). Subsequently, we derive binned PSDs, to which we fit a power-law function $\text{PSD}\propto f^{\beta}$ to infer the power-law slope of the PSDs. In Figs.~\ref{fig_mdot1_ls_psd}, \ref{fig_mdot2_ls_psd}, and \ref{fig_mdot3_ls_psd} we show Lomb-Scargle periodograms with the most prominent periodicity peaks (left panels) and PSDs with power-law fits (right panels) for isotropic cluster runs (I, II, and III). In Fig.~\ref{fig_mdot_d_ls_psd} we display the Lomb-Scargle periodogram with the most prominent peak (left panel) and the PSD with the best-fit power-law function (right panel) for the disk-like cluster run.

\section{Comparison of the IRS 13E X-ray surface-brightness profile with the mock cluster}
\label{appendix_surf_brightness}

\begin{figure*}[h!]
    \centering
    \includegraphics[width=\columnwidth]{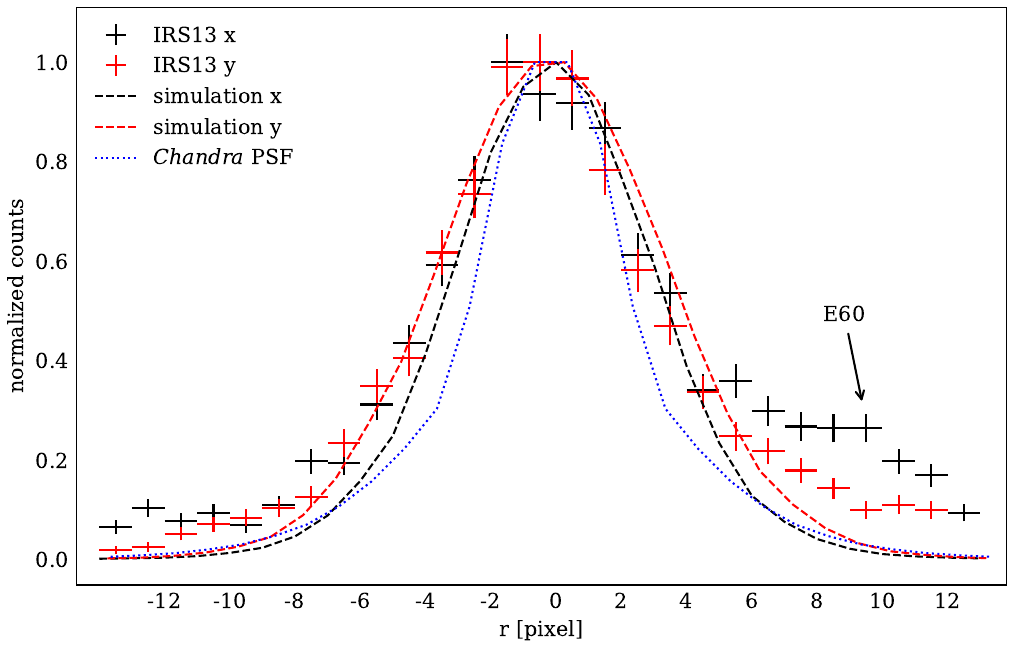}
    \caption{Comparison of the observed IRS 13E X-ray surface-brightness profiles (black and red points with errorbars in normalized units in two perpendicular directions $x$ and $y$ expressed in pixels) with the corresponding ones inferred from simulations presented in this study (black and red dashed lines). A dotted blue line depicts the Chandra PSF. The more extended western wing of IRS 13E can be attributed to the E60 source. One pixel is equivalent to $\sim 0.125$ arcsec or $\sim 5\times 10^{-3}\,{\rm pc}$ at the Galactic center.}
    \label{fig_Xray_profiles}
\end{figure*}

In this Appendix, we compare the IRS 13E X-ray surface-brightness profile with the one inferred from simulations. In particular, we consider the profiles in two directions: East-West ($x$) and North-South ($y$). In addition, we extract the profile of the Chandra point-spread function (PSF). In Fig.~\ref{fig_Xray_profiles}, we compare the $x-y$ surface-brightness profiles for the observed Chandra image as well as for the simulated one with the Chandra PSF. We see that the observed profile is comparable to the PSF close to the peak, while it is slightly extended towards the tails. The profile constructed from the simulated X-ray image of the compact cluster is consistent with the observed one, except for the wing to the west, where the observed profile is significantly more extended. This can be attributed to the surrounding diffuse X-ray emission and/or individual X-ray sources. This particular ``bump'' in the western tail (positive $x$-direction) was previously attributed to the source E60 \citep{2020MNRAS.492.2481W,2020ApJ...897..135Z}.

In terms of the image asymmetry, we compare the observed IRS 13E image with the simulated one using the asymmetry index defined as follows,
\begin{equation}
    A=\frac{\sum_{i}|I_i-I_{{\rm rot},i}|}{2\sum_i |I_{i}|}
    \label{eq_asymmetry_index}
\end{equation}
where we subtract the image rotated by $180^{\circ}$ from the actual image. The observed image has $A=0.21$ while the simulated one has $A=0.13$, which is consistent with the simulated image being more symmetric as is also visible in Fig.~\ref{fig_Xray_profiles}. As we already mentioned, in the observed image, there is a contribution at least from the source E60, which is not a part of the IRS 13E association and as such is not included in the simulated cluster. 

\end{appendix}

\end{document}